\documentclass[fleqn,usenatbib]{mnras}

\usepackage{newtxtext,newtxmath}

\usepackage[T1]{fontenc}

\usepackage{graphicx}	
\usepackage{amsmath}	

\title[The SMC West Halo in 8D]{The VISCACHA survey -- IV. The SMC West Halo in 8D}

\author[B. Dias et al.]
{B. Dias$^{1}$\thanks{E-mail: bdiasm@academicos.uta.cl},
M. C. Parisi$^{2,3}$,
M. Angelo$^{4}$,
F. Maia$^{5}$,
R. A. P. Oliveira$^{6}$,
S. O. Souza$^{6}$,
\newauthor
L. O. Kerber$^{7}$,
J. F. C. Santos Jr.$^{8}$,
A. P\'erez-Villegas$^{9}$,
D. Sanmartim$^{10}$, 
B. Quint$^{11}$, 
\newauthor
L. Fraga$^{12}$, 
B. Barbuy$^{6}$,
E. Bica$^{13}$,
O. J. Katime Santrich$^{7}$,
J. A. Hernandez-Jimenez$^{14}$,
\newauthor
D. Geisler$^{15,16,17}$,
D. Minniti$^{18}$,
B. J. De Bórtoli$^{19,20}$,
L. P. Bassino$^{19,20}$,
J. P. Rocha$^{7}$
\\
$^{1}$Instituto de Alta Investigaci\'on, Sede Esmeralda, Universidad de Tarapac\'a, Av. Luis Emilio Recabarren 2477, Iquique, Chile\\
$^{2}$Observatorio Astron\'omico, Universidad Nacional de C\'ordoba, Laprida 854, X5000BGR, C\'ordoba, Argentina.\\
$^{3}$Instituto de Astronom{\'\i}a Te\'orica y Experimental (CONICET-UNC), Laprida 854, X5000BGR, C\'ordoba, Argentina.\\
 $^{4}$Centro Federal de Educa\c c\~ao Tecnol\'ogica de Minas Gerais, Av. Monsenhor Luiz de Gonzaga, 103, 37250-000 Nepomuceno, MG, Brazil\\
$^{5}$Instituto de F\'isica, Universidade Federal do Rio de Janeiro, 21941-972, Rio de Janeiro, RJ, Brazil\\
$^{6}$Universidade de S\~ao Paulo, IAG, Rua do Mat\~ao 1226, Cidade Universit\'aria, S\~ao Paulo 05508-900, Brazil\\
$^{7}$Departamento de Ci\^encias Exatas e Tecnol\'ogicas, UESC, Rodovia Jorge Amado km 16, 45662-900, Brazil\\
$^{8}$Departamento de F\'isica, ICEx - UFMG, Av. Ant\^onio Carlos 6627, Belo Horizonte 31270-901, Brazil\\
$^{9}$Instituto de Astronomía, Universidad Nacional Autonóma de México, Apartado Postal 106, C.P. 22800, Ensenada, Baja California, Mexico\\
$^{10}$Carnegie Observatories, Las Campanas Observatory, Casilla 601, La Serena, Chile\\
$^{11}$Rubin Observatory Project Office, 950 N. Cherry Ave., Tucson, AZ 85719, USA\\
$^{12}$Laboratório Nacional de Astrofísica, Rua Estados Unidos 154, Itajubá 37504-364, Brazil\\
$^{13}$Departamento de Astronomia, IF - UFRGS, Av. Bento Gon\c calves 9500, Porto Alegre, RS, 91501-970, Brazil\\
$^{14}$IP\&D-Universidade do Vale do Paraíba-Av. Shishima Hifumi, 2911-12244-000 - São José dos Campos - SP-Brazil\\
$^{15}$Departamento de Física y Astronomía, Universidad de La Serena, Avenida Juan Cisternas 1200, La Serena, Chile\\
$^{16}$Instituto de Investigación Multidisciplinario en Ciencia y Tecnología, Universidad de La Serena Benavente 980, La Serena, Chile\\
$^{17}$Departmento de Astronomía, Universidad de Concepción, Casilla 160-C, Concepción, Chile\\
$^{18}$Departamento de Ciencias F\'isicas, Universidad Andres Bello, Fernandez Concha 700, Las Condes, Santiago, Chile\\
$^{19}$Facultad de Ciencias Astronómicas y Geofísicas de la Universidad Nacional de La Plata, and Instituto de Astrofísica de La Plata \\(CCT La Plata - CONICET, UNLP), Paseo del Bosque S/N, B1900FWA La Plata, Argentina\\
$^{20}$Consejo Nacional de Investigaciones Científicas y Técnicas, Godoy Cruz 2290, C1425FQB, Ciudad Autónoma de Buenos Aires, Argentina
}

\date{Accepted 2022 January 25. Received 2022 January 24; in original form 2021 December 13}

\pubyear{2022}

\begin{document}
\label{firstpage}
\pagerange{\pageref{firstpage}--\pageref{lastpage}}
\maketitle

\begin{abstract}
The structure of the Small Magellanic Cloud (SMC) is very complex, in particular in the periphery that suffers more from the interactions with the Large Magellanic Cloud (LMC). A wealth of observational evidence has been accumulated revealing tidal tails and bridges made up of gas, stars and star clusters. Nevertheless, a full picture of the SMC outskirts is only recently starting to emerge with a 6D phase-space map plus age and metallicity using star clusters as tracers. In this work, we continue our analysis of another outer region of the SMC, the so-called West Halo, and combined it with the previously analysed Northern Bridge. We use both structures to define the Bridge and Counter-bridge trailing and leading tidal tails. These two structures are moving away from each other, roughly in the SMC-LMC direction. The West Halo form a ring around the SMC inner regions that goes up to the background of the Northern Bridge shaping an extended layer of the Counter-bridge. Four old Bridge clusters were identified at distances larger than 8 kpc from the SMC centre moving towards the LMC, which is consistent with the SMC-LMC closest distance of 7.5 kpc when the Magellanic Bridge was formed about 150Myr ago; this shows that the Magellanic Bridge was not formed only by pulled gas, but it also removed older stars from the SMC during its formation. We also found age and metallicity radial gradients using projected distances on sky, which are vanished when we use the real 3D distances.
\end{abstract}




\section{Introduction}

The Small Magellanic Cloud (SMC) has a complex structure and large line-of-sight depth that makes it difficult to characterise its past evolution and trace interactions with the Large Magellanic Cloud (LMC) and the Milky Way \citep[e.g.][]{bekki+09,besla+07,beslaPhD,dias+16,niederhofer+18,zivick+18,deleo+20}. With the advent of large photometric, spectroscopic and astrometric surveys in the past decade or so, multiple efforts have been adding more information that helps constrain events in the past history of the SMC and the Magellanic System. For example, multiple bursts of star formation have been detected using star clusters and field stars \citep[e.g.][]{harris+04,piatti+11,parisi+14,rubele+18,bica+20}, although not all peak formation times coincide among different works, and they are still a matter of debate. For example, \citet{harris+04} found that $\sim$50\% of the SMC stellar mass was formed before $\sim 8.4$\,Gyr ago, followed by a quiescent period until about $\sim3$\,Gyr when multiple bursts of star formation started to take place, and they related it to a close encounter with the Milky Way. On the other hand, \citet{rubele+18} analysed an area 30\% larger and did not find any peak of star formation before $\sim8$\,Gyr ago, and concluded that 80\% of the SMC stars formed between 8 and 3.5\,Gyr ago with a peak around $\sim5$\,Gyr, which they also related with a close encounter with the Milky Way. Based on Hubble Space Telescope (HST) data analysis, \citet{cignoni+13} argued that interaction-triggered star formation is not the only mechanism to enhance star formation, and they explained the rise about $\sim7$\,Gyr ago as spontaneous star formation as has happened in some other isolated dwarf galaxies, although they could not rule out a minor merger, and they argued against a major merger as proposed by \citet{tsujimoto+09}. They only indicated one burst as being triggered by interactions with the LMC, a $\sim200$\,Myr old population located at the SMC wing. Based on star cluster ages, \citet{piatti+11} found peaks at $\sim2$\,Gyr and $\sim5$\,Gyr, later confirmed by \citet{parisi+14} and \citet{bica+20}. There are a couple of differences in these analysis, which include photometric depth, the SMC regions surveyed, and assumptions on distance and metallicity, but in all cases, there are star formation peaks related to the interactions of the SMC with other galaxies (Milky Way, LMC). Therefore, a higher age resolution is crucial to pinpoint the time-scales of the SMC interactions.

The ages and metallicities of red giant stars degenerate \citep[e.g.][]{cole+05,cioni+19}, which is a challenge for reaching age accuracy from field star colour-magnitude diagrams (CMD). Star clusters observed with deep photometry provide ages that are better constrained, as they can be derived in a self-consistent isochrone fit of age, metallicity, distance, and reddening \citep[e.g.][]{souza+20}. Adding spectroscopic metallicities as a prior for the isochrone fitting further enhances age accuracy. 

Besides age resolution, star clusters also provide a 3D map of the SMC, that is relevant because this galaxy spans about $\sim4$\,kpc perpendicular to the line-of-sight and has a depth of about $\sim20$\,kpc along the line-of-sight \citep[e.g.][]{nidever+13}. Therefore, before any strong conclusions on the origins of each SMC structure, it is crucial to assess projection effects. The VIsible Soar photometry of star Clusters in tApii and Coxi HuguA (VISCACHA) survey \citep{maia+19,dias+20} has been consistently observing star clusters in the outer regions of the SMC that are mostly affected by the interactions with the LMC. The uncertainties reached on age and distance are typically 4-20 and 1-6 per cent respectively \citep[e.g.][]{dias+21}, which are suitable for the aforementioned purposes.

\begin{figure}
    \centering
    \includegraphics[width=\columnwidth]{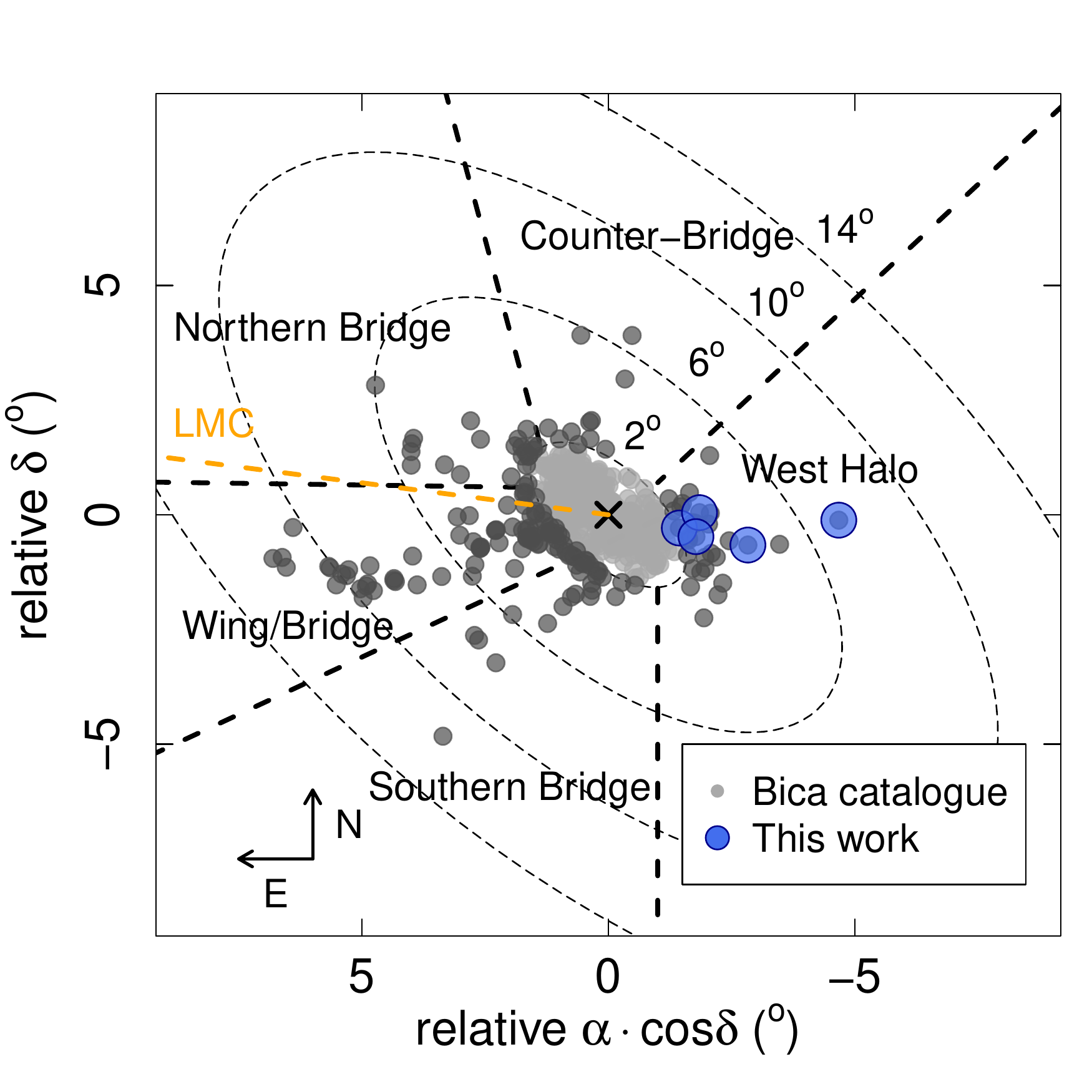}
    \caption{Projected distribution of SMC clusters from \citet{bica+20} catalogue. The thin dashed ellipses are aligned and concentric to the SMC projected Main Body and used as a proxy for the projected distance to the SMC centre. The distance $a$ is the semi-major axis of the ellipses indicated in degrees in the figure. The ellipses are tilted by $45^{\circ}$ and have an aspect ratio of $b/a=0.5$. Thick dashed lines split the regions outside $a>2^{\circ}$. The West Halo targets observed with SAMI/SOAR and GMOS/Gemini analysed in this work are marked with blue circles. The LMC direction is indicated by an orange dashed line.}
    \label{fig:SMCregions}
\end{figure}

\citet{dias+14} first introduced the framework of studying the outer star clusters of the SMC in different groups, split azimuthally in the plane of the sky (see Fig. \ref{fig:SMCregions}). The motivation was the complex dynamical evolution of this galaxy after many close encounters with the LMC that certainly affected the structure of the SMC stellar populations, in particular the outermost regions. At that time, the Magellanic Bridge was known to have gas and a young stellar content extending towards the LMC \citep[e.g.][]{hindman+63,putman+03,harris07}, that we call the Wing/Bridge in Fig. \ref{fig:SMCregions}. The other regions had some information but they were incomplete and not fully understood. Since then, many works have given more details on the external stellar populations and structure of the SMC. For example, a second branch of the Magellanic Bridge made up of an old stellar population has been characterised using RR Lyrae as tracers \citep[e.g.][]{jacyszyn+17,belokurov+17}, that we call the Southern Bridge in Fig. \ref{fig:SMCregions}. A Northern Bridge has been discovered as a third branch of the Magellanic Bridge, with star clusters also moving from the SMC towards the LMC \citep{dias+21}. In this same work, the first star cluster belonging to the Counter-Bridge was found, confirming the predictions by models \citep[e.g.][]{diaz+12} and partial evidence by observations \citep[e.g.][]{nidever+13}.

The West Halo was proposed by \citet{dias+16} as a structure moving away from the SMC, which was confirmed by proper motions \citep[PMs,][]{niederhofer+18,zivick+18,piatti21}.
\citet{niederhofer+18} showed that stars located in the inner SMC region towards the West Halo are already moving away from the SMC centre, based on PMs from the VISTA survey of the Magellanic Clouds system \citep[VMC, ][]{cioni+11}. \citet{zivick+18} corroborated these findings analysing outer fields observed with PMs from HST and Gaia. In particular, \citet{zivick+18} showed PM vectors in the West Halo area that are of the same order of those found in the well-known Magellanic Bridge, which means that the transverse motion on the sky has similar magnitude, and could be related somehow. Accordingly, \citet{tatton+21} discussed the possibility that the West Halo is actually the beginning of the tidal counterpart of the Magellanic Bridge, which warps behind the SMC towards the Northeast.

In this paper we use VISCACHA data to derive distances and ages;
GMOS/Gemini spectra to derive radial velocities (RVs) and metallicities; and Gaia Early Data Release 3 \citep[EDR3, ][]{gaiaedr3} to get PMs for five clusters in the West Halo. Following the analysis framework started by \citet{dias+21}, we end up with a full phase-space vector for all clusters analysed plus age and metallicity. We discuss the results in the context of the interactions with the LMC and the formation of the tidal structures.

The paper is organised as follows. In Sect. 2 we describe the observations, whereas the analysis is presented in Sect. 3. The results are discussed in Sect. 4, and conclusions can be found in Sect. 5.

\section{Observations}

The sample selection is based on GMOS/Gemini spectroscopic observations of $\sim10-40$ selected giant stars per field centred at each of five West Halo clusters, spanning a large range of distances from the SMC centre (see Fig. \ref{fig:SMCregions}). All five clusters are part of the VISCACHA sample and naturally covered by the all-sky Gaia data. This sample is the starting point to trace the 6D structure plus age and metallicity of this region, which is presumably a tidal tail.

\subsection{Photometry from the VISCACHA survey with SAMI/SOAR}
\label{sec:photometry}

We use the photometry in V and I filters obtained with SAMI/SOAR \citep{tokovinin+16} within the VISCACHA survey (see Table \ref{tab:loggmos}). Data reduction, analysis, photometry, and completeness are discussed in detail by \citet[][Paper I]{maia+19}. Briefly, PSF photometry was performed in all images with an IDL code developed by our group based on {\sc starfinder} \citep{diolaiti+00}, and photometric calibrations were based on Stetson standard star fields observed in the same nights as the science observations. Typically, the 50\% photometric completeness level reaches V$\sim24$\,mag in the cluster outskirts, and V$\sim23$\,mag within the cluster core radius. Finally, a probability of each star to belong to a given cluster is statistically estimated by comparing the density of stars in the CMD built from the cluster stellar sample and from a nearby comparison field. Membership is assigned based on the local CMD overdensity and distance relative to the cluster centre \citep{maia+10}.

\subsection{Spectroscopic follow-up with GMOS/Gemini-S}

The five selected clusters are all older than $1-2$\,Gyr, which is a criterion to derive metallicities from CaII triplet (CaT) lines \citep[e.g.][]{cole+04,dias+20b}.
We selected red giant branch (RGB) stars in the GMOS/Gemini field centred at each of the five clusters based on GMOS/Gemini pre-images obtained with $r$ and $i$ filters. The spectra were taken using the R831 grating combined with the CaT filter and $1.0"$ width slits (see Table \ref{tab:loggmos}), covering a final wavelength range between $7347-9704\,\text{\normalfont\AA}$ , centred at $8540\,\text{\normalfont\AA}$, which includes the three CaT lines at 8498, 8542, 8662$\,\text{\normalfont\AA}$, and a spectral resolution of $R\approx2,000$ and SNR ranging from $\sim 25-90$. 
The pre-images were reduced by Gemini staff using a default pipeline. The PSF photometry was performed using DAOPHOT \citep{stetson+87} to produce $\ r,(r-i)\ $ CMDs that were the basis of the RGB stars selection. No photometric calibration was performed as only relative magnitudes were required.
The magnitudes relative to the horizontal branch/red clump level were used in the CaT-[Fe/H] calibration following the recipes described in detail in \citet{dias+20b} and summarised below.

\begin{table}
\caption{Log of observations.}
\label{tab:loggmos}
\centering
\setlength{\tabcolsep}{3pt}
\begin{tabular}{llcccc}
\hline
cluster       & date             & grism/filter & exp.time (s) & airmass & FWHM (") \\
\hline
\multicolumn{6}{c}{SAMI/SOAR photometry}\\
\hline
NGC~152    & 2016-11-04 & V, I   & 4x200, 4x300   & 1.37  & 0.71, 0.45  \\
AM~3       & 2016-11-04 & V, I   & 6x200, 6x300   & 1.37  & 0.51, 0.38  \\
Lindsay~2  & 2019-10-05 & V, I   & 3x400, 3x600   & 1.57  & 0.81, 0.74   \\
Kron~7     & 2016-09-27 & V, I   & 6x200, 6x300   & 1.37  & 0.64, 0.49  \\
Kron~8     & 2021-07-10 & V, I   & 3x400, 3x600   & 1.44  & 0.86, 0.61  \\
\hline
\multicolumn{6}{c}{GMOS/Gemini Pre-images}\\
\hline
NGC~152  & 2017-08-22 & r, i   & 3x60, 3x60   &  1.50       & 0.75, 0.65 \\
AM~3         & 2017-09-17 & r, i   & 3x60, 3x60   & 1.36         & 1.00, 0.90 \\
Lindsay~2 & 2017-08-22  & r, i   & 3x60, 3x60   &  1.42        & 0.77, 0.64 \\
Kron~7     & 2017-08-22 & r, i   & 3x60, 3x60   &   1.43       & 0.69, 0.68 \\
Kron~8     & 2017-08-22 & r, i   & 3x60, 3x60   &   1.47      & 0.75, 0.68 \\
\hline
\multicolumn{6}{c}{GMOS/Gemini Multi-object spectroscopy}\\
\hline
NGC~152  & 2017-10-21  & R831+CaT & 3x805          & 1.37        &  1.0 \\
AM~3        & 2017-10-21  & R831+CaT & 3x805          & 1.37        & 0.8 \\
Lindsay~2 & 2017-12-16  & R831+CaT & 3x805          & 1.47        & 0.8 \\
Kron~7      & 2017-10-25  & R831+CaT & 3x805          & 1.37        & 0.8 \\
Kron~8      & 2017-10-25  & R831+CaT & 3x805          & 1.45        & 0.8 \\
\hline
\end{tabular}
\raggedright
Notes: The FWHM for the GMOS/Gemini pre-images and SAMI/SOAR images were measured on the reduced and combined images. The FWHM for the spectroscopic observation is a reference in the V band.
\end{table}

The equivalent width of the CaT lines is very sensitive to surface gravity, temperature and metallicity, therefore it is necessary to remove these additional effects from the equivalent widths before converting them into metallicities. A convenient and robust proxy for gravity and temperature is the magnitude relative to the horizontal branch level \citep[][and references therein]{dias+20b}. 
It was important to have this procedure in mind when we selected the RGB stars to be observed in each cluster, as they should cover a range of at least one magnitude. Another empirical fact is that stars below the horizontal branch have less sensitivity to gravity on the CaT lines which makes it difficult to correct for this effect, not to mention that the relative faintness of such stars produces lower S/N. For this reason we preferred stars above the horizontal branch. In summary, the choice of stars was based on the compromise between the magnitude range and the distribution of slits in the mask without overlapping each other, maximising the number of stars within a given cluster tidal radius. In addition, we also selected stars outside the tidal radius of each cluster that were more likely to be field stars as reference for the membership selection. In the case of crowded fields, we proceeded with a membership probability calculation based on the photometry to boost the probability of observing a cluster star and not a foreground field star.

The GMOS/Gemini MOS data were reduced using the scripts\footnote{\url{http://drforum.gemini.edu/topic/gmos-mos-guidelines-part-1/}} developed by M. Angelo. The scripts are all based on default Gemini IRAF package v1.14. In a few words, after bias and flat field correction and cosmic ray cleaning (via LACOSMIC IRAF task, as described in \citealp{vandokkum01}), the spectra were extracted based on each slit position, arc lamps were used to find the pixel-wavelength solution, whereas skylines were used to find the absolute zero-point in wavelength, that is a crucial step for reaching accurate RV measurements. Different exposures were combined after extraction using the sum of the flux per pixel of the 1D spectra, and the final spectra were continuum normalised, as no flux calibration is required for the CaT technique.

\section{Analysis}

In this Section, we describe the determination of the parameters from the photometric and spectroscopic data and also how we used Gaia EDR3 PMs to complement this study.

\subsection{Radial velocities and metallicities from CaT}
\label{sec:gmosanalysis}

The first step in the analysis was to measure the RV of each star. The RV is used to estimate membership to a given cluster in contrast with field stars, and it is required to shift all spectra to the rest frame before fitting line profiles to derive metallicities. The RV was derived by cross-correlation with a set of synthetic templates from Paula Coelho's library \citep{coelho+14}, degraded in spectral resolution to properly match that of GMOS/Gemini spectra. The template atmospheric parameters span the ranges $4750 \leq {\rm T_{eff} (K)} \leq 5250$ and log($g$)=1, representative of RGB stars. We also allowed variations in the metallicity of the theoretical spectra, which span the range $-1.3 \leq {\rm[Fe/H]} \leq -0.5$. A total of 10 templates have been obtained from these specifications. The final RV is the mean of all measured RV, whereas the average of the individual RV errors give a mean value for the uncertainty.

The CaT lines are widely used to analyse RGB stars in star clusters (e.g. \citealp{armandroff+88}, \citealp{rutledge+97a}, \citealp{cole+04}, \citealp{saviane+12}, \citealp{vasquez+15}, \citealp{vasquez+18}, \citealp{dias+20b}). They are strong in the near infrared, therefore not expensive for telescope time and useful, even for star cluster stars at the distance of the Magellanic Clouds, to derive RVs and metallicities with precision of $\sim$1-5km/s and $\sim$0.05-0.15 dex, respectively \citep[e.g.][]{parisi+09,parisi+15,dias+21}.

\begin{figure}
\includegraphics[width=\columnwidth]{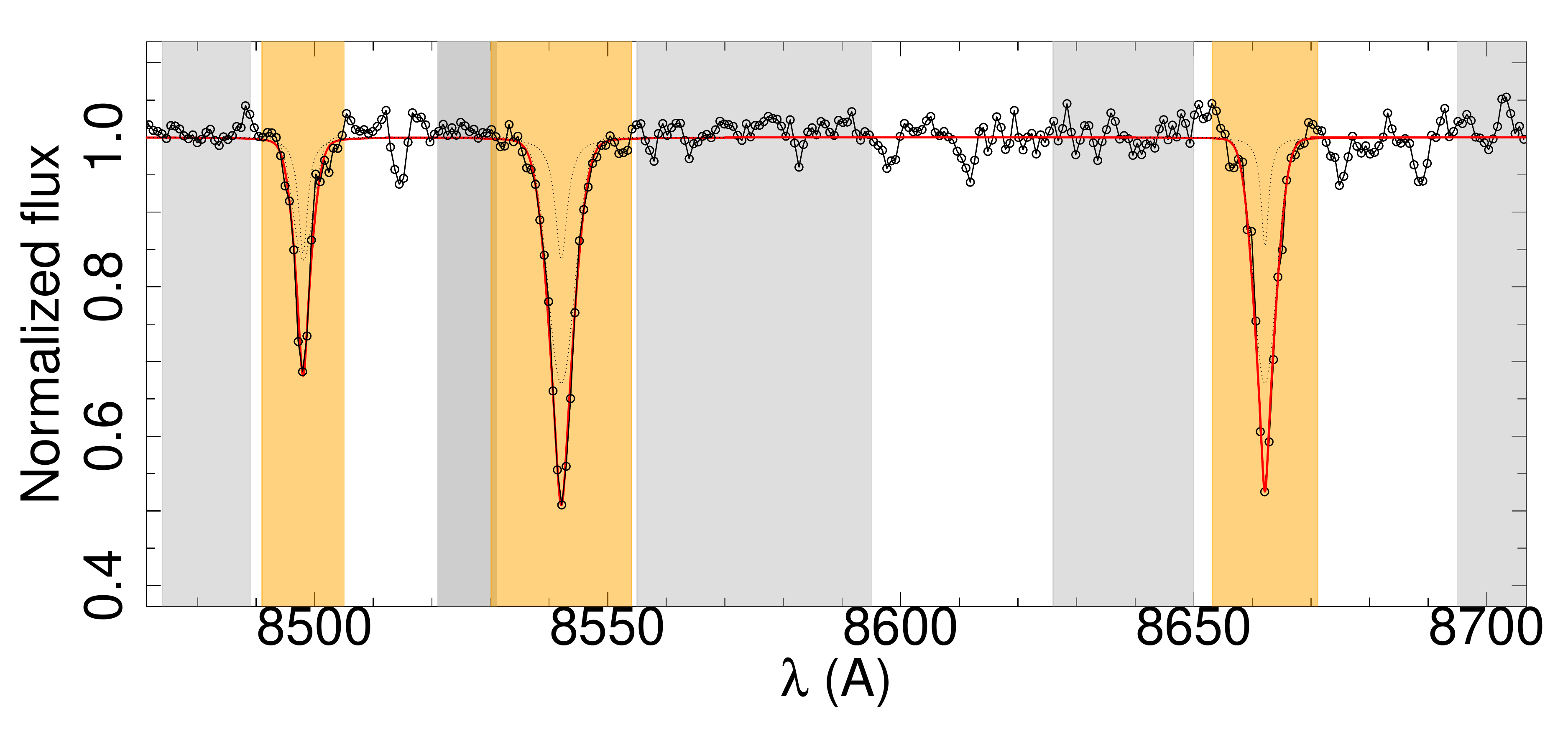}
\caption{CaT fitting with Gaussian+Lorentzian profile for an example star from this work. The observed spectrum is shown as black lines and dots, whereas the red line is the best fit. Dashed lines show the Gaussian and Lorentzian components separately. Grey and orange regions highlight the continua and line regions adopted for the profile fitting. }
\label{fig:CaTfit}
\end{figure}

The philosophy behind using CaT lines to derive metallicities is similar to that applied in spectral indices, consisting in the definition of a passband and two local continua within which the flux is computed. This procedure is very sensitive to S/N as it considers the flux pixel to pixel. In the case of CaT, the strategy is slightly adapted, where a Gaussian plus a Lorentzian profile is fitted to each line and the equivalent width (EW) of the fitted function is assumed as the quantity later converted into metallicity. There is a vast literature on the topic (see \citealp{dias+20b} and references therein);
therefore, we limit ourselves to a general description of our assumptions in this paper. An example of line fitting is shown in Fig. \ref{fig:CaTfit}.

We adopted the recipes from \citet{cole+04} in order to be consistent with the analysis by \citet{parisi+09,parisi+15,dias+21}. Specifically, we use $\rm{\sum EW = EW_{8498} + EW_{8542} + EW_{8662}}$ with a line profile fitted by Gaussian+Lorentzian function, bandpasses and continua windows defined by \citet{armandroff+88}. The total EW is fitted against relative magnitude to derive the reduced EW ${\rm W'}$, which is the proxy for metallicity, following the equation ${\rm \sum EW = W' + \beta_m (m-m_{HB})}$, where $\rm{m}$ is the magnitude on a given filter. The only difference is that \citet{cole+04} used $(V-V_{\rm HB})$ as a proxy for luminosity which led to a slope $\beta_V = 0.73$ and we use $(r-r_{\rm HB})$ that has an equivalent $\beta_r = 0.67$ following the conversion by \citet{dias+20b}. These authors showed that their calibration is in excellent agreement with that derived by \citet{cole+04}, therefore our metallicities are on the same scale as in our previous works. The membership determination is described in detail in Appendix \ref{app:gmosobs}. Kron\,8 was a difficult case for membership assessment because there were only two member stars. Therefore we joined our sample with that from \cite{parisi+15} to constrain the RV and [Fe/H] of this cluster (see Appendix \ref{app:gmosobs} for more details.).


\subsection{CMD isochrone fitting}

To obtain the fundamental parameters for the analysed clusters $V,(V-I)$ CMDs from the VISCACHA photometry, we used the SIRIUS code \citep{souza+20} with the PARSEC isochrones \citep{bressan+12} dataset. This code uses a Bayesian approach based on the Markov chain Monte Carlo (MCMC) sampling method, which depends on building a likelihood function with as much information as possible to reach an isochrone fitting with physical meaning. For example, in the case of low-mass star clusters analysed here, a lower number of RGB stars is in high contrast with the well-populated main sequence. Therefore, a higher weight is necessary for the RGB stars in order to take into account the effect of the initial mass function. Similar to the well-known prior from RR Lyrae stars, the red clump (RC) magnitude can be employed as a prior to constrain the distance modulus parameter space. The RC prior is less robust that using RR Lyrae. However, it allows us to restrict the region of HB/RC isochrone in the CMD. Also, we used the spectroscopic metallicity from CaT as a prior for the isochrone fit. The employment of the priors described above reduces degeneracy between the parameters, increasing the precision on distance and age, which are the main fundamental parameters for the present analysis. 

\begin{figure*}
    \centering
    \includegraphics[width=0.32\textwidth]{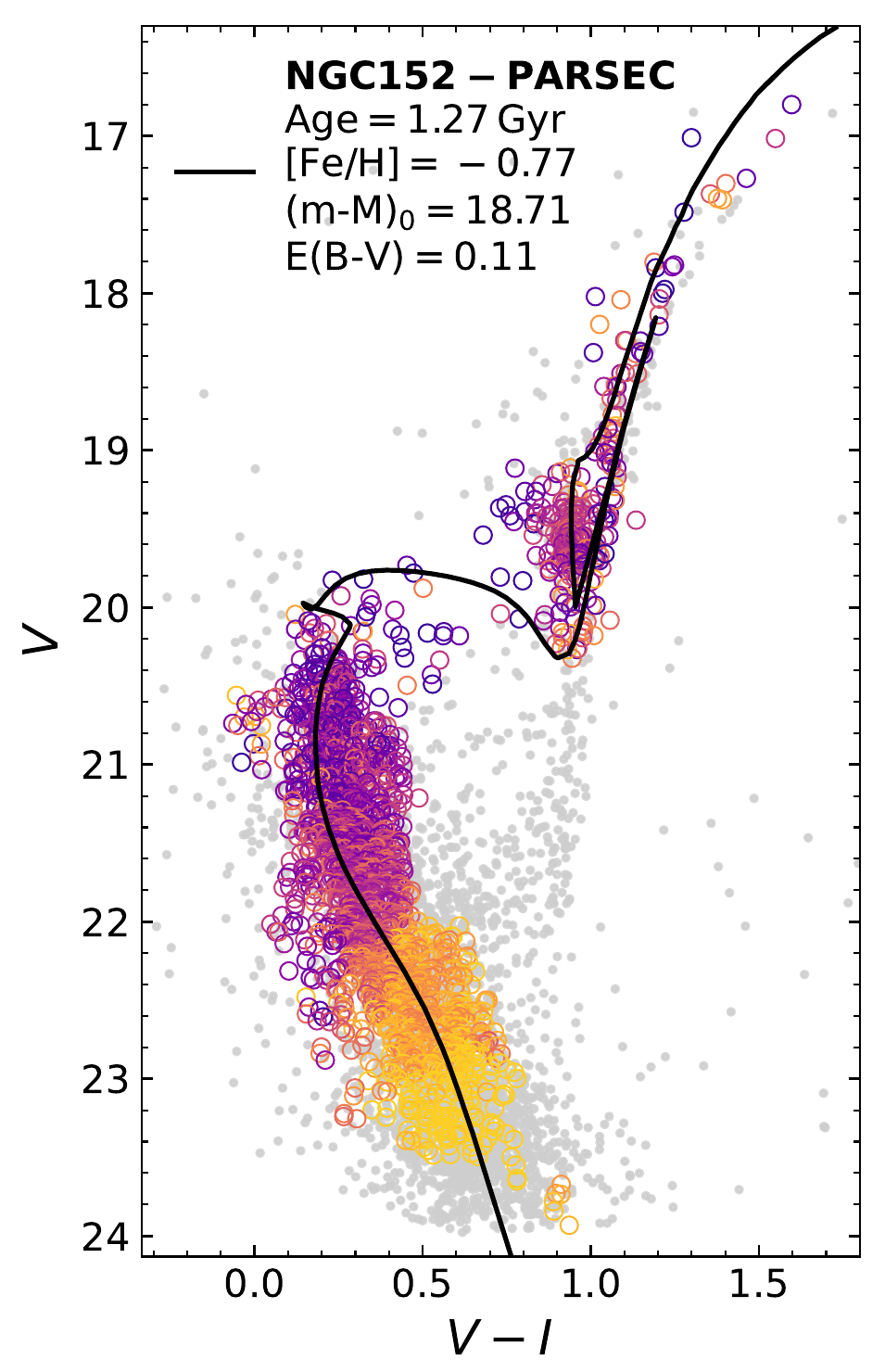}
    \includegraphics[width=0.32\textwidth]{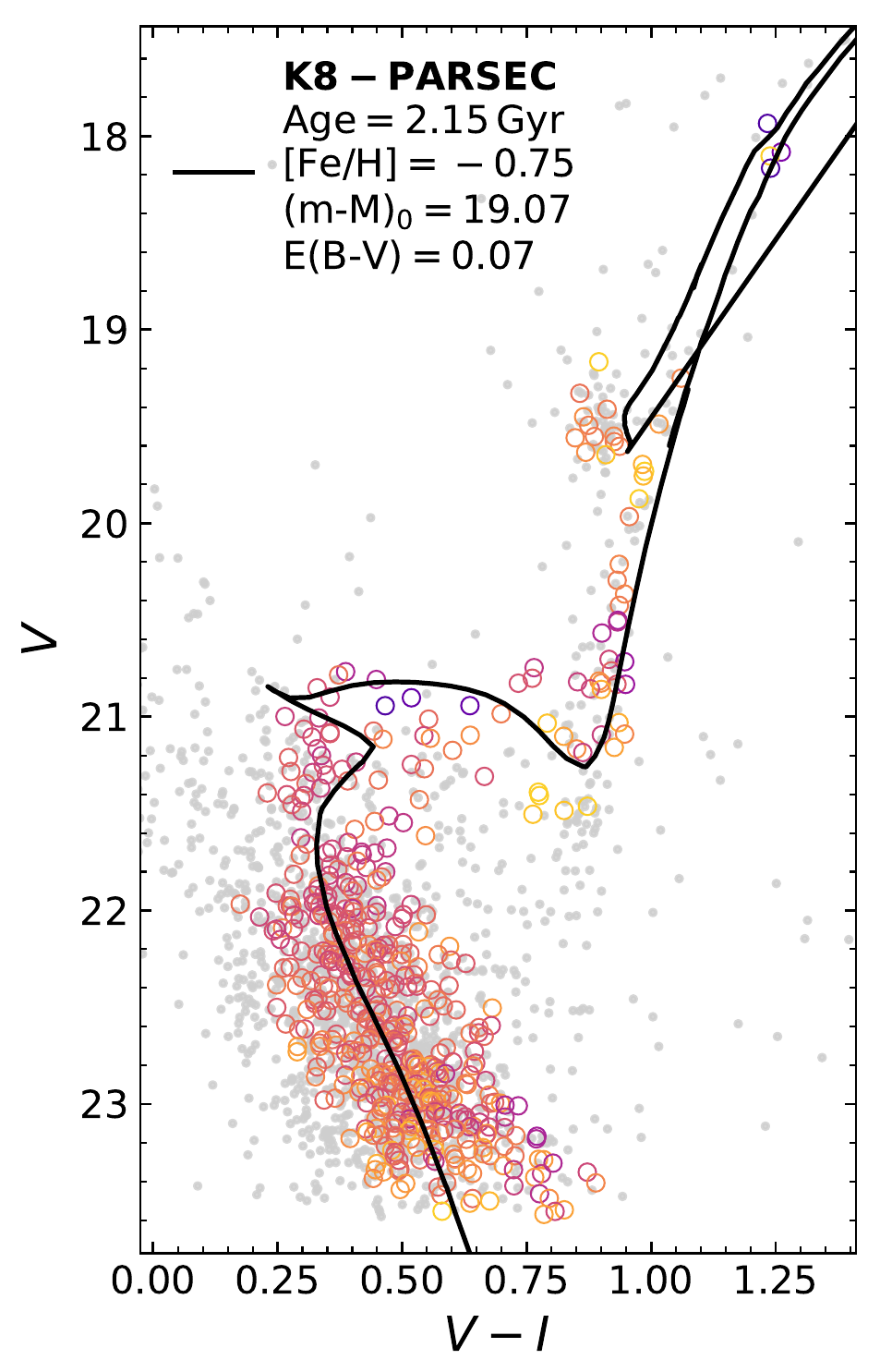}
    \includegraphics[width=0.32\textwidth]{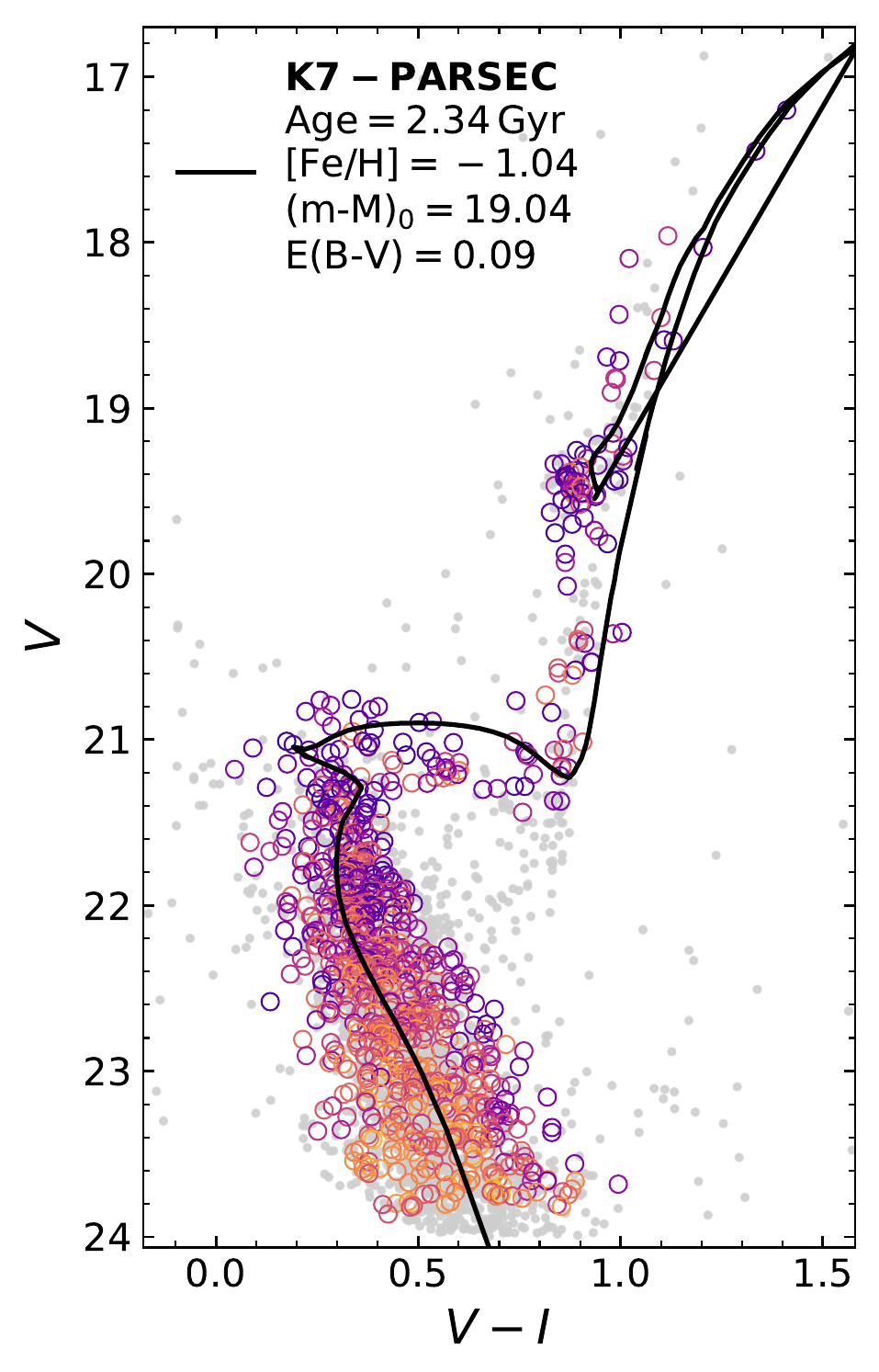}
    \includegraphics[width=0.32\textwidth]{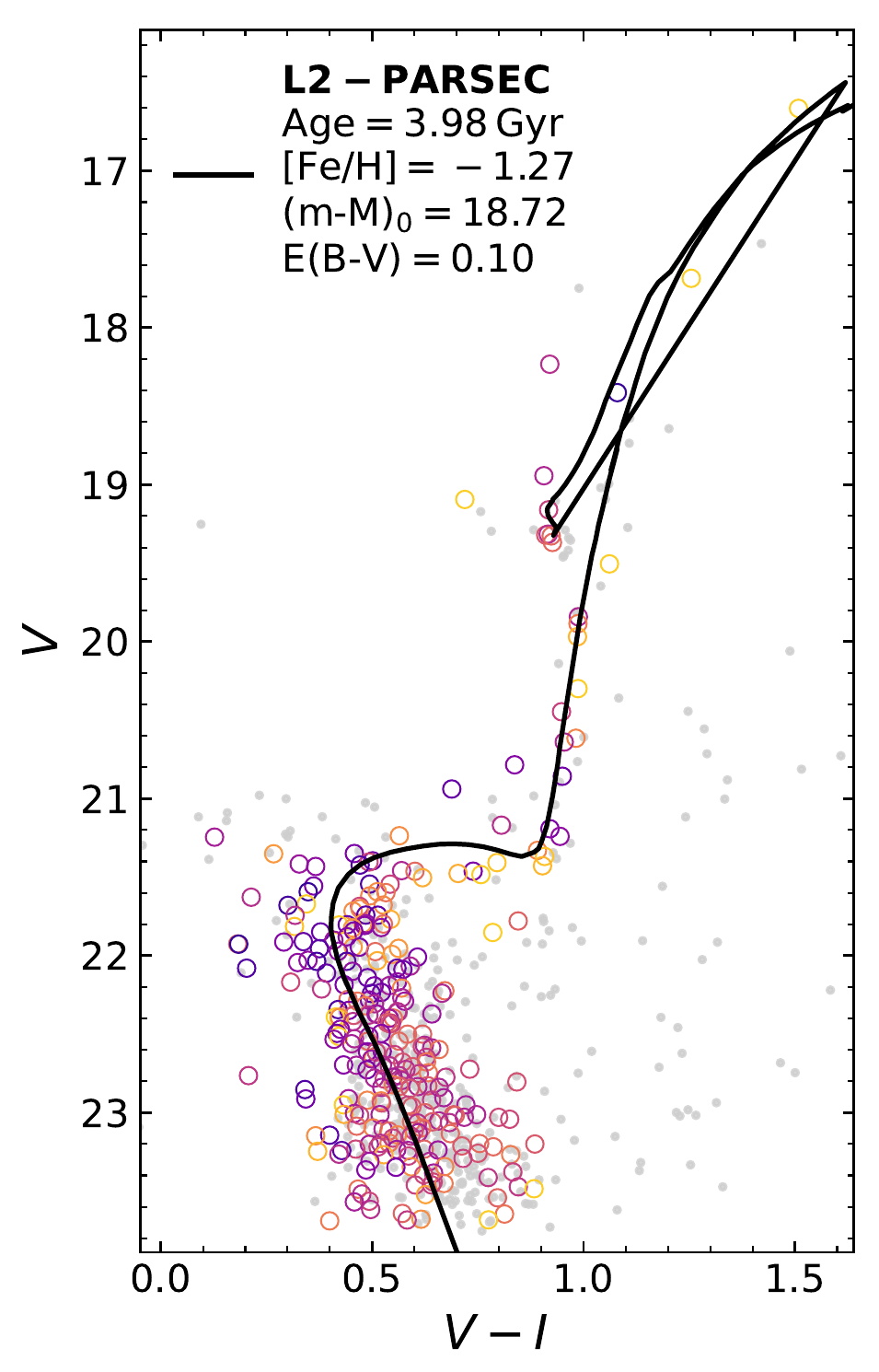}
    \includegraphics[width=0.32\textwidth]{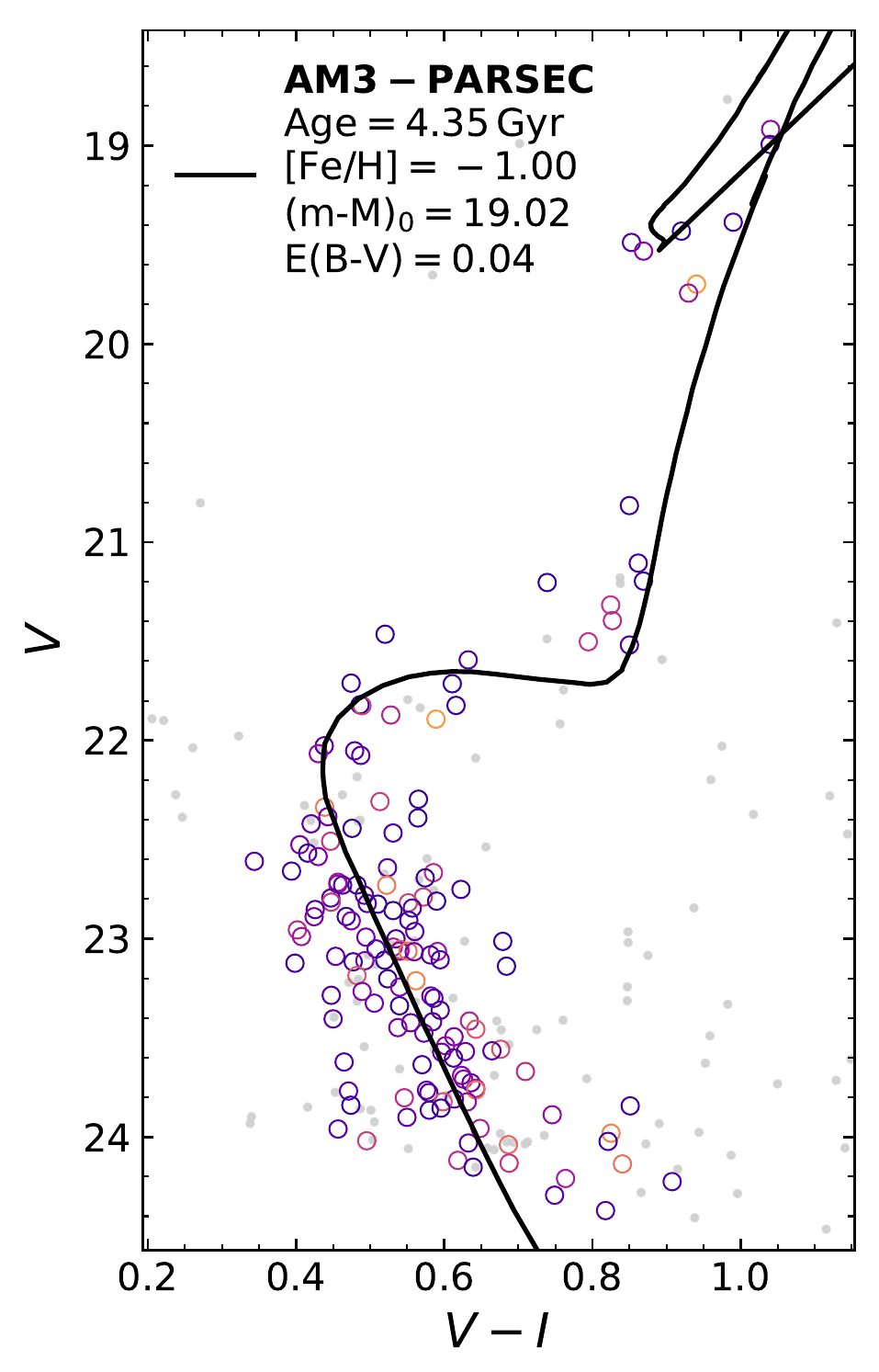}
    \hspace{2mm}
    \includegraphics[height=8.9cm]{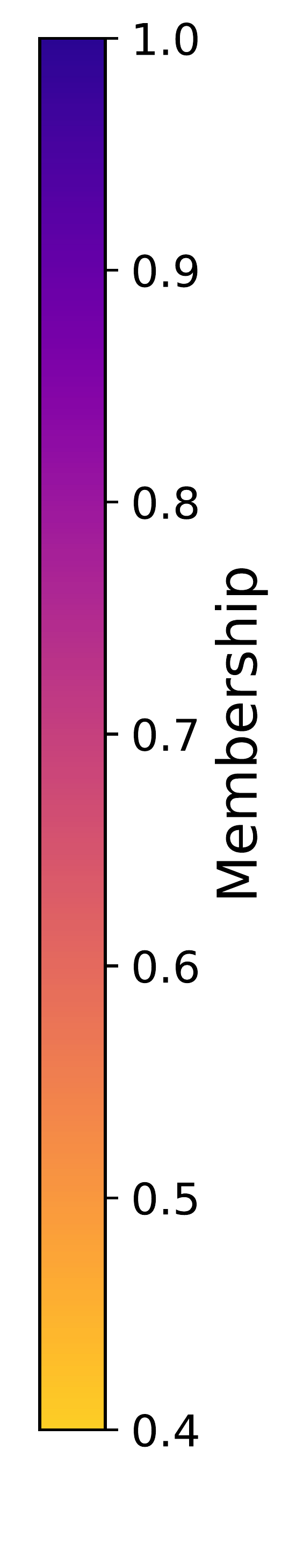}
    \caption{CMDs with the best PARSEC isochrone statistically fitted using the SIRIUS code. We adopted priors in metallicity from spectroscopy and on the RC magnitude from the photometry. The CMDs are cleaned as explained in Section \ref{sec:photometry} and the point colours represent the membership probability to belong to each cluster. Grey points in the background are all stars with membership probability smaller than 40\%, i.e., most likely field star contamination.}
    \label{fig:CMDfit}
\end{figure*}

Figure \ref{fig:CMDfit} presents the CMDs of the five sample clusters with the best-fit isochrone derived in the isochrone fitting, which corresponds to the $50^{th}$ percentile of the posterior distribution. Figure \ref{fig:posterior} in the Appendix \ref{app:sirius} shows the corner plots, with the posterior distributions of the four free parameters (age, metallicity, distance modulus and reddening) in the diagonal panels, and the correlations between each two parameters in the other panels. The dashed lines in the corners plots correspond to the median and 1-$\sigma$ level (i.e. $16^{th}$ and $84^{th}$ percentiles) of each parameter. These are the adopted final parameters and uncertainties listed in Table \ref{tab:results}.

\begin{table*}
    \caption{Derived parameters for the star clusters. (1) cluster name; (2,3) $ (\alpha,\delta) $ coordinates from \citet{bica+20}; (4) projected angular distance from the SMC centre $a$ following the definition by \citet{dias+14}; (5) number of member stars and observed stars corresponding to the GMOS/Gemini spectroscopy; (6,7) $RV_{hel}$ and [Fe/H]$_{\rm CaT}$ from GMOS/Gemini spectra; (8,9,10,11) age, [Fe/H]$_{\rm CMD}$, E(B-V), distance from VISCACHA CMD isochrone fitting; (12,13) ( $\mu_{\alpha}\cdot cos(\delta)$, $\mu_{\delta}$ ) PMs from Gaia EDR3.}
    \label{tab:results}
    \centering
    \footnotesize
    \begin{tabular}{lrrcrll}
    \hline
    \noalign{\smallskip}
    \multicolumn{1}{c}{Cluster} &
    \multicolumn{1}{c}{$\alpha_{J2000}$} &
    \multicolumn{1}{c}{$\delta_{J2000}$} &
    \multicolumn{1}{c}{a} &
    \multicolumn{1}{c}{N$_{mem}$/N$_{obs}$} &
    \multicolumn{1}{c}{${\rm RV_{hel}}$} &
    \multicolumn{1}{c}{[Fe/H]$_{\rm CaT}$} \\
    \multicolumn{1}{c}{} &
    \multicolumn{1}{c}{(hh:mm:ss.s)} &
    \multicolumn{1}{c}{(dd:mm:ss)} &
    \multicolumn{1}{c}{(deg)} &
    \multicolumn{1}{c}{} &
    \multicolumn{1}{c}{(${\rm km\ s^{-1}}$)} &
    \multicolumn{1}{c}{} \\
    \noalign{\smallskip}
    \multicolumn{1}{c}{(1)} &
    \multicolumn{1}{c}{(2)} &
    \multicolumn{1}{c}{(3)} &
    \multicolumn{1}{c}{(4)} &
    \multicolumn{1}{c}{(5)} &
    \multicolumn{1}{c}{(6)} &
    \multicolumn{1}{c}{(7)} \\
    \noalign{\smallskip}
    \hline
    \noalign{\smallskip}
    NGC~152      & 00:32:56.3 & $-$73:06:57 & 2.035 & 6 / 36   & $176.4\pm1.4(2.6)$   & $-0.75\pm0.08(0.11)$ \\
    Kron~8$^{a}$ & 00:28:01.9 & $-$73:18:12 & 2.432 & 2 / 34   & $194.8\pm3.3(0.8)$   & $-0.84\pm0.12(0.16)$ \\
                 &            &             &       & 4 / 56   & $198.1\pm2.4(4.1)$   & $-0.76\pm0.07(0.13)$ \\
    Kron~7       & 00:27:45.2 & $-$72:46:53 & 2.970 & 10 / 28   & $145.4\pm1.1(5.2)$   & $-1.09\pm0.05(0.13)$ \\
    Lindsay~2    & 00:12:55.0 & $-$73:29:12 & 3.939 & 4 / 26   & $171.4\pm1.5(5.2)$   & $-1.28\pm0.08(0.09)$ \\
    AM~3         & 23:48:59.0 & $-$72:56:42 & 7.283 & 4 / 11   & $157.0\pm1.9(1.1)$   & $-1.00\pm0.09(0.08)$ \\
    \noalign{\smallskip}
    \hline \hline
    \noalign{\smallskip}
    \multicolumn{1}{c}{Cluster} &
    \multicolumn{1}{c}{Age} &
    \multicolumn{1}{c}{[Fe/H]$_{\rm CMD}$} &
    \multicolumn{1}{c}{E(B-V)} &
    \multicolumn{1}{c}{d} &
    \multicolumn{1}{c}{$\mu_{\alpha}\cdot {\rm cos}(\delta)$} &
    \multicolumn{1}{c}{$\mu_{\delta}$} \\
    \multicolumn{1}{c}{} &
    \multicolumn{1}{c}{(Gyr)} &
    \multicolumn{1}{c}{} &
    \multicolumn{1}{c}{(mag)} &
    \multicolumn{1}{c}{(kpc)} &
    \multicolumn{1}{c}{(${\rm mas\ yr^{-1}}$)} &
    \multicolumn{1}{c}{(${\rm mas\ yr^{-1}}$)} \\
    \noalign{\smallskip}
    \multicolumn{1}{c}{(cont.)} &
    \multicolumn{1}{c}{(8)} &
    \multicolumn{1}{c}{(9)} &
    \multicolumn{1}{c}{(10)} &
    \multicolumn{1}{c}{(11)} &
    \multicolumn{1}{c}{(12)} &
    \multicolumn{1}{c}{(13)} \\
    \noalign{\smallskip}
    \hline
    \noalign{\smallskip}
    NGC~152      & $1.27^{+0.04}_{-0.26}$   & $-0.77^{+0.07}_{-0.21}$    & $0.11^{+0.07}_{-0.04}$   & $55.2^{+1.8}_{-1.5}$  &   $0.19\pm 0.08(0.21)$    & $-1.24\pm 0.07(0.03)$  \\
    \noalign{\smallskip}
    Kron~8$^{a}$ & $2.15^{+0.21}_{-0.21}$   & $-0.75^{+0.07}_{-0.07}$   & $0.07^{+0.04}_{-0.05}$    & $65.2^{+3.4}_{-3.2}$  &  $0.48\pm 0.13(--)$       & $-1.17\pm 0.11(--)$    \\
                 &                          &                           &                           &                       &  $0.54\pm 0.06(0.04)$     & $-1.24\pm 0.06(0.04)$  \\
    \noalign{\smallskip}
    Kron~7       & $2.34^{+0.20}_{-0.08}$   & $-1.04^{+0.05}_{-0.05}$   & $0.09^{+0.03}_{-0.04}$    & $64.3^{+2.4}_{-2.3}$  &  $0.67\pm0.08(0.23)$      & $-1.19\pm0.08(0.08)$   \\
    \noalign{\smallskip}
    Lindsay~2    & $3.98^{+0.37}_{-0.55}$   & $-1.27^{+0.10}_{-0.08}$   & $0.10^{+0.05}_{-0.05}$    & $55.5^{+2.9}_{-2.7}$  &  $0.67\pm0.13(0.07)$      & $-1.51\pm0.11(0.15)$   \\
    \noalign{\smallskip}
    AM~3         & $4.4^{+1.3}_{-1.4}$      & $-1.00^{+0.10}_{-0.10}$    & $0.04^{+0.04}_{-0.07}$    & $63.7^{+4.2}_{-3.7}$  &  $0.46\pm 0.13(0.09)$     & $-1.16\pm 0.13(0.03)$  \\
    \noalign{\smallskip}
    \hline
    \end{tabular}
    \\Notes: $^a$ The first row contains Kron\,8 results from GMOS/Gemini and SAMI/SOAR, exactly as done for the other clusters. The second line is an update of the CaT results of RV, [Fe/H] based on the joint sample of GMOS/Gemini+FORS2/VLT-ESO, and as a consequence the PMs are also updated. We adopt the joint sample results for these parameters. See text for details.
\end{table*}


\subsection{Comparison with previous investigations}

Clusters of the present sample have been previously investigated  and their parameters derived by different works using diverse data and analysis techniques. A summary of the most relevant previous determinations can be seen in Table \ref{tab:biblio}. Our analysis is self-consistent and homogeneous, which is a requirement to properly analyse the 3D structure of the SMC. The goal of this comparison with the literature is to show that the parameters derived here are reasonable. In general, our age, metallicity and distance determinations agree with the values reported by previous works. The RV derived by \cite{song+21} for NGC\,152 based on high-resolution spectroscopy is in agreement with the RV derived in this work. The RV found by \citet{parisi+15} for Kron\,8 is shifted to RV = $204.4\pm1.3{\rm km\ s^{-1}}$ applying the offset defined in Appendix \ref{app:lit}, which is compatible with RV = $198.1\pm2.4(4.1){\rm km\ s^{-1}}$ found here.  The RV of Kron\,7 previously measured by \cite{dacosta+98} becomes RV = $138.6\pm5.0{\rm km\ s^{-1}}$ with the offset defined in Appendix \ref{app:lit} and it is compatible with the RV = $145.4\pm1.1(5.2){\rm km\ s^{-1}}$ found here. The distance of NGC\,152 derived by \citet{crowl+01} results in d = 65\,kpc after the correction explained in Appendix \ref{app:lit} which disagrees with the shorter distance derived in this work. This particular cluster presents an extended main sequence turnoff which can be the source of dispersion in some parameters depending on the analysis. Our metallicity values for NGC\,152 and Kron\,8 are in excellent agreement with previous spectroscopic studies \citep{parisi+15,song+21}. \cite{dacosta+98} derived [Fe/H] = $-0.81\pm0.04$ in the metallicity scale of \cite{carretta+97} which can be transformed into the scale of \cite{carretta+09} with a relation provided in that paper resulting in [Fe/H] = $-0.92\pm0.04$ for Kron\,7, which is similar to the metallicity [Fe/H] = $-1.09\pm0.05(0.13)$ found in the present work. As far as we are aware, this is the first time that AM\,3 and Lindsay\,2 are analysed using spectroscopy of individual stars.

\begin{table*}
\caption{Cluster parameters from the literature}
\label{tab:biblio}
\centering
\setlength{\tabcolsep}{3pt}
\begin{tabular}{lcccccl}
\hline
    Cluster & ${\rm RV_{hel}}$           & Age & [Fe/H] & d & Method & Reference \\
            & (${\rm km\ s^{-1}}$) & (Gyr) &        & (Kpc)    &        &            \\
    \hline
    NGC~152      &  ---                  &1.4$\pm$0.2    & -0.94$\pm$0.15 & 61.0$\pm5.5$* &  photometry & \citet{crowl+01}\\
                 & 172.4$^{+0.5}_{-0.9}$ & ---             & -0.73 $\pm$ 0.11   & ---  & high-resolution spectroscopy & \citet{song+21}\\
                 &            ---          & 1.23$\pm$0.07 & -0.87$\pm$0.07 & 60.0$\pm$2.9 & photometry & \citet{dias+16}\\
                 \\
    Kron~8      &  208$\pm$1.3$^*$    & ---        & -0.70$\pm$0.04 & --- & CaT spectroscopy & \citet{parisi+15}\\
                  &       ---              & 2.94$\pm$0.31& -1.12$\pm$0.15 & 69.8$\pm$2.3 & photometry & \citet{dias+16}\\
                  \\
    Kron~7       &           ---            & 3.5$\pm$0.5& ---&--- & Integrated spectroscopy & \citet{piatti+05}\\
                 & 132$\pm$5$^*$ & 3.5$\pm$1  &  -0.81$\pm$0.04 &  --- & CaT spectroscopy& \citet{dacosta+98}\\
                 &        ---             & 3 & -0.8$\pm$0.04;-1.3$\pm$0.3&--- & photometry & \citet{livanou+13}\\
                 \\
    Lindsay~2    &        ---              & 4.0$^{+0.9}_{-0.7}$& -1.4$^{+0.2}_{-0.2}$&54.4$^{+1.5}_{-1.5}$ & photometry& \citet{dias+14}\\
                 \\
    AM~3         &            ---           & 6.0$\pm$0.15& -1.25$\pm$0.25&--- & photometry & \citet{piatti11} \\
                 &         ---              & 4.9$^{+2.1}_{-1.5}$ & -0.8$^{+0.2}_{-0.6}$& 63.1$^{+1.8}_{-1.7}$& photometry & \citet{dias+14}\\
                 &          ---             & 5.48$^{+0.46}_{-0.74}$ & -1.36$^{+0.31}_{-0.25}$ & 64.8$^{+2.1}_{-2.0}$ & photometry & \citet{maia+19}\\
    \hline
    \end{tabular}
    \\Notes: *We apply a correction defined in Appendix \ref{app:lit}, see the text for details.
\end{table*}


\subsection{Proper motions}

\begin{figure*}
\includegraphics[width=\textwidth]{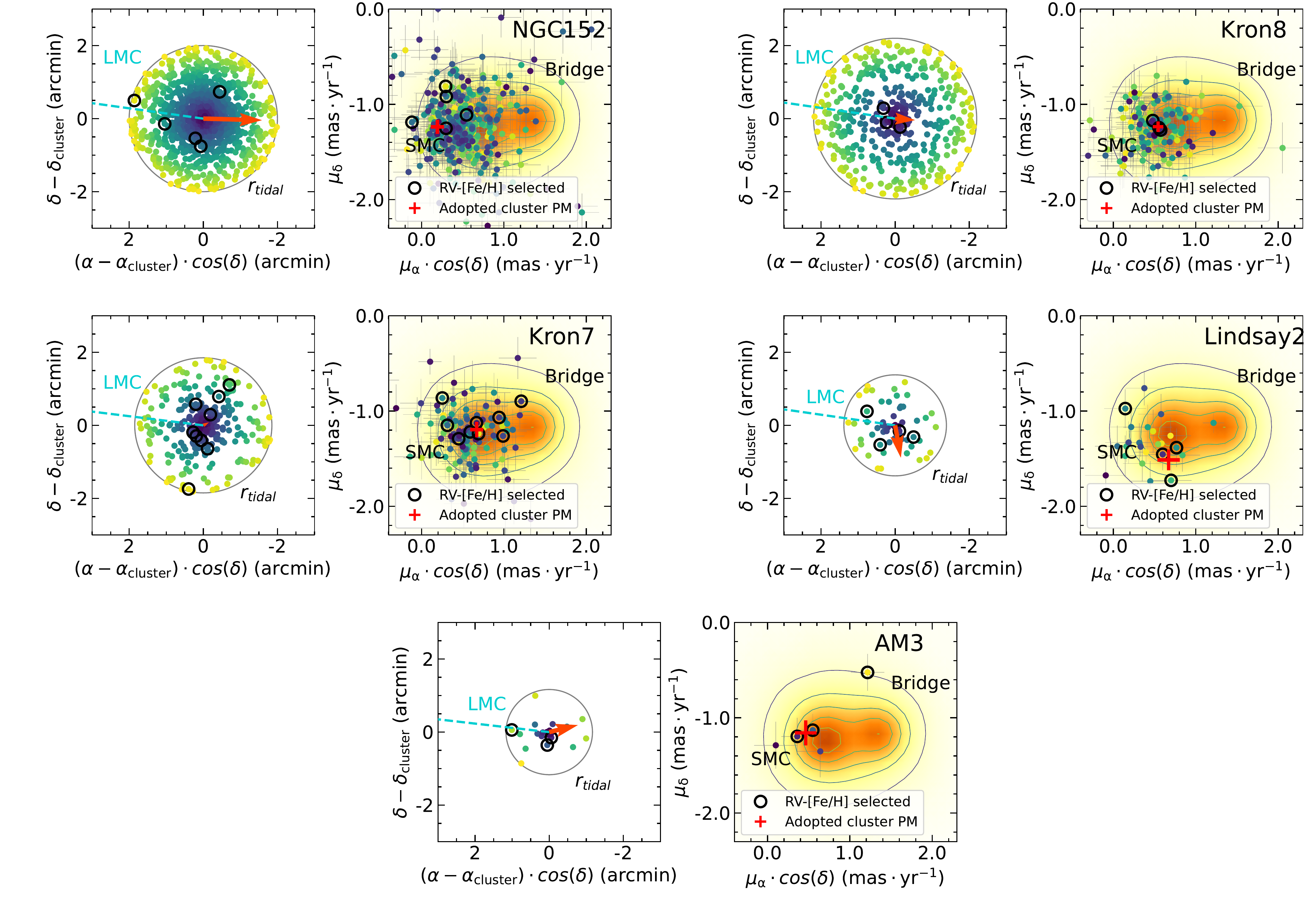}
\caption{Spatial distribution of cluster stars and Vector point diagram (VPD) for the 5 clusters using Gaia EDR3 data. Only stars within each cluster radius (from Paper II) are shown with colours representing relative distance from the respective cluster centre. The VPD smoothed orange colour shows the SMC and Magellanic Bridge regions for reference. Black circles indicate the final selected members from spectroscopy and the average is shown as red arrow on sky and red cross on the VPD. For the case of Kron\,8 the member stars are from the joint sample of GMOS/Gemini+FORS2/VLT-ESO, see text and Appendix \ref{app:gmosobs} for details.}
\label{fig:vpd}
\end{figure*}

Gaia EDR3 data was downloaded for a $7'$ region around each cluster centre coordinates to cover the entire GMOS/Gemini FOV and cluster size adopted from \citet{bica+20} as a reference. Following the selection criteria by \cite{vasiliev+18} with a more relaxed constraint on PM errors, we selected only stars with $\sigma_{\mu_{\alpha}} < 0.3\ {\rm mas\cdot yr^{-1}}$, $\sigma_{\mu_{\delta}} < 0.3\ {\rm mas\cdot yr^{-1}}$, and $\pi < 3\cdot\sigma_{\pi}$, i.e., parallax consistent with zero; moreover, only stars within the cluster radius were used. We show in Fig. \ref{fig:vpd} the on-sky distribution of stars available in Gaia EDR3 colour-coded by their distance from the cluster centre, and with PM vectors only if they comply with the selection criteria. We also show the PM distribution of the selected good-quality stars overplotted on a smoothed density map of the PM distribution of SMC and Bridge stars as a reference to guide the eye, as done in \citet{dias+21}. Cluster member stars are not selected using PMs, instead, the member stars from spectroscopic selection are identified among the good-quality Gaia PMs (identified in Fig. \ref{fig:vpd}). The cluster weighted mean PMs are the member's PM average using their uncertainties as weights after one $\sigma$-clipping loop.

\section{Discussion}

\subsection{The 3D distribution of West Halo clusters}

\begin{figure}
    \centering
    \includegraphics[width=\columnwidth]{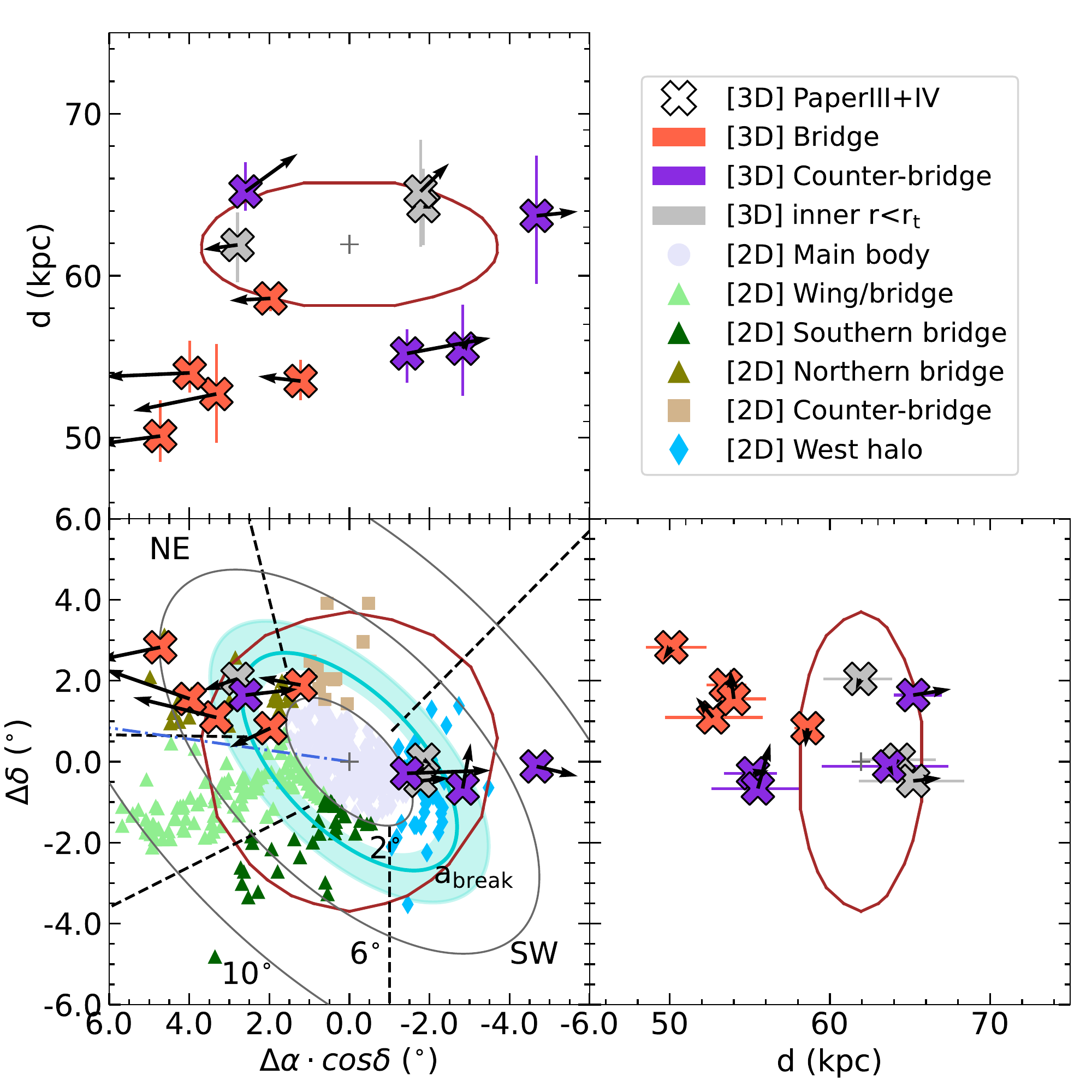}
    \caption{3D distribution of star clusters from this paper and Paper III. All catalogued SMC clusters \citep{bica+20} are displayed on the sky plane in the lower left panel with different colours and symbols and split by the ellipses and dashed lines following the 2D projected regions defined in \citet{dias+14} and Paper III and labelled in Fig. \ref{fig:SMCregions}. The break radius from Paper III is marked as blue ellipse with the respective 1$\sigma$ shaded area. A blue dot dashed line indicates the direction to the LMC. The top and right panels show only the cluster from this paper and Paper III with the line-of-sight distance information. The brown circles and ellipses are the projections of a sphere of radius 4\,kpc around the SMC centre for reference of the putative tidal radius. The velocity vectors are shown to make the classification of Bridge and Counter-bridge clusters more evident.}
    \label{fig:3Dpos}
\end{figure}

The 3D distribution of the five clusters from this work and the seven from \cite[][hereafter Paper III]{dias+21} are shown in Fig. \ref{fig:3Dpos} in three projected planes. In the sky plane we show the distribution of all SMC clusters catalogued by \cite{bica+20}, colour-coded with the classification in groups by \cite{dias+14,dias+21}. The ellipses aligned to the SMC Main Body used as a proxy for the distance from the SMC centre are drawn; in addition, the break radius of the SMC surface brightness profile (Paper III) is highlighted. The radial surface brightness (or mass) profile of galaxies are classified as type I when a single exponential can describe the profile, type II when at some break radius the slope gets steeper with a downbending brightness, and type III when at some break radius the slope gets flatter with a upbending brightness; in particular the origin of a type III profile is diverse and it is under discussion \citep[][]{martin-navarro+12}. The SMC profile using star clusters as proxies for the mass (Paper III) reveals the the SMC is a type III galaxy.
One possibility for the formation of the upbending type III profile is the accretion of stars (or gas to build extended star-forming discs) during a galaxy merger; given the many SMC-LMC interactions it would not be unnatural to think about a merger origin for the SMC profile, in fact the SMC disc plus spheroid stellar component could be explained by a merger \citep[e.g.][]{tsujimoto+09}. On the other hand, type III galaxies are usually more massive than the SMC \citep[see e.g.][and references therein]{pfeffer+21}.

Interestingly, the theoretical estimate of the SMC tidal radius of $r_{t}\sim4.5^{\circ}$ or $r_{t}\sim5~$kpc \citep{massana+20} is only $1\sigma$ larger than the break radius of $a_{break}=3.4^{\circ +1.0}_{~\ -0.6}$ (Paper III). In fact, the calculation by \cite{massana+20} was intended to match the break radius they found using field stars, but they did not discuss the uncertainties. They adopted a lighter mass for the LMC, whereas there are estimates for the LMC mass ranging from $1.4-1.9 \times 10^{11} M_{\odot}$ \citep{shipp+22}. \cite{massana+20} also changed a factor 3 by 2 in their Eq. 9 because they assume a flat rotation curve for the LMC, which is only valid outside $\sim 4$\,kpc \citep{vdM+02}. Combining only these two sources of uncertainty, the SMC tidal radius ranges from $r_t = 4.0-5.2$\,kpc according to their Eq. 9. Hence, we adopt $r_t=4$\,kpc as a putative SMC tidal radius to classify clusters as inner or outer in 3D space hereafter because this slightly smaller radius seems to agree better with the break radius of the star clusters distribution. We show in Fig. \ref{fig:3Dpos} how a sphere of 4~kpc around the SMC centre is seen projected in those panels.

The outer clusters ($r > r_{t}$) are classified as Bridge or Counter-bridge clusters, depending on their position and direction of their velocity vector, similarly to what was done in Paper III. Outer foreground Eastern clusters located outside the tidal radius ($r > r_{t}$), with line of sight distance closer than the SMC (${\rm d_{los} < d_{SMC}}$) and East from the SMC ($\Delta\alpha\cdot cos\delta > 0^{\circ}$) are classified as Bridge clusters. Outer foreground Western clusters located at $r > r_{t}$, ${\rm d_{los} < d_{SMC}}$ and $\Delta\alpha\cdot cos\delta < 0^{\circ}$ plus all outer background clusters located at $r > r_{t}$ and ${\rm d_{los} > d_{SMC}}$ are classified as Counter-bridge clusters. This 3D classification is based on full phase-space vector and therefore supersedes the initial classification in 2D projected groups, that are still used here to refer to sky regions. This classification is clear in Fig. \ref{fig:3Dpos} where Bridge clusters form a stripe moving towards the LMC and the Counter-bridge clusters are leaving the SMC not as a line, but with clusters departing from the entire half of the SMC opposite to the Bridge. 

There are no catalogued West Halo clusters more distant than the ones present here on the sky plane, but there may be other distant clusters along the line-of-sight that might help to trace the extension of the Counter-bridge. The full sample clusters from the VISCACHA survey will help answering this question.
 This interpretation of the sample clusters forming a Bridge and Counter-bridge structure seems reasonable, given the past history of interactions between LMC and SMC, and that they are currently moving away from each other. The homogeneous distances for the sample clusters were crucial for the current interpretations of the tidal structures in the SMC. The starting point of this study was the SMC projected regions on the sky plane by \cite{dias+14}, including the West Halo, that is the focus of the current work. \citet{dias+16} found the first evidence that the West Halo was possibly moving away from the SMC, which was confirmed using proper motions from Gaia DR1 and HST \citep{zivick+18}, VMC \citep{niederhofer+18} and Gaia EDR3 \citep{piatti21}. We now reveal the 3D structure of the West Halo where its large line-of-sight depth shows that the sky motion is combined to a line-of-sight motion away from us, that seems to be part of a larger structure defined as the Counter-bridge.

\subsection{Literature compilation and $\mathcal{N}$-body simulations}

We present a literature compilation of SMC clusters with available distance, radial velocities and proper motions in Appendix \ref{app:lit}. As these parameters come from heterogeneous studies, data and analysis, we use our homogeneous sample with self-consistent parameters as reference to anchor the literature parameters, and check whether the trends of different relations agree. We also compare the combined sample of our results and literature compilation against the results from $\mathcal{N}$-body simulations of \cite{diaz+12} and \cite{besla+12}.

\begin{figure*}
    \centering
    \includegraphics[width=\columnwidth]{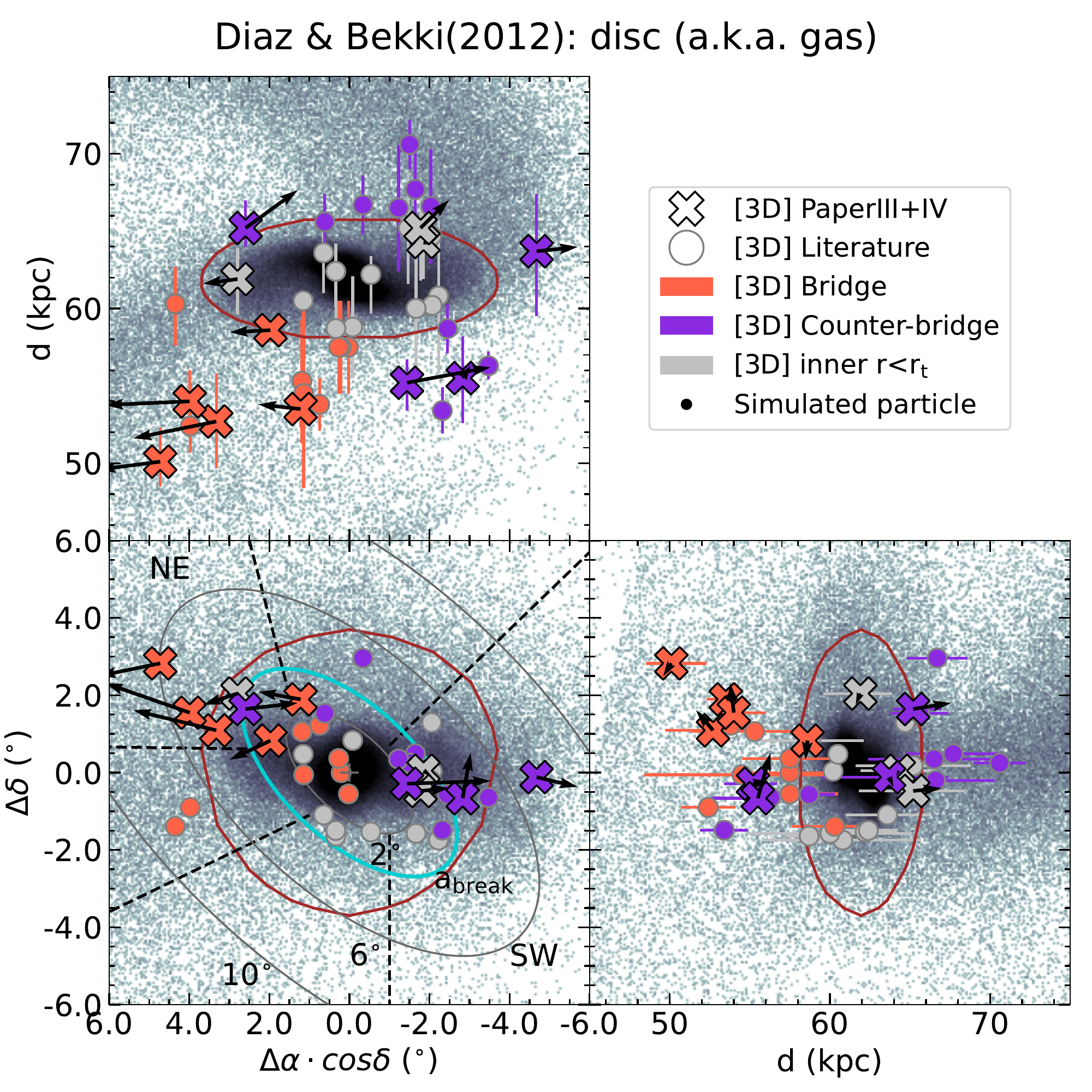}
    \includegraphics[width=\columnwidth]{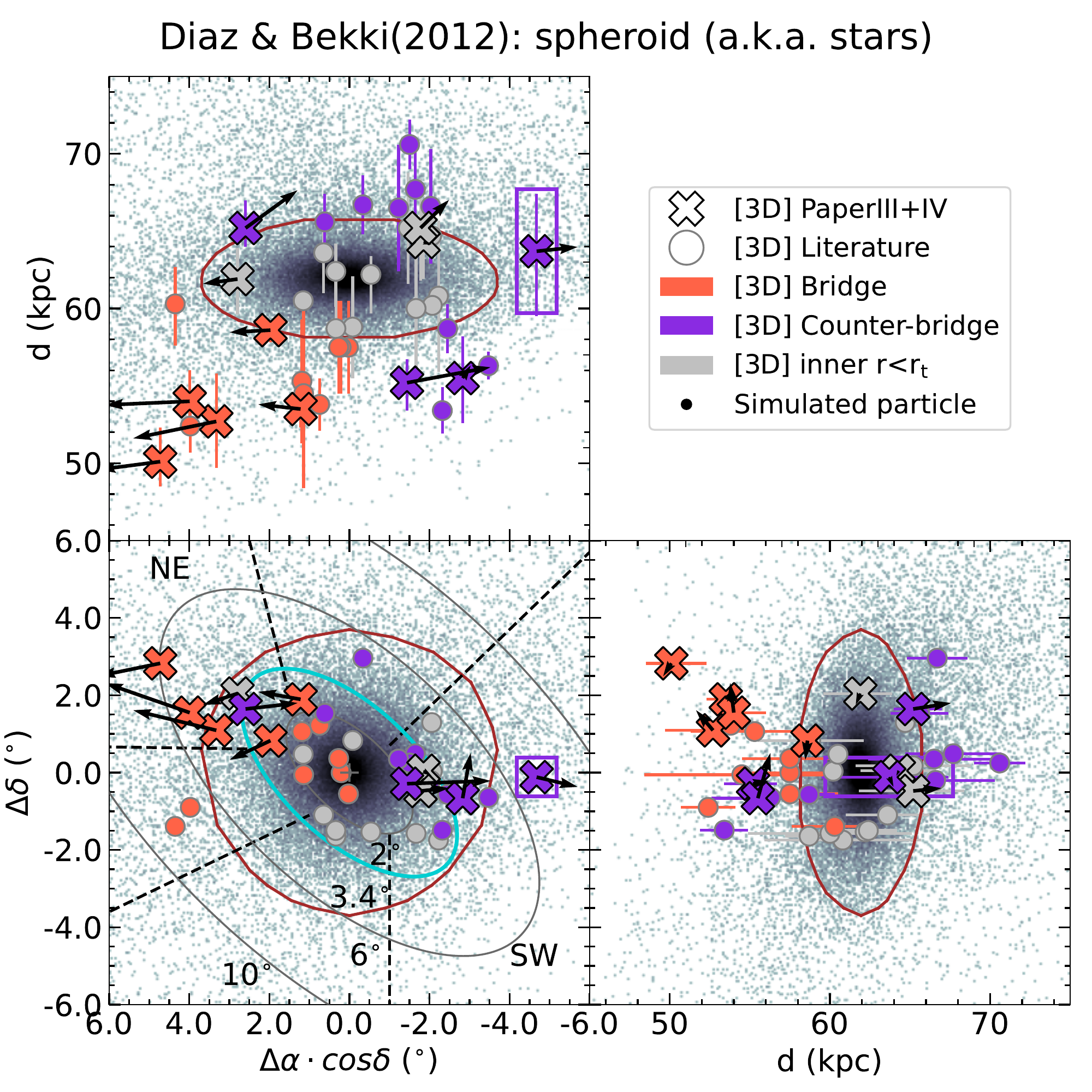}
    \includegraphics[width=\columnwidth]{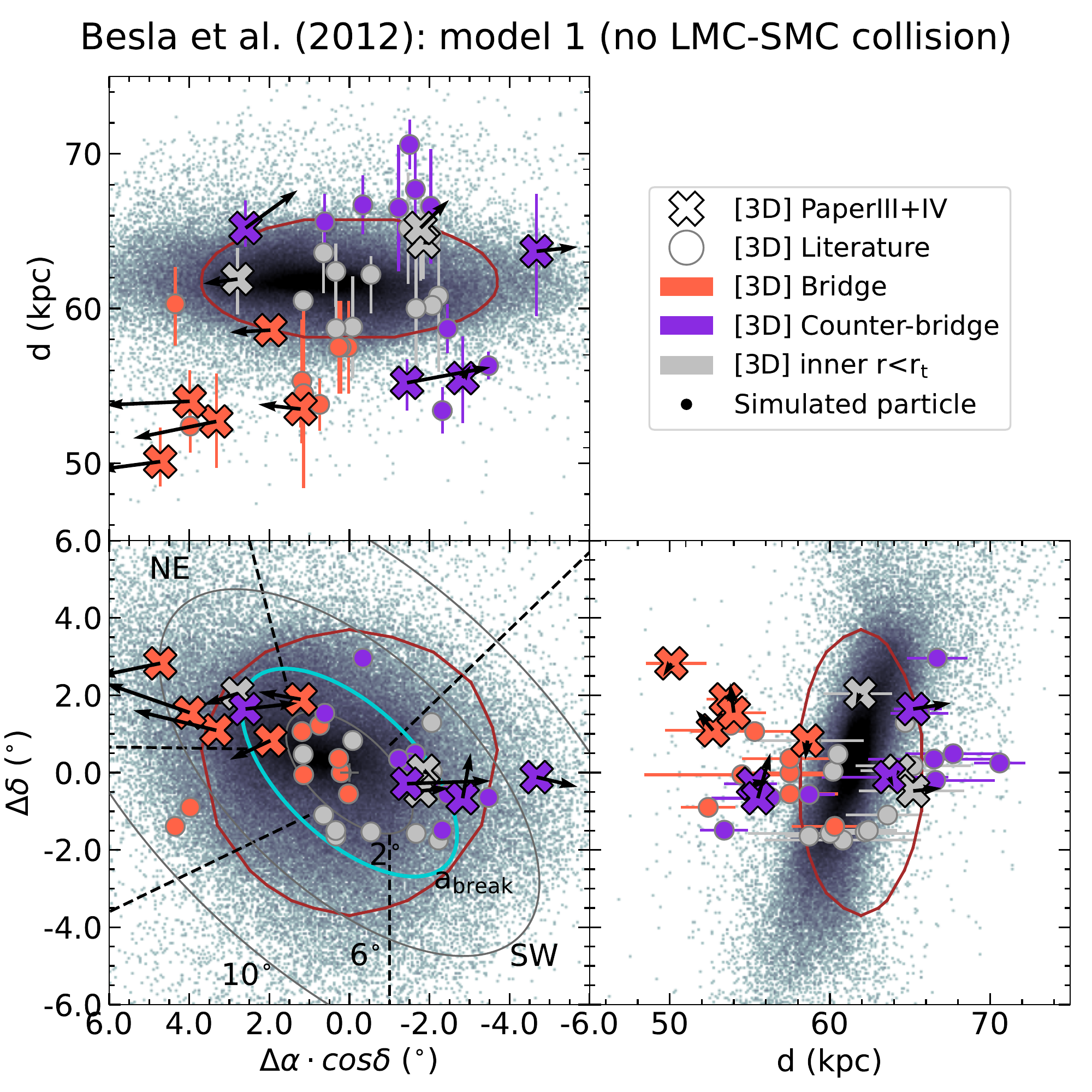}
    \includegraphics[width=\columnwidth]{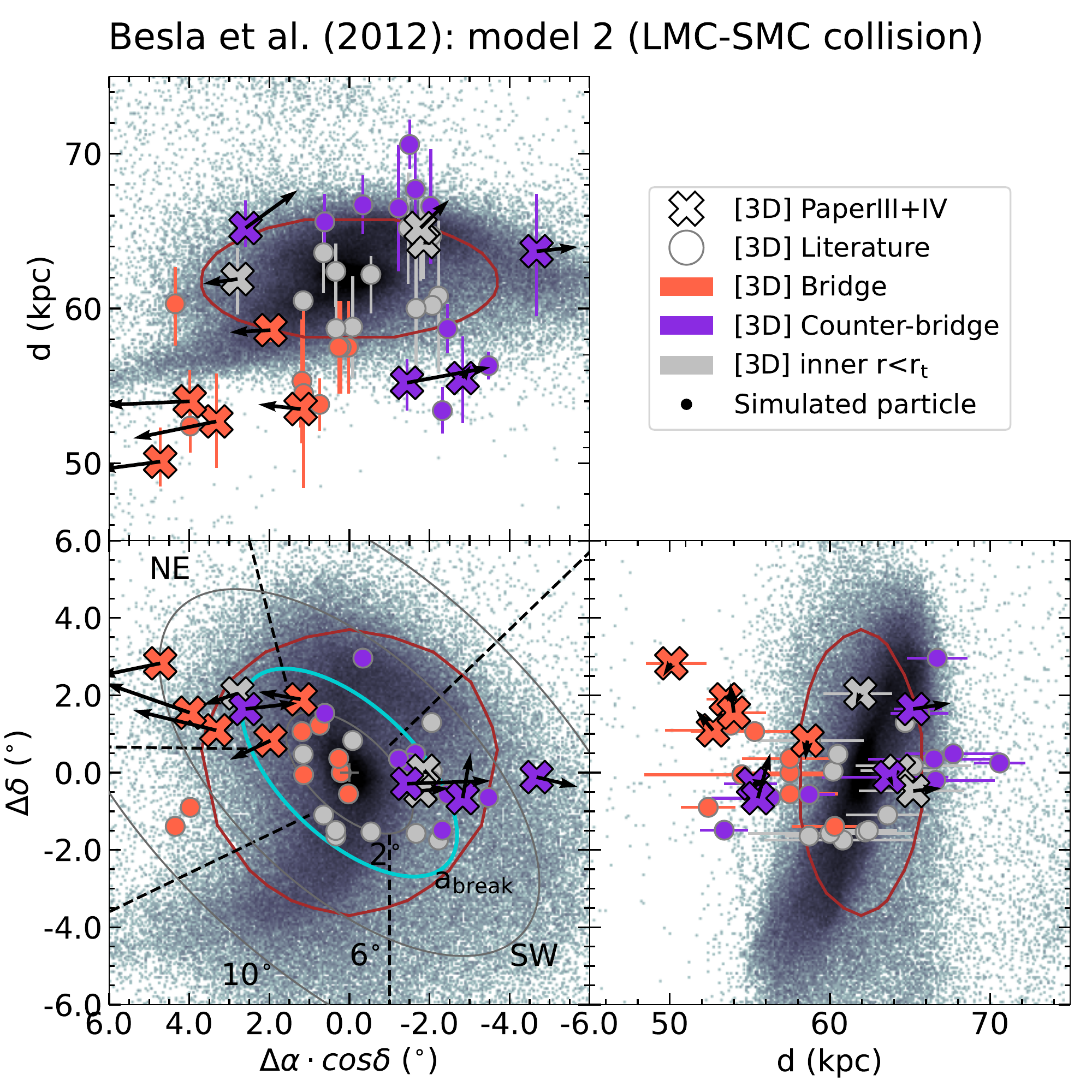}
    \caption{Similar to Fig. \ref{fig:3Dpos} now including literature compilation from Appendix \ref{app:lit} and the $\mathcal{N}$-body simulations of \citet{diaz+12} and \citet{besla+12}. The top panels show simulation results from \citet{diaz+12} where the particles represent gas and stars on the left and right panels, respectively. The bottom panels show the simulation results from \citet{besla+12} with particles representing stars in both panels, but the left panel is for their model 1 without LMC-SMC collision whereas the right panel is for their model 2 with a direct collision between LMC and SMC. The reference lines are the same as in Fig. \ref{fig:3Dpos}. The purple rectangles indicate the region around AM\,3 that is used in Fig.\ref{fig:3Dmov}.}
    \label{fig:3Dposlitsim}
\end{figure*}

Our sample clusters are located at the Northern Bridge and West Halo sky regions whereas the literature compilation sample contains clusters spread out. There are a number of clusters that occupy the Main body sky region, but when checking their line-of-sight distances they reveal to be split into inner ($r < r_t$), foreground ($r > r_t$ and ${\rm d_{los} < d_{SMC}}$) and background ($r > r_t$ and ${\rm d_{los} > d_{SMC}}$) groups, proving that the SMC tidal tails are not exclusively features on the outskirts of the projected distribution on sky as seen in Fig. \ref{fig:SMCregions}, but along the line-of-sight even along the Main Body. Therefore, a full analysis of the SMC must contain 3D information.

\cite{diaz+12} used $\mathcal{N}$-body simulations to reproduce the gaseous Magellanic Stream and they concluded that it was formed about 2\,Gyr ago when the LMC and SMC became a strongly interacting pair. Before that, LMC and SMC were independent satellites of the Milky Way. Traditionally, the SMC is assumed as rotating disc, but \cite{diaz+12} added a non-rotating spheroid as a second SMC component in their simulations representing an older stellar component subject to the same forces as the disc (gas) particles. This setup is very convenient to compare with our cluster sample with clusters older than about 1\,Gyr. In order to allow a direct comparison of the SMC 3D structure of simulations and observations, we have shifted the SMC centre from the simulations to match the optical centre adopted in this work. We show in the upper panels of Fig. \ref{fig:3Dposlitsim} the best models discussed in \cite{diaz+12} with separate components of the initial disc and spheroid particles. The disc represents the gas distribution and it would be related to recent in situ star formation along the Bridge and Counter-bridge; the spheroid represents the older stellar population composed by stars already formed at the initial conditions of the simulations 5\,Gyr ago and it would be related to tidally stripped stars towards the Bridge and Counter-bridge. The simulation results on the top-left panel clearly shows that the Counter-bridge is a broad gas structure that has its starting region between the West Halo and Northern Bridge on the background of the SMC, goes farther away and bends to the East, whereas keeping the declination roughly constant. The Counter-bridge clusters from our sample seem to confine the start of the Counter-bridge in these simulations, and the literature clusters in the background seem to be aligned with the beginning of the Counter-bridge tail from the simulations. The velocity vectors also seem to support the beginning of the simulated Counter-bridge, nevertheless a full phase-space information on clusters with line-of-sight distances larger than $\sim$66-70\,kpc are required to constrain the extension of the tail. See also next section for more details on the velocities. 

The Bridge density is clearer in the $\Delta\alpha$-d projection where there is a ubiquitous tail getting closer to us as it moves away from SMC centre, but in the sky plane the Bridge is more sparse, which is consistent with the existence of three tails of the Bridge on the sky plane as shown in Fig. \ref{fig:SMCregions}. Nevertheless, the star clusters are around 8\,kpc closer than what the simulations predicted. The upper right panel shows the same for the older stellar component, that should in principle agree better with the star clusters, because they are all older than about 1\,Gyr. However the simulation results do not show clear density regions for the Bridge and Counter-bridge as it is the case of the gas. Therefore, the motions of these simulation particles and clusters must be compared for a full understanding that will be discussed in the next section.

\cite{besla+12} performed $\mathcal{N}$-body simulations showing that the Magellanic Stream was formed from the interaction between the LMC and SMC alone, without any influence of the Milky Way, which is another evidence supporting the scenario where the binary pair LMC+SMC are in their first infall towards the Milky Way. Although both canonical scenario exemplified by \cite{diaz+12} and the new scenario exemplified by \cite{besla+12} are able to reproduce the gaseous Magellanic Bridge, the final distribution of SMC stars is not the same in both simulations. Therefore, it is very useful to use our sample clusters to compare with the simulated particles, specially because the simulated particle mass in \cite{besla+12} is $2.6\times10^3M_{\odot}$, which is closer to the cluster masses (see \citealp{santos+20}, hereafter Paper II) than to individual field stars. As in the previous case we also shifted the simulated particles from \cite{besla+12} to match the adopted centre. We show in the bottom panels of Fig. \ref{fig:3Dposlitsim} the two independent models discussed in \cite{besla+12}: model 1 does not present any collisions between LMC and SMC, whereas model 2 present collisions between the galaxies. Model 2 with collision clearly shows the formation of a Bridge and Counter-bridge in the SMC, whereas the isolated SMC does not form these structures. Nevertheless, the simulated Bridge matches the Southern Bridge sky region made of old stars \citep{belokurov+17} where there is still no clusters analysed. Along the line-of-sight, the simulated Bridge appears in the foreground of the SMC, but 4-5\,kpc more distant than the Bridge clusters. A sample of Southern bridge clusters homogeneously analysed as done here should help examine this difference further in the future. The simulated Counter-bridge seems to start in the North of the SMC bending to the West and then South reaching the West Halo region, the whole tail being at a similar line-of-sight distance. In this case, the West Halo cluster classified as Counter-bridge (AM\,3) would be located towards the end of the tail, as opposed to being at the beginning of the tail as indicated by the upper panels of Fig.\ref{fig:3Dposlitsim}. The velocity vectors for the background Counter-bridge clusters seem to follow the simulated tail on the $\Delta\alpha$-d plane, however the foreground Counter-bridge clusters do not agree. It was not expected that the simulations by \cite{besla+12} would reproduce all details from the SMC structure because the simulations did not use the SMC structure as constraints.

\subsection{The 3D motion}

\begin{figure*}
    \centering
    \includegraphics[width=0.9\textwidth]{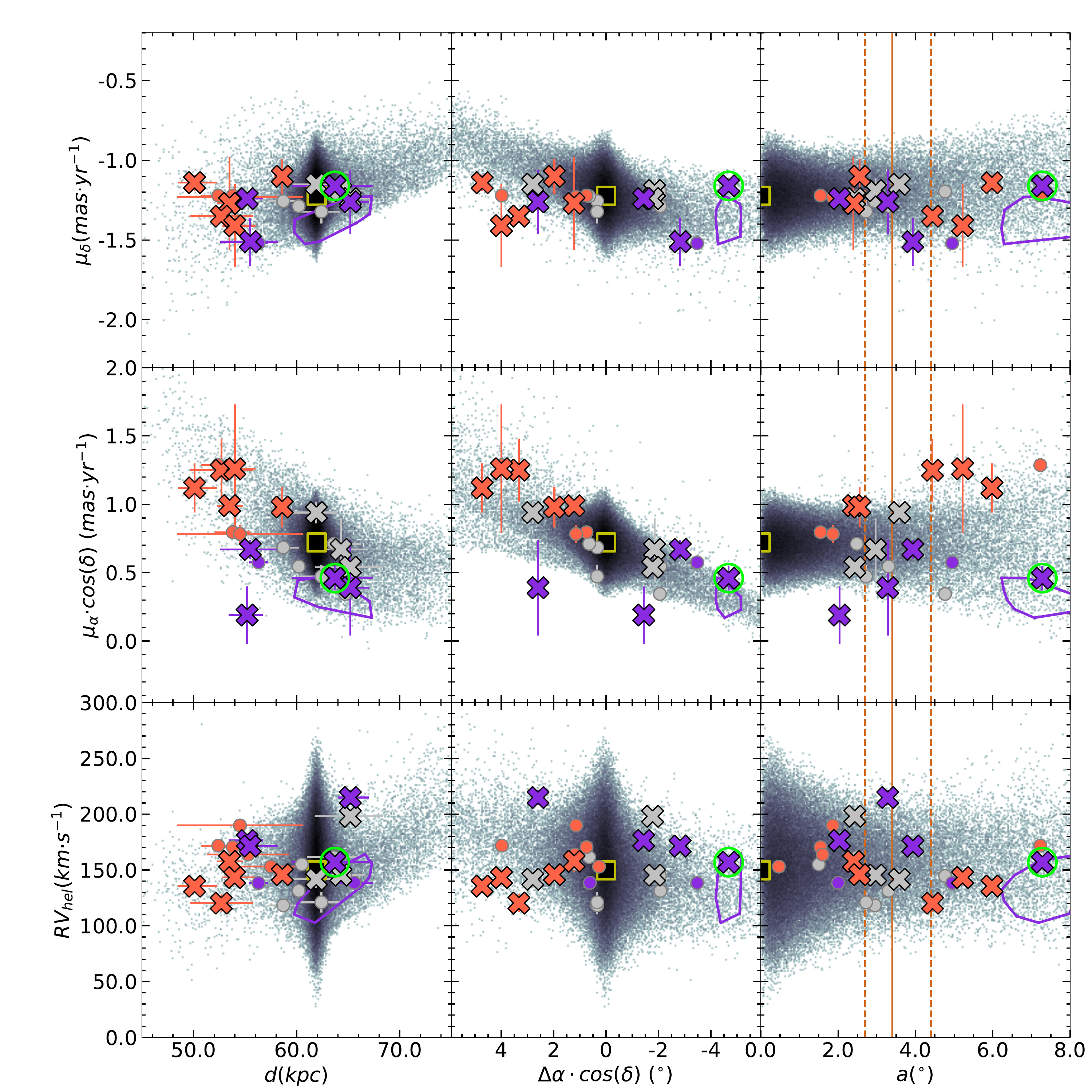}
    \caption{3D motion of star clusters from this paper, Paper III, and. literature compilation over the simulated particles from the spheroid component of the best model of \citet{diaz+12}. Colours and symbols are the same as in Fig. \ref{fig:3Dposlitsim}. The purple contours contain the simulated particles within the rectangles shown in the upper panels of Fig. \ref{fig:3Dposlitsim}, which is the region of the cluster AM\,3 represented by the purple cross close to the contour in all panels.}
    \label{fig:3Dmov}
\end{figure*}

We show in Fig. \ref{fig:3Dmov} position versus velocity distribution using observable parameters in different directions, namely, line-of-sight distance, relative right ascension, projected angular distance $a$ to the SMC centre, radial velocity, proper motions in right ascension and declination. The cluster sample is composed by the objects from this work, from Paper III and from the literature compilation whenever the information is available (see Tables  \ref{tab:results} and \ref{tab:lit}). We also show the simulated particles from the spheroid component from the best model by \cite{diaz+12}, limited to those particles contained within the limits of the upper right triple panels of Fig. \ref{fig:3Dposlitsim}. The corrected literature results (see Appendix \ref{app:lit} for details) seem to follow similar trends as the homogeneous results from the present work combined to those from Paper III. The simulations were anchored on mean RV and proper motions for the SMC slightly different from those adopted here, therefore we shifted the simulated particles to match the observations for a more straightforward comparison. The simulation and observation trends are similar, with some differences discussed below.

The left three panels of Fig. \ref{fig:3Dmov} show good agreement of simulations with observations, in particular revealing an increasing differential velocity with respect to the SMC along the line-of-sight and along the East-West direction, with no detectable trend along the North-South direction. The foreground Bridge clusters are moving towards us and towards East, i.e., the LMC, whereas both foreground and background Counter-bridge clusters are moving West and away from us, i.e., away from the LMC. One apparent disagreement between simulations and observations is that the simulations show a higher concentration of background stars in comparison with foreground stars which is not observed. On the other hand, background clusters seem to be closer to the SMC than the foreground clusters, possibly indicating that the simulated Counter-bridge is more extended than the observations reveal. A full phase-space information of more distant clusters is required to confirm this finding.

The middle three panels of Fig. \ref{fig:3Dmov} shows how the velocities behave to the East and West of the SMC. Bridge clusters are moving relatively faster from the SMC to the East along the line-of-sight and along the East-West. direction, with apparently no motion relative to the SMC along the North-South direction. This is reproduced only in the central panel by the simulations, whereas the top and bottom panels show different or opposite trends. The differences could be due to dispersion and low-number statistics, or could be related to a mismatch of the final orientation of the SMC in the simulations with observations, or simply that new simulations using the SMC structure as a constrain are required.

The right three panels of Fig. \ref{fig:3Dmov} show how the velocities change with increasing angular distance from the SMC centre on sky; it can be thought as if the middle panels folded. The most distant clusters, outside the projected break radius $a>3.4^{\circ}$ seem to be more concentrated in one stripe, whereas the simulations show two stripes for the upper and bottom panels. The middle panel shows two stripes in the simulations, one filled with Bridge clusters and another one filled with Counter-bridge clusters.

In the previous section, AM\,3 was highlighted as a case cluster to check whether it represents the beginning or the end of the Counter-bridge, supporting the simulations of \cite{diaz+12} or \cite{besla+12}, respectively, depending on the kinematics. Unfortunately, only the simulations of \cite{diaz+12} provided kinematics, therefore, no direct comparisons between the two simulations can be done, but at least we can compare in detail with one of them. Simulated stars were selected in the region of AM\,3 as shown in Fig. \ref{fig:3Dposlitsim} in a purple box within 4\,kpc in line-of-sight distance, and within 0.5$^{\circ}$ in $\Delta\delta$ and $\Delta\alpha\ cos\delta$ from the AM\,3 position. The selected particles are circled by a purple contour in all panels of Fig. \ref{fig:3Dmov}, where AM\,3 is highlighted with a green circle. In all cases, the simulated particles around AM\,3 have kinematics slightly offset with respect to the observed kinematics of AM\,3. This cluster is too far from the SMC towards the West Halo sky region, and it is isolated, making it an excellent probe of the Counter-bridge. 
\cite{tatton+21} also argued that the West Halo region had characteristics of a Counter-bridge structure moving away from the LMC. Moreover, \cite{maia+19} discovered that AM\,3 is being currently dissolved, with a clear lack of low-mass stars. Internal forces would naturally lead to cluster dissolution, given its low-mass (${\rm log(}M/M_{\odot}{\rm )}=2.90\pm0.30$, Paper II) and older age ($4.4^{+1.3}_{-1.4}$\,Gyr from the present work), as discussed by \cite{chandar+10}. Nevertheless, external forces could have boosted the cluster dissolution, in particular if the Jacobi radius  of the cluster changed throughout its lifetime (e.g. Paper II), which happens for example when the host galaxy changes its potential, i.e., when there is a close encounter with another galaxy \citep{miholics+14}. In summary, AM\,3 seems to be a very convenient target to probe the Counter-bridge tidal tail and be used to constrain future simulations.

We present additional plots in Appendix \ref{app:kinematics} to make the comparison between observations and simulations clear region by region, as there is some overlap in Fig. \ref{fig:3Dmov}. One feature that becomes very clear is that the inner clusters and simulated stars reveal some opposite outward motion as can be seen in particular in the central panel of Fig. \ref{fig:3Dmovinner} and in the vectors of Fig. \ref{fig:3Dpos}. This means that the tidal disruption of the SMC begins within the putative SMC tidal radius of 4\,kpc. In fact, \cite{deleo+20} analysed spectra for 3,000 RGB stars to get RV and combined with proper motions from Gaia DR2 and their results were compatible with the SMC bound stellar population being restricted to about $\sim2$\,kpc from the SMC centre. Doing the exercise of assuming the SMC tidal radius as 2\,kpc instead of 4\,kpc as we have done, it makes all the inner SMC simulated particles to be within the break radius $a<3.4^{\circ}$. This is consistent with the 3D structure of the SMC based on red clump (RC) and RR Lyrae stars, i.e., the old population component analysed by \citet{subramanian+12}, who found a triaxial ellipsoid with axes ratio 1:1.33:1.61 within $3^{\circ}$ from the SMC centre. The extended Bridge cluster population and simulated particles keep the same trends as shown in Fig. \ref{fig:3Dmovfore}, whereas the extended Counter-bridge cluster population shows better constrained trends connecting the points that were already in Fig. \ref{fig:3Dmovback}. In conclusion, our data also shows that the tidal disruption of the SMC starts in the very inner regions, even before reaching the putative tidal radius  of 4\,kpc. This is expected, because the Jacobi radius of the SMC shrinks during a close encounter with the LMC leaving the inner stellar content more susceptible to be tidally stripped out from the SMC main body. Now the SMC is moving away from the LMC and its Jacobi radius is apparently increasing faster than the speed of tidal removing of the SMC stars along the Bridge and Counter-bridge, generating this complex structure of tails moving away from the SMC centre, starting within the SMC tidal radius.

\subsection{The age and metallicity radial gradients}

Previous works have analysed radial gradients of age and metallicity of star clusters in the SMC using their projected distance from the SMC centre 
(Fig.\ref{fig:SMCregions}
and \citealp[e.g.][]{piatti+11b,parisi+15,dias+16}). There is a large dispersion in metallicities that prevent any smooth single radial gradient to be defined. \cite{dias+16} proposed to analyse the gradients per region and concluded that the age and metallicity gradients are well behaved if the West Halo clusters were analysed separately. In all cases, the gradients may be diluted due to projection effects, which are relevant in the case of the SMC. Now that we have the 3D spatial distribution, we are able to find the real distance of the clusters to the SMC centre and provide a clearer view on gradients.

\begin{figure*}
    \centering
    \includegraphics[width=\textwidth]{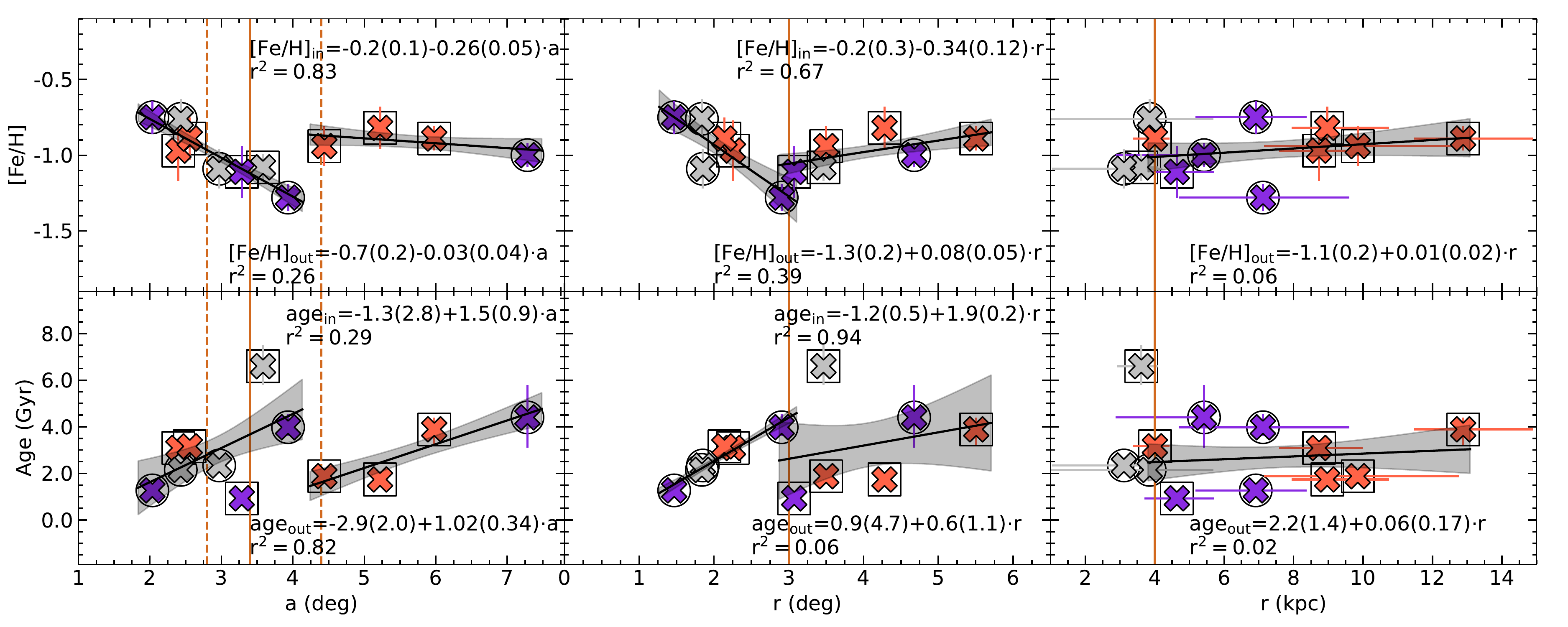}
    \caption{Age and metallicity radial gradients. The left panels use the projected semi-major axis of the ellipses from Fig.\ref{fig:SMCregions} as distance indicator, the middle panels use the projected radial angular distance, and the right panels use the real 3D distance in kpc. Colours and symbols are the same as in Fig. \ref{fig:3Dpos}. Empty circles and squares indicate the clusters from this work (West Halo) and from Paper III (Northern Bridge). From left to right, vertical lines indicate the break radius $a=3.4^{\circ +1.0}_{~\ -0.6}$ from Paper III, the old stellar spheroid radius $r\sim3^{\circ}$ from \citet{subramanian+12} and the putative tidal radius adopted in this work $r\sim4$\,kpc. The vertical dashed lines on the left panels indicate the uncertainties.}
    \label{fig:agemetgrad}
\end{figure*}

We show the age and metallicity radial gradients in Fig. \ref{fig:agemetgrad} using three different indicators for the distance from the SMC centre. The semi-major axis $a$(deg) of the ellipses from Fig. \ref{fig:SMCregions}, the on-sky projected angular distance $r$(deg), and the physical 3D distance $r$(kpc) calculated from the Cartesian coordinates defined 
by Eqs. 1,2,3,5 from \cite{vdM+01}. We adopted the optical centre of the SMC as $(\alpha_{\rm J2000},\delta_{\rm J2000},{\rm d})=(00^{\rm h}52^{\rm m}45.0^{\rm s},-72^{\circ}49'43'',61.94~{\rm kpc})$ \citep{crowl+01,degrijs+15} to convert the sky coordinates and line of sight distances into a Cartesian system centred at the SMC with z increasing towards us, x increasing towards West and y increasing towards North, i.e., the sky plane is z=0.
The inner clusters in each panel are those to the left of the vertical lines. In the left panels, the cut is at the break radius defined in Paper III, in the middle panels the cut is at the limit of the old stellar spheroid defined by \cite{subramanian+12}, and the right panels show the putative SMC tidal radius adopted in this work.

\citet{parisi+15,dias+16,bica+20} detected a negative metallicity gradient within $\sim 4^{\circ}$ from the SMC centre, with a change of slope for those clusters outside this radius. The present sample is small, nevertheless a similar behaviour is found in the projected gradients on the left and middle panels of Fig. \ref{fig:agemetgrad}. \citet{parisi+15} and Parisi et al. (submitted) found a small group of metal-poor clusters within $\sim 4^{\circ}$ that do not match the main gradient. Such clusters are not present in the current small sample, which hampers any further detailed discussion on internal gradients. There is a selection effect here, because the VISCACHA sample clusters were selected based on the external stripes (possible tidal tails) defined on the projected distribution on sky (see Fig. \ref{fig:SMCregions}), therefore we expected to find mostly external clusters anyway. Despite the small sample, the four West Halo clusters within $a < 4.4^{\circ}$ indicated by circles in Fig. \ref{fig:agemetgrad} show well behaved gradients as those found by \cite{dias+16} in age and metallicity for the same region. Specifically, the slopes as a function of $a$ for the inner clusters are $-0.30\pm0.06\ {\rm dex\ deg^{-1}}$ and $1.35\pm0.20\ {\rm Gyr\ deg^{-1}}$ for metallicity and age; these results are similar to the slopes found by \cite{dias+16} for West Halo clusters, $-0.34\pm0.21\ {\rm dex\ deg^{-1}}$ and $1.9\pm0.6\ {\rm Gyr\ deg^{-1}}$.

Perhaps the most interesting result is the comparison between projected (left and middle panels) and real 3D gradients (right panel) in Fig. \ref{fig:agemetgrad}. Anything that looks like a gradient of age and metallicity on sky disappears when we adopt the real 3D distances from the SMC centre. The uncertainties in $r$ come from the uncertainties in line of sight distance from the CMD fits, which are about $\sim 2$\,kpc on average. A larger sample and smaller uncertainties in individual distances may reveal more details, but from this initial small sample it certainly looks like that age and metallicity gradients in the SMC periphery would be a projection effect. Whether the SMC has a surviving metallicity gradient after many interactions with the LMC, tidal disruption and possibly radial migration, the trends would be found only in the inner SMC regions \citep[see e.g.][]{dobbie+14}.

Looking to the most distant clusters in the right panels of Fig. \ref{fig:agemetgrad}, we notice that
there are four Bridge clusters (in red) that are located in the Northern Bridge sky area (squares) with distances larger than $\sim8\,kpc$, i.e., an extended stripe of clusters pointing towards the LMC. \citet{zivick+18} found that the Magellanic Bridge was formed $\sim150$\,Myr ago with an impact parameter of $\sim7.5$\,kpc. Therefore, we conclude that these four clusters with ages between 1.7-3.9\,Gyr are a genuine old stellar counterpart of the Magellanic Bridge, supporting that the Bridge was formed not only by pulled gas, but also by older stars. In fact, the velocity vectors in Fig. \ref{fig:3Dpos} show that the Bridge clusters are moving towards the LMC direction, but not in a straight line; therefore the SMC-LMC past orbit and the results of tidal effects must be traced by future models for a more detailed comparison.

\section{Summary and conclusions}

We presented the full phase-space vectors for five star clusters located at the SMC West Halo sky region, i.e., the opposite side of the SMC with respect to the LMC. These clusters are roughly aligned with the SMC-LMC direction and are moving away from the SMC. In order to have a broader picture, we combined these results with the clusters from the Northern Bridge, i.e., the Northern foreground region towards the East pointing to the LMC and moving towards the LMC. The 3D distribution and motion of the clusters were used in combination with a putative tidal radius of 4\,kpc to classify the clusters as Bridge or Counter-bridge, because both sky regions analysed have a large line-of-sight depth. The Bridge is an extended tidal tail pointing to the LMC region and moving towards the LMC, whereas the Counter-bridge is an extended ring on the boundaries of the SMC opposing to the Bridge that is moving away from the SMC. More distant clusters are required to trace the extension of the Counter-bridge tail. The West Halo was first introduced by \cite{dias+16} as a structure moving away from the SMC, later confirmed by PMs \citep{niederhofer+18,zivick+18,piatti21} and proposed to be a part of the Counter-bridge by \cite{tatton+21}, and it is now confirmed to be aligned with at least one of the three branches of the Magellanic Bridge. Therefore, we conclude that the SMC West Halo region contains a leading tidal tail, currently bound to the SMC, and it is part of a larger structure called the Counter-Bridge, which is the predicted tidal counterpart of the Magellanic Bridge.
Furthermore, we found signatures of tidal disruption well within the SMC tidal radius, in agreement to the conclusions by \cite{subramanian+12} and \cite{deleo+20}, although the closest clusters in our sample are $\sim 3$\,kpc away from the SMC centre.

A comparison with a compilation of parameters from the literature showed that, in general, the literature parameters follow similar trends in 3D space and motion to our sample, even though the literature compilation includes heterogeneous data and analysis. The combined homogeneous plus literature parameters were compared to the models by \cite{diaz+12} and \cite{besla+12}, but no perfect match was found to any of the simulations. This is not totally unexpected because the simulations were intended to reproduce only the Magellanic Stream, and not all of the details of the SMC structure. The Bridge clusters seem to be 5-8\,kpc closer to us than the simulated Bridge in both cases, whereas the Counter-bridge has a different shape and orientation. The observed clusters could help constrain interaction simulations, however the Counter-bridge clusters are relatively close to the SMC. On the other hand, the foreground Counter-bridge clusters (foreground West Halo clusters) are moving West which could be compatible with \cite{diaz+12} but not with \cite{besla+12}. Future simulations must be able to reproduce the SMC structure and kinematics in a self-consistent way to further constrain the extension of the LMC-SMC interactions as \citet{zivick+19,zivick+21} started to do.

The age and metallicity radial gradients were analysed using different distance indicators. The angular projected distances confirm the negative metallicity gradient and a positive age gradient for the inner regions combined with a constant and wide age and metallicity distribution for the external clusters found by e.g. \citet{piatti+11,parisi+15,bica+20}. The use of physical 3D radial distance erases the projected gradients. The physical radial distance reveals four Bridge clusters with ages within 1.7-3.9\,Gyr that are aligned towards the LMC and located more than 8\,kpc away from the SMC centre. This distance coincides with the impact parameter of 7.5\,kpc of the LMC-SMC encounter that created the Magellanic Bridge about 150\,Myr ago \citep{zivick+18}. We conclude that these clusters are a genuine piece of the older stellar counterpart of the Magellanic Bridge, supporting that the Bridge was not only formed by removing gas from the SMC, but also removing older stars including clusters.

\section*{Acknowledgements}

MCP and BD thank Sergio Vasquez for providing his script to measure EW of CaT lines.
MA and BD thank German Gimeno for useful discussions on the GMOS/Gemini data reduction process.
BD thanks Jonathan Diaz and Gurtina Besla for kindly providing their respective simulation results from \cite{diaz+12} and \cite{besla+12} for direct comparison with our observational results.
BD thanks the Gemini SOS personnel for the pre-image data reduction.\\
This research was partially supported by the Argentinian
institution SECYT (Universidad Nacional de Córdoba),
Consejo Nacional de Investigaciones Científicas y Técnicas de la República Argentina, Agencia
Nacional de Promoción Científica y Tecnológica, and Universidad Nacional de La Plata (Argentina).\\
D.G. gratefully acknowledge financial support from the Direcci\'on de Investigaci\'on y Desarrollo de
la Universidad de La Serena through the Programa de Incentivo a la Investigaci\'on de
Acad\'emicos (PIA-DIDULS).\\
D.M. gratefully acknowledges support by the ANID BASAL project FB210003.\\
Based on observations obtained at the Southern Astrophysical Research (SOAR) telescope, which is a joint project of the Minist\'{e}rio da Ci\^{e}ncia, Tecnologia e Inova\c{c}\~{o}es (MCTI/LNA) do Brasil, the US National Science Foundation’s NOIRLab, the University of North Carolina at Chapel Hill (UNC), and Michigan State University (MSU).\\
Based on observations obtained at the international Gemini Observatory, a program of NSF’s NOIRLab, which is managed by the Association of Universities for Research in Astronomy (AURA) under a cooperative agreement with the National Science Foundation. on behalf of the Gemini Observatory partnership: the National Science Foundation (United States), National Research Council (Canada), Agencia Nacional de Investigaci\'{o}n y Desarrollo (Chile), Ministerio de Ciencia, Tecnolog\'{i}a e Innovaci\'{o}n (Argentina), Minist\'{e}rio da Ci\^{e}ncia, Tecnologia, Inova\c{c}\~{o}es e Comunica\c{c}\~{o}es (Brazil), and Korea Astronomy and Space Science Institute (Republic of Korea).\\
This work presents results from the European Space Agency (ESA) space mission Gaia. Gaia data are being processed by the Gaia Data Processing and Analysis Consortium (DPAC). Funding for the DPAC is provided by national institutions, in particular the institutions participating in the Gaia MultiLateral Agreement (MLA). The Gaia mission website is https://www.cosmos.esa.int/gaia. The Gaia archive website is https://archives.esac.esa.int/gaia.


\section*{Data availability}

The data underlying this article are available in the NOIRLab Astro Data Archive (https://astroarchive.noao.edu/) and in the Gemini Observatory Archive (https://archive.gemini.edu/).



\bibliographystyle{mnras}
\bibliography{bibliography} 

\begin{thebibliography}{}
\makeatletter
\relax
\def\mn@urlcharsother{\let\do\@makeother \do\$\do\&\do\#\do\^\do\_\do\%\do\~}
\def\mn@doi{\begingroup\mn@urlcharsother \@ifnextchar [ {\mn@doi@}
  {\mn@doi@[]}}
\def\mn@doi@[#1]#2{\def\@tempa{#1}\ifx\@tempa\@empty \href
  {http://dx.doi.org/#2} {doi:#2}\else \href {http://dx.doi.org/#2} {#1}\fi
  \endgroup}
\def\mn@eprint#1#2{\mn@eprint@#1:#2::\@nil}
\def\mn@eprint@arXiv#1{\href {http://arxiv.org/abs/#1} {{\tt arXiv:#1}}}
\def\mn@eprint@dblp#1{\href {http://dblp.uni-trier.de/rec/bibtex/#1.xml}
  {dblp:#1}}
\def\mn@eprint@#1:#2:#3:#4\@nil{\def\@tempa {#1}\def\@tempb {#2}\def\@tempc
  {#3}\ifx \@tempc \@empty \let \@tempc \@tempb \let \@tempb \@tempa \fi \ifx
  \@tempb \@empty \def\@tempb {arXiv}\fi \@ifundefined
  {mn@eprint@\@tempb}{\@tempb:\@tempc}{\expandafter \expandafter \csname
  mn@eprint@\@tempb\endcsname \expandafter{\@tempc}}}

\bibitem[\protect\citeauthoryear{{Armandroff} \& {Zinn}}{{Armandroff} \&
  {Zinn}}{1988}]{armandroff+88}
{Armandroff} T.~E.,  {Zinn} R.,  1988, \mn@doi [\aj] {10.1086/114792}, \href
  {https://ui.adsabs.harvard.edu/abs/1988AJ.....96...92A} {96, 92}

\bibitem[\protect\citeauthoryear{{Bekki} \& {Chiba}}{{Bekki} \&
  {Chiba}}{2009}]{bekki+09}
{Bekki} K.,  {Chiba} M.,  2009, \mn@doi [\pasa] {10.1071/AS08020}, \href
  {https://ui.adsabs.harvard.edu/abs/2009PASA...26...48B} {26, 48}

\bibitem[\protect\citeauthoryear{{Belokurov}, {Erkal}, {Deason}, {Koposov}, {De
  Angeli}, {Evans}, {Fraternali}  \& {Mackey}}{{Belokurov}
  et~al.}{2017}]{belokurov+17}
{Belokurov} V.,  {Erkal} D.,  {Deason} A.~J.,  {Koposov} S.~E.,  {De Angeli}
  F.,  {Evans} D.~W.,  {Fraternali} F.,   {Mackey} D.,  2017, \mn@doi [\mnras]
  {10.1093/mnras/stw3357}, \href
  {https://ui.adsabs.harvard.edu/abs/2017MNRAS.466.4711B} {466, 4711}

\bibitem[\protect\citeauthoryear{{Besla}}{{Besla}}{2011}]{beslaPhD}
{Besla} G.,  2011, PhD thesis, Harvard University

\bibitem[\protect\citeauthoryear{{Besla}, {Kallivayalil}, {Hernquist},
  {Robertson}, {Cox}, {van der Marel}  \& {Alcock}}{{Besla}
  et~al.}{2007}]{besla+07}
{Besla} G.,  {Kallivayalil} N.,  {Hernquist} L.,  {Robertson} B.,  {Cox} T.~J.,
   {van der Marel} R.~P.,   {Alcock} C.,  2007, \mn@doi [\apj]
  {10.1086/521385}, \href
  {https://ui.adsabs.harvard.edu/abs/2007ApJ...668..949B} {668, 949}

\bibitem[\protect\citeauthoryear{{Besla}, {Kallivayalil}, {Hernquist}, {van der
  Marel}, {Cox}  \& {Kere{\v{s}}}}{{Besla} et~al.}{2012}]{besla+12}
{Besla} G.,  {Kallivayalil} N.,  {Hernquist} L.,  {van der Marel} R.~P.,  {Cox}
  T.~J.,   {Kere{\v{s}}} D.,  2012, \mn@doi [\mnras]
  {10.1111/j.1365-2966.2012.20466.x}, \href
  {https://ui.adsabs.harvard.edu/abs/2012MNRAS.421.2109B} {421, 2109}

\bibitem[\protect\citeauthoryear{{Bica}, {Westera}, {Kerber}, {Dias}, {Maia},
  {Santos}, {Barbuy}  \& {Oliveira}}{{Bica} et~al.}{2020}]{bica+20}
{Bica} E.,  {Westera} P.,  {Kerber} L. d.~O.,  {Dias} B.,  {Maia} F.,  {Santos}
  Jo{\~a}o F.~C. J.,  {Barbuy} B.,   {Oliveira} R. A.~P.,  2020, \mn@doi [\aj]
  {10.3847/1538-3881/ab6595}, \href
  {https://ui.adsabs.harvard.edu/abs/2020AJ....159...82B} {159, 82}

\bibitem[\protect\citeauthoryear{{Bressan}, {Marigo}, {Girardi}, {Salasnich},
  {Dal Cero}, {Rubele}  \& {Nanni}}{{Bressan} et~al.}{2012}]{bressan+12}
{Bressan} A.,  {Marigo} P.,  {Girardi} L.,  {Salasnich} B.,  {Dal Cero} C.,
  {Rubele} S.,   {Nanni} A.,  2012, \mn@doi [\mnras]
  {10.1111/j.1365-2966.2012.21948.x}, \href
  {https://ui.adsabs.harvard.edu/abs/2012MNRAS.427..127B} {427, 127}

\bibitem[\protect\citeauthoryear{{Carretta} \& {Gratton}}{{Carretta} \&
  {Gratton}}{1997}]{carretta+97}
{Carretta} E.,  {Gratton} R.~G.,  1997, \mn@doi [\aaps] {10.1051/aas:1997116},
  \href {https://ui.adsabs.harvard.edu/abs/1997A&AS..121...95C} {121, 95}

\bibitem[\protect\citeauthoryear{{Carretta}, {Bragaglia}, {Gratton}, {D'Orazi}
  \& {Lucatello}}{{Carretta} et~al.}{2009}]{carretta+09}
{Carretta} E.,  {Bragaglia} A.,  {Gratton} R.,  {D'Orazi} V.,   {Lucatello} S.,
   2009, \mn@doi [\aap] {10.1051/0004-6361/200913003}, \href
  {https://ui.adsabs.harvard.edu/abs/2009A&A...508..695C} {508, 695}

\bibitem[\protect\citeauthoryear{{Chandar}, {Fall}  \& {Whitmore}}{{Chandar}
  et~al.}{2010}]{chandar+10}
{Chandar} R.,  {Fall} S.~M.,   {Whitmore} B.~C.,  2010, \mn@doi [\apj]
  {10.1088/0004-637X/711/2/1263}, \href
  {https://ui.adsabs.harvard.edu/abs/2010ApJ...711.1263C} {711, 1263}

\bibitem[\protect\citeauthoryear{{Cignoni}, {Cole}, {Tosi}, {Gallagher},
  {Sabbi}, {Anderson}, {Grebel}  \& {Nota}}{{Cignoni}
  et~al.}{2013}]{cignoni+13}
{Cignoni} M.,  {Cole} A.~A.,  {Tosi} M.,  {Gallagher} J.~S.,  {Sabbi} E.,
  {Anderson} J.,  {Grebel} E.~K.,   {Nota} A.,  2013, \mn@doi [\apj]
  {10.1088/0004-637X/775/2/83}, \href
  {https://ui.adsabs.harvard.edu/abs/2013ApJ...775...83C} {775, 83}

\bibitem[\protect\citeauthoryear{{Cioni} et~al.,}{{Cioni}
  et~al.}{2011}]{cioni+11}
{Cioni} M. R.~L.,  et~al., 2011, \mn@doi [\aap] {10.1051/0004-6361/201016137},
  \href {https://ui.adsabs.harvard.edu/abs/2011A&A...527A.116C} {527, A116}

\bibitem[\protect\citeauthoryear{{Cioni} et~al.,}{{Cioni}
  et~al.}{2019}]{cioni+19}
{Cioni} M. . R.~L.,  et~al., 2019, \mn@doi [The Messenger]
  {10.18727/0722-6691/5128}, \href
  {https://ui.adsabs.harvard.edu/abs/2019Msngr.175...54C} {175, 54}

\bibitem[\protect\citeauthoryear{{Coelho}}{{Coelho}}{2014}]{coelho+14}
{Coelho} P.~R.~T.,  2014, \mn@doi [\mnras] {10.1093/mnras/stu365}, \href
  {https://ui.adsabs.harvard.edu/abs/2014MNRAS.440.1027C} {440, 1027}

\bibitem[\protect\citeauthoryear{{Cole}, {Smecker-Hane}, {Tolstoy}, {Bosler}
  \& {Gallagher}}{{Cole} et~al.}{2004}]{cole+04}
{Cole} A.~A.,  {Smecker-Hane} T.~A.,  {Tolstoy} E.,  {Bosler} T.~L.,
  {Gallagher} J.~S.,  2004, \mn@doi [\mnras]
  {10.1111/j.1365-2966.2004.07223.x}, \href
  {https://ui.adsabs.harvard.edu/abs/2004MNRAS.347..367C} {347, 367}

\bibitem[\protect\citeauthoryear{{Cole}, {Tolstoy}, {Gallagher}  \&
  {Smecker-Hane}}{{Cole} et~al.}{2005}]{cole+05}
{Cole} A.~A.,  {Tolstoy} E.,  {Gallagher} John~S. I.,   {Smecker-Hane} T.~A.,
  2005, \mn@doi [\aj] {10.1086/428007}, \href
  {https://ui.adsabs.harvard.edu/abs/2005AJ....129.1465C} {129, 1465}

\bibitem[\protect\citeauthoryear{{Crowl}, {Sarajedini}, {Piatti}, {Geisler},
  {Bica}, {Clari{\'a}}  \& {Santos}}{{Crowl} et~al.}{2001}]{crowl+01}
{Crowl} H.~H.,  {Sarajedini} A.,  {Piatti} A.~E.,  {Geisler} D.,  {Bica} E.,
  {Clari{\'a}} J.~J.,   {Santos} Jo{\~a}o F.~C. J.,  2001, \mn@doi [\aj]
  {10.1086/321128}, \href
  {https://ui.adsabs.harvard.edu/abs/2001AJ....122..220C} {122, 220}

\bibitem[\protect\citeauthoryear{{Da Costa} \& {Hatzidimitriou}}{{Da Costa} \&
  {Hatzidimitriou}}{1998}]{dacosta+98}
{Da Costa} G.~S.,  {Hatzidimitriou} D.,  1998, \mn@doi [\aj] {10.1086/300340},
  \href {https://ui.adsabs.harvard.edu/abs/1998AJ....115.1934D} {115, 1934}

\bibitem[\protect\citeauthoryear{{De Leo}, {Carrera}, {No{\"e}l}, {Read},
  {Erkal}  \& {Gallart}}{{De Leo} et~al.}{2020}]{deleo+20}
{De Leo} M.,  {Carrera} R.,  {No{\"e}l} N. E.~D.,  {Read} J.~I.,  {Erkal} D.,
  {Gallart} C.,  2020, \mn@doi [\mnras] {10.1093/mnras/staa1122}, \href
  {https://ui.adsabs.harvard.edu/abs/2020MNRAS.495...98D} {495, 98}

\bibitem[\protect\citeauthoryear{{Dias} \& {Parisi}}{{Dias} \&
  {Parisi}}{2020}]{dias+20b}
{Dias} B.,  {Parisi} M.~C.,  2020, \mn@doi [\aap]
  {10.1051/0004-6361/202039055}, \href
  {https://ui.adsabs.harvard.edu/abs/2020A&A...642A.197D} {642, A197}

\bibitem[\protect\citeauthoryear{{Dias}, {Kerber}, {Barbuy}, {Santiago},
  {Ortolani}  \& {Balbinot}}{{Dias} et~al.}{2014}]{dias+14}
{Dias} B.,  {Kerber} L.~O.,  {Barbuy} B.,  {Santiago} B.,  {Ortolani} S.,
  {Balbinot} E.,  2014, \mn@doi [\aap] {10.1051/0004-6361/201322092}, \href
  {https://ui.adsabs.harvard.edu/abs/2014A&A...561A.106D} {561, A106}

\bibitem[\protect\citeauthoryear{{Dias}, {Kerber}, {Barbuy}, {Bica}  \&
  {Ortolani}}{{Dias} et~al.}{2016}]{dias+16}
{Dias} B.,  {Kerber} L.,  {Barbuy} B.,  {Bica} E.,   {Ortolani} S.,  2016,
  \mn@doi [\aap] {10.1051/0004-6361/201527558}, \href
  {https://ui.adsabs.harvard.edu/abs/2016A&A...591A..11D} {591, A11}

\bibitem[\protect\citeauthoryear{{Dias} et~al.,}{{Dias} et~al.}{2020}]{dias+20}
{Dias} B.,  et~al., 2020, in {Bragaglia} A.,  {Davies} M.,  {Sills} A.,
  {Vesperini} E.,  eds,  IAU Symposium Vol. 351, IAU Symposium. pp 89--92
  (\mn@eprint {arXiv} {1909.02566}), \mn@doi{10.1017/S174392131900694X}

\bibitem[\protect\citeauthoryear{{Dias} et~al.,}{{Dias} et~al.}{2021}]{dias+21}
{Dias} B.,  et~al., 2021, \mn@doi [\aap] {10.1051/0004-6361/202040015}, \href
  {https://ui.adsabs.harvard.edu/abs/2021A&A...647L...9D} {647, L9}

\bibitem[\protect\citeauthoryear{{Diaz} \& {Bekki}}{{Diaz} \&
  {Bekki}}{2012}]{diaz+12}
{Diaz} J.~D.,  {Bekki} K.,  2012, \mn@doi [\apj] {10.1088/0004-637X/750/1/36},
  \href {https://ui.adsabs.harvard.edu/abs/2012ApJ...750...36D} {750, 36}

\bibitem[\protect\citeauthoryear{{Diolaiti}, {Bendinelli}, {Bonaccini},
  {Close}, {Currie}  \& {Parmeggiani}}{{Diolaiti} et~al.}{2000}]{diolaiti+00}
{Diolaiti} E.,  {Bendinelli} O.,  {Bonaccini} D.,  {Close} L.,  {Currie} D.,
  {Parmeggiani} G.,  2000, \mn@doi [\aaps] {10.1051/aas:2000305}, \href
  {https://ui.adsabs.harvard.edu/abs/2000A&AS..147..335D} {147, 335}

\bibitem[\protect\citeauthoryear{{Dobbie}, {Cole}, {Subramaniam}  \&
  {Keller}}{{Dobbie} et~al.}{2014}]{dobbie+14}
{Dobbie} P.~D.,  {Cole} A.~A.,  {Subramaniam} A.,   {Keller} S.,  2014, \mn@doi
  [\mnras] {10.1093/mnras/stu910}, \href
  {https://ui.adsabs.harvard.edu/abs/2014MNRAS.442.1663D} {442, 1663}

\bibitem[\protect\citeauthoryear{{Gaia Collaboration} et~al.,}{{Gaia
  Collaboration} et~al.}{2021}]{gaiaedr3}
{Gaia Collaboration} et~al., 2021, \mn@doi [\aap]
  {10.1051/0004-6361/202039657}, \href
  {https://ui.adsabs.harvard.edu/abs/2021A&A...649A...1G} {649, A1}

\bibitem[\protect\citeauthoryear{{Glatt} et~al.,}{{Glatt}
  et~al.}{2008}]{glatt+08}
{Glatt} K.,  et~al., 2008, \mn@doi [\aj] {10.1088/0004-6256/136/4/1703}, \href
  {https://ui.adsabs.harvard.edu/abs/2008AJ....136.1703G} {136, 1703}

\bibitem[\protect\citeauthoryear{{Goudfrooij} et~al.,}{{Goudfrooij}
  et~al.}{2014}]{goudfrooij+14}
{Goudfrooij} P.,  et~al., 2014, \mn@doi [\apj] {10.1088/0004-637X/797/1/35},
  \href {https://ui.adsabs.harvard.edu/abs/2014ApJ...797...35G} {797, 35}

\bibitem[\protect\citeauthoryear{{Harris}}{{Harris}}{2007}]{harris07}
{Harris} J.,  2007, \mn@doi [\apj] {10.1086/511816}, \href
  {https://ui.adsabs.harvard.edu/abs/2007ApJ...658..345H} {658, 345}

\bibitem[\protect\citeauthoryear{{Harris} \& {Zaritsky}}{{Harris} \&
  {Zaritsky}}{2004}]{harris+04}
{Harris} J.,  {Zaritsky} D.,  2004, \mn@doi [\aj] {10.1086/381953}, \href
  {https://ui.adsabs.harvard.edu/abs/2004AJ....127.1531H} {127, 1531}

\bibitem[\protect\citeauthoryear{{Hindman}, {Kerr}  \& {McGee}}{{Hindman}
  et~al.}{1963}]{hindman+63}
{Hindman} J.~V.,  {Kerr} F.~J.,   {McGee} R.~X.,  1963, \mn@doi [Australian
  Journal of Physics] {10.1071/PH630570}, \href
  {https://ui.adsabs.harvard.edu/abs/1963AuJPh..16..570H} {16, 570}

\bibitem[\protect\citeauthoryear{{Jacyszyn-Dobrzeniecka}
  et~al.,}{{Jacyszyn-Dobrzeniecka} et~al.}{2017}]{jacyszyn+17}
{Jacyszyn-Dobrzeniecka} A.~M.,  et~al., 2017, \mn@doi [\actaa]
  {10.32023/0001-5237/67.1.1}, \href
  {https://ui.adsabs.harvard.edu/abs/2017AcA....67....1J} {67, 1}

\bibitem[\protect\citeauthoryear{{Livanou}, {Dapergolas}, {Kontizas},
  {Nordstr{\"o}m}, {Kontizas}, {Andersen}, {Dirsch}  \& {Karampelas}}{{Livanou}
  et~al.}{2013}]{livanou+13}
{Livanou} E.,  {Dapergolas} A.,  {Kontizas} M.,  {Nordstr{\"o}m} B.,
  {Kontizas} E.,  {Andersen} J.,  {Dirsch} B.,   {Karampelas} A.,  2013,
  \mn@doi [\aap] {10.1051/0004-6361/201220926}, \href
  {https://ui.adsabs.harvard.edu/abs/2013A&A...554A..16L} {554, A16}

\bibitem[\protect\citeauthoryear{{Maia}, {Corradi}  \& {Santos}}{{Maia}
  et~al.}{2010}]{maia+10}
{Maia} F.~F.~S.,  {Corradi} W.~J.~B.,   {Santos} J.~F.~C. J.,  2010, \mn@doi
  [\mnras] {10.1111/j.1365-2966.2010.17034.x}, \href
  {https://ui.adsabs.harvard.edu/abs/2010MNRAS.407.1875M} {407, 1875}

\bibitem[\protect\citeauthoryear{{Maia} et~al.,}{{Maia} et~al.}{2019}]{maia+19}
{Maia} F. F.~S.,  et~al., 2019, \mn@doi [\mnras] {10.1093/mnras/stz369}, \href
  {https://ui.adsabs.harvard.edu/abs/2019MNRAS.484.5702M} {484, 5702}

\bibitem[\protect\citeauthoryear{{Mart{\'\i}n-Navarro}
  et~al.,}{{Mart{\'\i}n-Navarro} et~al.}{2012}]{martin-navarro+12}
{Mart{\'\i}n-Navarro} I.,  et~al., 2012, \mn@doi [\mnras]
  {10.1111/j.1365-2966.2012.21929.x}, \href
  {https://ui.adsabs.harvard.edu/abs/2012MNRAS.427.1102M} {427, 1102}

\bibitem[\protect\citeauthoryear{{Massana} et~al.,}{{Massana}
  et~al.}{2020}]{massana+20}
{Massana} P.,  et~al., 2020, \mn@doi [\mnras] {10.1093/mnras/staa2451}, \href
  {https://ui.adsabs.harvard.edu/abs/2020MNRAS.498.1034M} {498, 1034}

\bibitem[\protect\citeauthoryear{{Miholics}, {Webb}  \& {Sills}}{{Miholics}
  et~al.}{2014}]{miholics+14}
{Miholics} M.,  {Webb} J.~J.,   {Sills} A.,  2014, \mn@doi [\mnras]
  {10.1093/mnras/stu1951}, \href
  {https://ui.adsabs.harvard.edu/abs/2014MNRAS.445.2872M} {445, 2872}

\bibitem[\protect\citeauthoryear{{Milone} et~al.,}{{Milone}
  et~al.}{2018}]{milone+18}
{Milone} A.~P.,  et~al., 2018, \mn@doi [\mnras] {10.1093/mnras/sty661}, \href
  {https://ui.adsabs.harvard.edu/abs/2018MNRAS.477.2640M} {477, 2640}

\bibitem[\protect\citeauthoryear{{Nidever}, {Monachesi}, {Bell}, {Majewski},
  {Mu{\~n}oz}  \& {Beaton}}{{Nidever} et~al.}{2013}]{nidever+13}
{Nidever} D.~L.,  {Monachesi} A.,  {Bell} E.~F.,  {Majewski} S.~R.,
  {Mu{\~n}oz} R.~R.,   {Beaton} R.~L.,  2013, \mn@doi [\apj]
  {10.1088/0004-637X/779/2/145}, \href
  {https://ui.adsabs.harvard.edu/abs/2013ApJ...779..145N} {779, 145}

\bibitem[\protect\citeauthoryear{{Niederhofer} et~al.,}{{Niederhofer}
  et~al.}{2018}]{niederhofer+18}
{Niederhofer} F.,  et~al., 2018, \mn@doi [\aap] {10.1051/0004-6361/201833144},
  \href {https://ui.adsabs.harvard.edu/abs/2018A&A...613L...8N} {613, L8}

\bibitem[\protect\citeauthoryear{{Parisi}, {Grocholski}, {Geisler},
  {Sarajedini}  \& {Clari{\'a}}}{{Parisi} et~al.}{2009}]{parisi+09}
{Parisi} M.~C.,  {Grocholski} A.~J.,  {Geisler} D.,  {Sarajedini} A.,
  {Clari{\'a}} J.~J.,  2009, \mn@doi [\aj] {10.1088/0004-6256/138/2/517}, \href
  {https://ui.adsabs.harvard.edu/abs/2009AJ....138..517P} {138, 517}

\bibitem[\protect\citeauthoryear{{Parisi} et~al.,}{{Parisi}
  et~al.}{2014}]{parisi+14}
{Parisi} M.~C.,  et~al., 2014, \mn@doi [\aj] {10.1088/0004-6256/147/4/71},
  \href {https://ui.adsabs.harvard.edu/abs/2014AJ....147...71P} {147, 71}

\bibitem[\protect\citeauthoryear{{Parisi}, {Geisler}, {Clari{\'a}},
  {Villanova}, {Marcionni}, {Sarajedini}  \& {Grocholski}}{{Parisi}
  et~al.}{2015}]{parisi+15}
{Parisi} M.~C.,  {Geisler} D.,  {Clari{\'a}} J.~J.,  {Villanova} S.,
  {Marcionni} N.,  {Sarajedini} A.,   {Grocholski} A.~J.,  2015, \mn@doi [\aj]
  {10.1088/0004-6256/149/5/154}, \href
  {https://ui.adsabs.harvard.edu/abs/2015AJ....149..154P} {149, 154}

\bibitem[\protect\citeauthoryear{{Pfeffer}, {Bekki}, {Forbes}, {Couch}  \&
  {Koribalski}}{{Pfeffer} et~al.}{2022}]{pfeffer+21}
{Pfeffer} J.~L.,  {Bekki} K.,  {Forbes} D.~A.,  {Couch} W.~J.,   {Koribalski}
  B.~S.,  2022, \mn@doi [\mnras] {10.1093/mnras/stab2934}, \href
  {https://ui.adsabs.harvard.edu/abs/2022MNRAS.509..261P} {509, 261}

\bibitem[\protect\citeauthoryear{{Piatti}}{{Piatti}}{2011a}]{piatti11}
{Piatti} A.~E.,  2011a, \mn@doi [\mnras] {10.1111/j.1745-3933.2011.01105.x},
  \href {https://ui.adsabs.harvard.edu/abs/2011MNRAS.416L..89P} {416, L89}

\bibitem[\protect\citeauthoryear{{Piatti}}{{Piatti}}{2011b}]{piatti+11b}
{Piatti} A.~E.,  2011b, \mn@doi [\mnras] {10.1111/j.1745-3933.2011.01145.x},
  \href {https://ui.adsabs.harvard.edu/abs/2011MNRAS.418L..69P} {418, L69}

\bibitem[\protect\citeauthoryear{{Piatti}}{{Piatti}}{2021}]{piatti21}
{Piatti} A.~E.,  2021, \mn@doi [\aap] {10.1051/0004-6361/202140643}, \href
  {https://ui.adsabs.harvard.edu/abs/2021A&A...650A..52P} {650, A52}

\bibitem[\protect\citeauthoryear{{Piatti}, {Santos}, {Clari{\'a}}, {Bica},
  {Ahumada}  \& {Parisi}}{{Piatti} et~al.}{2005}]{piatti+05}
{Piatti} A.~E.,  {Santos} J.~F.~C. J.,  {Clari{\'a}} J.~J.,  {Bica} E.,
  {Ahumada} A.~V.,   {Parisi} M.~C.,  2005, \mn@doi [\aap]
  {10.1051/0004-6361:20052982}, \href
  {https://ui.adsabs.harvard.edu/abs/2005A&A...440..111P} {440, 111}

\bibitem[\protect\citeauthoryear{{Piatti}, {Clari{\'a}}, {Bica}, {Geisler},
  {Ahumada}  \& {Girardi}}{{Piatti} et~al.}{2011}]{piatti+11}
{Piatti} A.~E.,  {Clari{\'a}} J.~J.,  {Bica} E.,  {Geisler} D.,  {Ahumada}
  A.~V.,   {Girardi} L.,  2011, \mn@doi [\mnras]
  {10.1111/j.1365-2966.2011.18627.x}, \href
  {https://ui.adsabs.harvard.edu/abs/2011MNRAS.417.1559P} {417, 1559}

\bibitem[\protect\citeauthoryear{{Putman}, {Staveley-Smith}, {Freeman},
  {Gibson}  \& {Barnes}}{{Putman} et~al.}{2003}]{putman+03}
{Putman} M.~E.,  {Staveley-Smith} L.,  {Freeman} K.~C.,  {Gibson} B.~K.,
  {Barnes} D.~G.,  2003, \mn@doi [\apj] {10.1086/344477}, \href
  {https://ui.adsabs.harvard.edu/abs/2003ApJ...586..170P} {586, 170}

\bibitem[\protect\citeauthoryear{{Rich}, {Shara}, {Fall}  \& {Zurek}}{{Rich}
  et~al.}{2000}]{rich+00}
{Rich} R.~M.,  {Shara} M.,  {Fall} S.~M.,   {Zurek} D.,  2000, \mn@doi [\aj]
  {10.1086/301156}, \href
  {https://ui.adsabs.harvard.edu/abs/2000AJ....119..197R} {119, 197}

\bibitem[\protect\citeauthoryear{{Rubele} et~al.,}{{Rubele}
  et~al.}{2018}]{rubele+18}
{Rubele} S.,  et~al., 2018, \mn@doi [\mnras] {10.1093/mnras/sty1279}, \href
  {https://ui.adsabs.harvard.edu/abs/2018MNRAS.478.5017R} {478, 5017}

\bibitem[\protect\citeauthoryear{{Rutledge}, {Hesser}, {Stetson}, {Mateo},
  {Simard}, {Bolte}, {Friel}  \& {Copin}}{{Rutledge}
  et~al.}{1997}]{rutledge+97a}
{Rutledge} G.~A.,  {Hesser} J.~E.,  {Stetson} P.~B.,  {Mateo} M.,  {Simard} L.,
   {Bolte} M.,  {Friel} E.~D.,   {Copin} Y.,  1997, \mn@doi [\pasp]
  {10.1086/133958}, \href
  {https://ui.adsabs.harvard.edu/abs/1997PASP..109..883R} {109, 883}

\bibitem[\protect\citeauthoryear{{Santos} Jo{\~a}o F.~C. et~al.,}{{Santos}
  et~al.}{2020}]{santos+20}
{Santos} Jo{\~a}o F.~C. J.,  et~al., 2020, \mn@doi [\mnras]
  {10.1093/mnras/staa2425}, \href
  {https://ui.adsabs.harvard.edu/abs/2020MNRAS.498..205S} {498, 205}

\bibitem[\protect\citeauthoryear{{Saviane}, {Da Costa}, {Held}, {Sommariva},
  {Gullieuszik}, {Barbuy}  \& {Ortolani}}{{Saviane} et~al.}{2012}]{saviane+12}
{Saviane} I.,  {Da Costa} G.~S.,  {Held} E.~V.,  {Sommariva} V.,  {Gullieuszik}
  M.,  {Barbuy} B.,   {Ortolani} S.,  2012, \mn@doi [\aap]
  {10.1051/0004-6361/201118138}, \href
  {https://ui.adsabs.harvard.edu/abs/2012A&A...540A..27S} {540, A27}

\bibitem[\protect\citeauthoryear{{Shipp} et~al.,}{{Shipp}
  et~al.}{2021}]{shipp+22}
{Shipp} N.,  et~al., 2021, arXiv e-prints, \href
  {https://ui.adsabs.harvard.edu/abs/2021arXiv210713004S} {p. arXiv:2107.13004}

\bibitem[\protect\citeauthoryear{{Song}, {Mateo}, {Bailey}, {Walker},
  {Roederer}, {Olszewski}, {Reiter}  \& {Kremin}}{{Song}
  et~al.}{2021}]{song+21}
{Song} Y.-Y.,  {Mateo} M.,  {Bailey} John~I. I.,  {Walker} M.~G.,  {Roederer}
  I.~U.,  {Olszewski} E.~W.,  {Reiter} M.,   {Kremin} A.,  2021, \mn@doi
  [\mnras] {10.1093/mnras/stab1065}, \href
  {https://ui.adsabs.harvard.edu/abs/2021MNRAS.504.4160S} {504, 4160}

\bibitem[\protect\citeauthoryear{{Souza}, {Kerber}, {Barbuy},
  {P{\'e}rez-Villegas}, {Oliveira}  \& {Nardiello}}{{Souza}
  et~al.}{2020}]{souza+20}
{Souza} S.~O.,  {Kerber} L.~O.,  {Barbuy} B.,  {P{\'e}rez-Villegas} A.,
  {Oliveira} R.~A.~P.,   {Nardiello} D.,  2020, \mn@doi [\apj]
  {10.3847/1538-4357/ab6a0f}, \href
  {https://ui.adsabs.harvard.edu/abs/2020ApJ...890...38S} {890, 38}

\bibitem[\protect\citeauthoryear{{Stetson}}{{Stetson}}{1987}]{stetson+87}
{Stetson} P.~B.,  1987, \mn@doi [\pasp] {10.1086/131977}, \href
  {https://ui.adsabs.harvard.edu/abs/1987PASP...99..191S} {99, 191}

\bibitem[\protect\citeauthoryear{{Subramanian} \& {Subramaniam}}{{Subramanian}
  \& {Subramaniam}}{2012}]{subramanian+12}
{Subramanian} S.,  {Subramaniam} A.,  2012, \mn@doi [\apj]
  {10.1088/0004-637X/744/2/128}, \href
  {https://ui.adsabs.harvard.edu/abs/2012ApJ...744..128S} {744, 128}

\bibitem[\protect\citeauthoryear{{Tatton} et~al.,}{{Tatton}
  et~al.}{2021}]{tatton+21}
{Tatton} B.~L.,  et~al., 2021, \mn@doi [\mnras] {10.1093/mnras/staa3857}, \href
  {https://ui.adsabs.harvard.edu/abs/2021MNRAS.504.2983T} {504, 2983}

\bibitem[\protect\citeauthoryear{{Tokovinin}, {Cantarutti}, {Tighe},
  {Schurter}, {Martinez}, {Thomas}  \& {van der Bliek}}{{Tokovinin}
  et~al.}{2016}]{tokovinin+16}
{Tokovinin} A.,  {Cantarutti} R.,  {Tighe} R.,  {Schurter} P.,  {Martinez} M.,
  {Thomas} S.,   {van der Bliek} N.,  2016, \mn@doi [\pasp]
  {10.1088/1538-3873/128/970/125003}, \href
  {https://ui.adsabs.harvard.edu/abs/2016PASP..128l5003T} {128, 125003}

\bibitem[\protect\citeauthoryear{{Tsujimoto} \& {Bekki}}{{Tsujimoto} \&
  {Bekki}}{2009}]{tsujimoto+09}
{Tsujimoto} T.,  {Bekki} K.,  2009, \mn@doi [\apjl]
  {10.1088/0004-637X/700/2/L69}, \href
  {https://ui.adsabs.harvard.edu/abs/2009ApJ...700L..69T} {700, L69}

\bibitem[\protect\citeauthoryear{{Vasiliev}}{{Vasiliev}}{2018}]{vasiliev+18}
{Vasiliev} E.,  2018, \mn@doi [\mnras] {10.1093/mnrasl/sly168}, \href
  {https://ui.adsabs.harvard.edu/abs/2018MNRAS.481L.100V} {481, L100}

\bibitem[\protect\citeauthoryear{{V{\'a}squez}, {Zoccali}, {Hill}, {Gonzalez},
  {Saviane}, {Rejkuba}  \& {Battaglia}}{{V{\'a}squez}
  et~al.}{2015}]{vasquez+15}
{V{\'a}squez} S.,  {Zoccali} M.,  {Hill} V.,  {Gonzalez} O.~A.,  {Saviane} I.,
  {Rejkuba} M.,   {Battaglia} G.,  2015, \mn@doi [\aap]
  {10.1051/0004-6361/201526534}, \href
  {https://ui.adsabs.harvard.edu/abs/2015A&A...580A.121V} {580, A121}

\bibitem[\protect\citeauthoryear{{V{\'a}squez} et~al.,}{{V{\'a}squez}
  et~al.}{2018}]{vasquez+18}
{V{\'a}squez} S.,  et~al., 2018, \mn@doi [\aap] {10.1051/0004-6361/201833525},
  \href {https://ui.adsabs.harvard.edu/abs/2018A&A...619A..13V} {619, A13}

\bibitem[\protect\citeauthoryear{{Zivick} et~al.,}{{Zivick}
  et~al.}{2018}]{zivick+18}
{Zivick} P.,  et~al., 2018, \mn@doi [\apj] {10.3847/1538-4357/aad4b0}, \href
  {https://ui.adsabs.harvard.edu/abs/2018ApJ...864...55Z} {864, 55}

\bibitem[\protect\citeauthoryear{{Zivick} et~al.,}{{Zivick}
  et~al.}{2019}]{zivick+19}
{Zivick} P.,  et~al., 2019, \mn@doi [\apj] {10.3847/1538-4357/ab0554}, \href
  {https://ui.adsabs.harvard.edu/abs/2019ApJ...874...78Z} {874, 78}

\bibitem[\protect\citeauthoryear{{Zivick}, {Kallivayalil}  \& {van der
  Marel}}{{Zivick} et~al.}{2021}]{zivick+21}
{Zivick} P.,  {Kallivayalil} N.,   {van der Marel} R.~P.,  2021, \mn@doi [\apj]
  {10.3847/1538-4357/abe1bb}, \href
  {https://ui.adsabs.harvard.edu/abs/2021ApJ...910...36Z} {910, 36}

\bibitem[\protect\citeauthoryear{{de Grijs} \& {Bono}}{{de Grijs} \&
  {Bono}}{2015}]{degrijs+15}
{de Grijs} R.,  {Bono} G.,  2015, \mn@doi [\aj] {10.1088/0004-6256/149/6/179},
  \href {https://ui.adsabs.harvard.edu/abs/2015AJ....149..179D} {149, 179}

\bibitem[\protect\citeauthoryear{{van Dokkum}}{{van
  Dokkum}}{2001}]{vandokkum01}
{van Dokkum} P.~G.,  2001, \mn@doi [\pasp] {10.1086/323894}, \href
  {https://ui.adsabs.harvard.edu/abs/2001PASP..113.1420V} {113, 1420}

\bibitem[\protect\citeauthoryear{{van der Marel} \& {Cioni}}{{van der Marel} \&
  {Cioni}}{2001}]{vdM+01}
{van der Marel} R.~P.,  {Cioni} M.-R.~L.,  2001, \mn@doi [\aj]
  {10.1086/323099}, \href
  {https://ui.adsabs.harvard.edu/abs/2001AJ....122.1807V} {122, 1807}

\bibitem[\protect\citeauthoryear{{van der Marel}, {Alves}, {Hardy}  \&
  {Suntzeff}}{{van der Marel} et~al.}{2002}]{vdM+02}
{van der Marel} R.~P.,  {Alves} D.~R.,  {Hardy} E.,   {Suntzeff} N.~B.,  2002,
  \mn@doi [\aj] {10.1086/343775}, \href
  {https://ui.adsabs.harvard.edu/abs/2002AJ....124.2639V} {124, 2639}

\makeatother
\end{thebibliography}



\appendix

\section{Membership selection}
\label{app:gmosobs}

For cluster membership analysis we use a combination of the distance of the targets to the cluster centre with their RVs and metallicities. We adopt the same cuts in RVs ($\pm$ 10km$\cdot$s$^{-1}$) and metallicity ($\pm$0.2 dex) as in our previous works \citep{parisi+09,parisi+15,dias+21}. Stars located at distances larger than the adopted cluster radius and with RV and metallicity values outside the aforementioned cuts are discarded as probable cluster members.

\begin{figure*}
    \centering
    $\begin{array}{cp{1cm}c}
    \includegraphics[width=0.8\columnwidth]{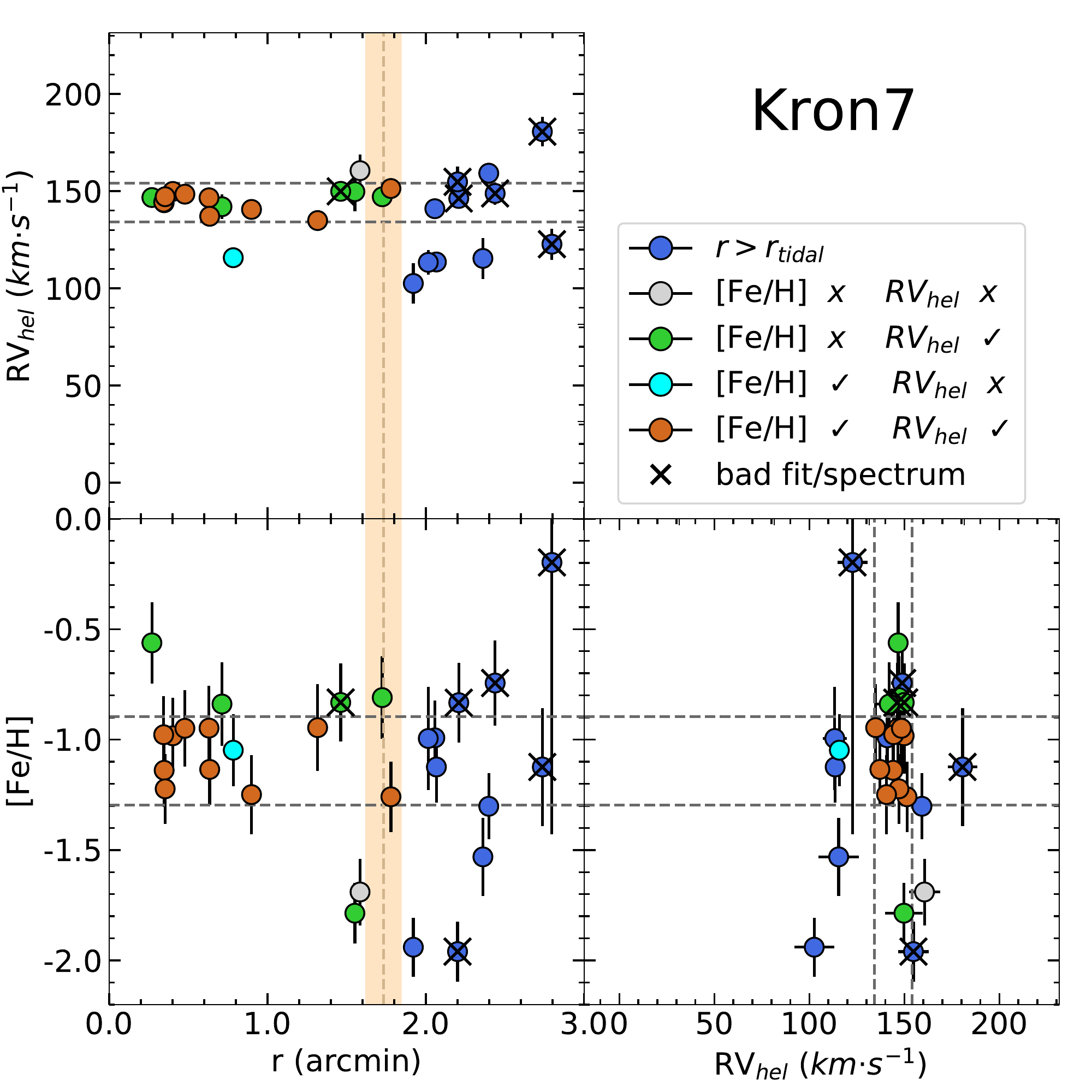} & & 
    \includegraphics[width=0.8\columnwidth]{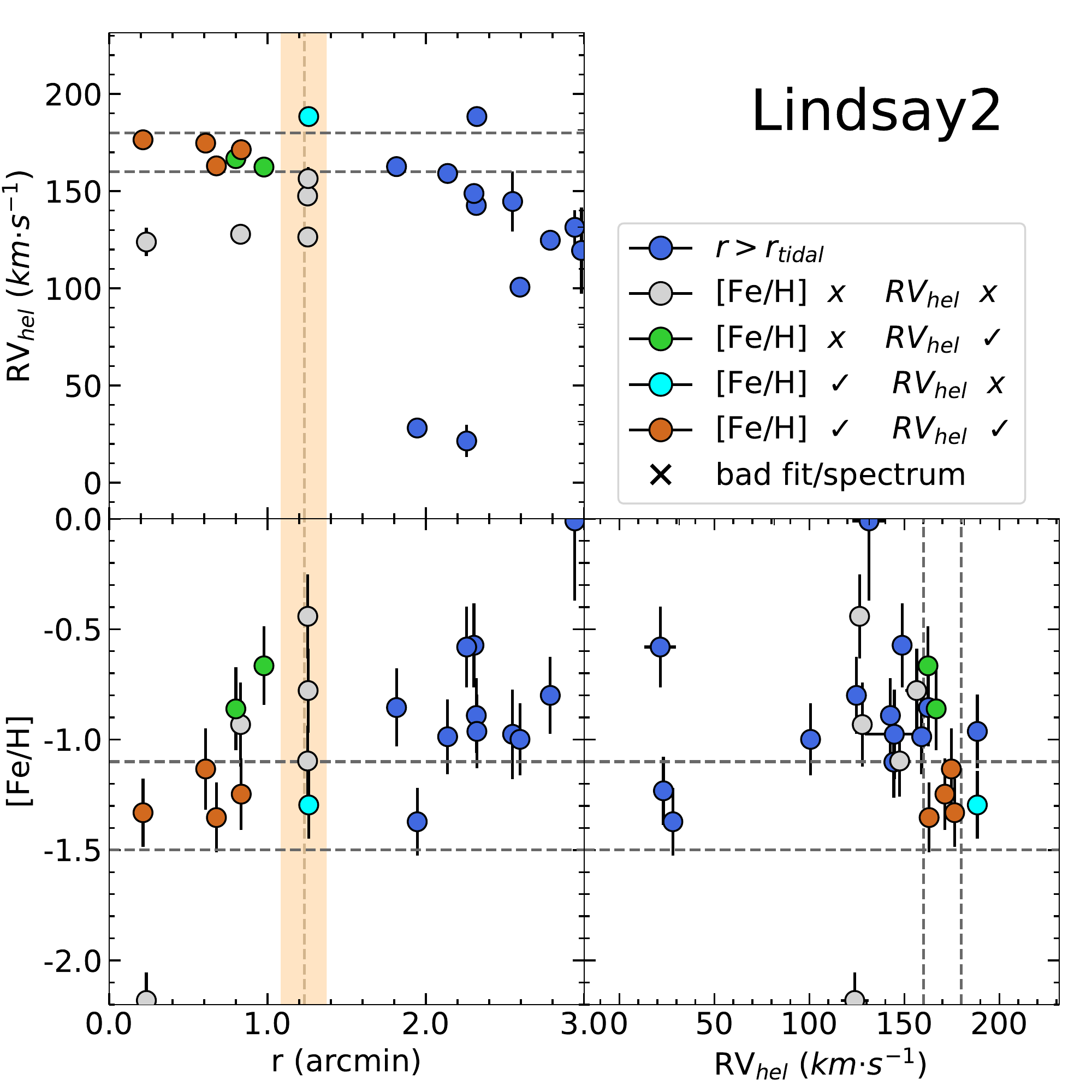}
    \end{array}$
    $\begin{array}{cp{1cm}c}
    \includegraphics[width=0.8\columnwidth]{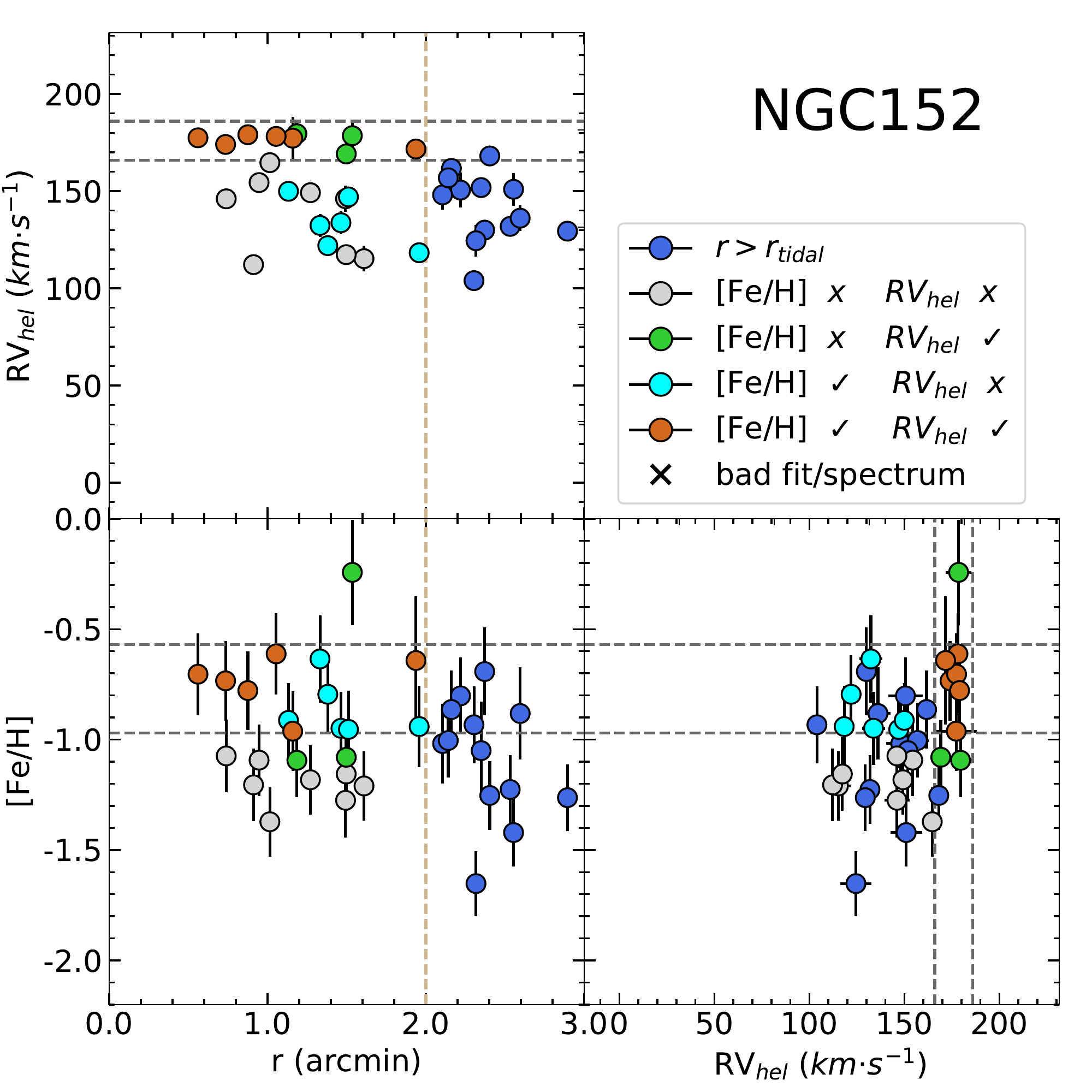} & & 
    \includegraphics[width=0.8\columnwidth]{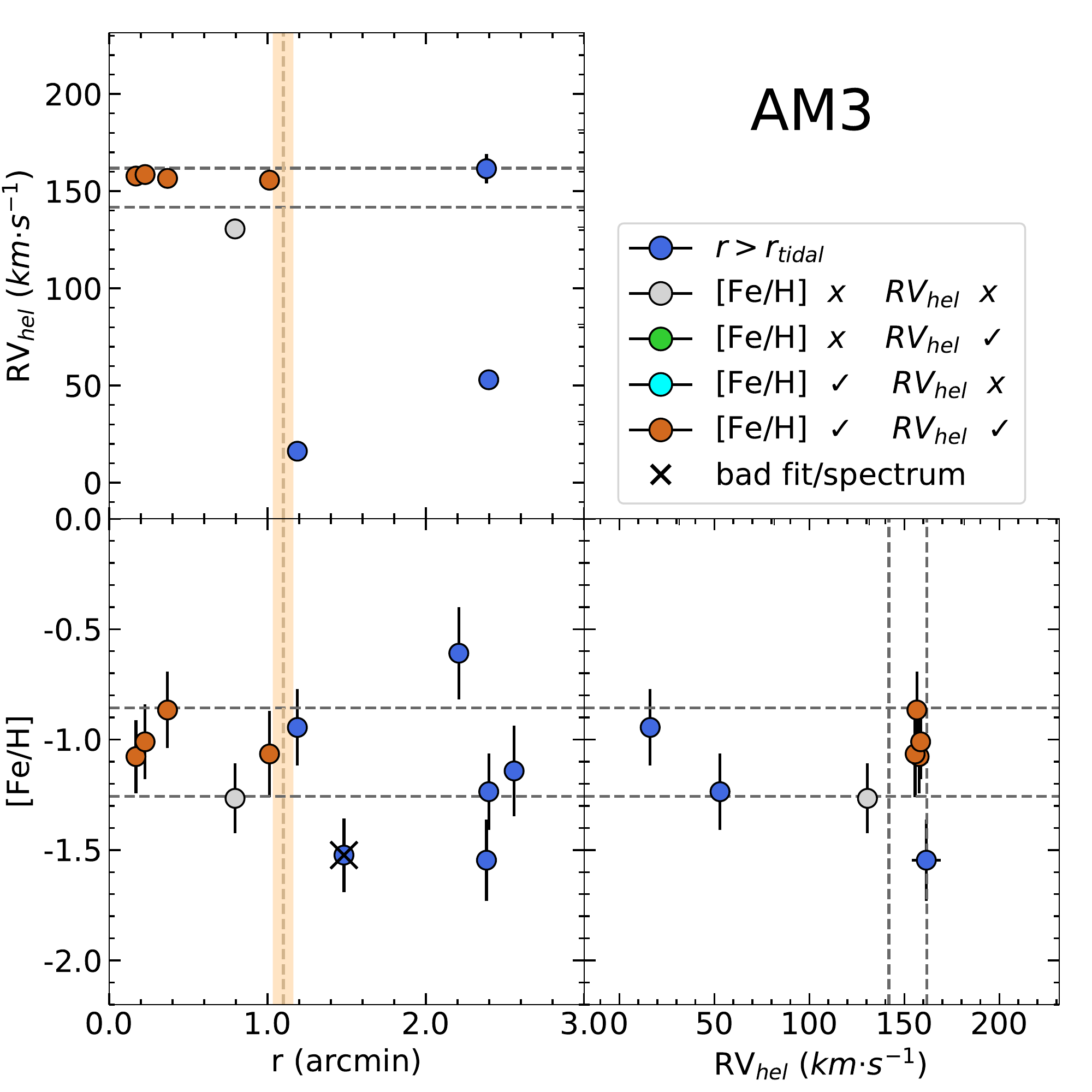}
    \end{array}$
    \caption{Membership selection of cluster stars with spectroscopic information. The shaded area marks the cluster tidal radius $\pm1\sigma$ from Paper II, otherwise only a line representing twice the visual radius by \citet{bica+20}. The limits in [Fe/H] and $RV_{hel}$ are 0.2 dex and 10km$\cdot$s$^{-1}$ around the group of innermost stars.}
    \label{fig:RVsel_appendix}
\end{figure*}

In the case of Kron\,8, it was not possible to identify member stars using the same strategy described for the other clusters, because there were only two member stars in the GMOS/Gemini sample. The membership selection for this particular cluster was done with the aid of the results by \citet{parisi+15} who also analysed Kron\,8 with exactly the same techniques employed here, but on a sample of RGB stars observed with FORS2, ESO. There are four stars in common between the GMOS/Gemini and FORS2/VLT-ESO samples with $\Delta{\rm[Fe/H]} = 0.03\pm0.13$ and $\Delta{\rm RV}=-6.6\pm7.2{\rm km\ s^{-2}}$. We consider that the metallicities are in the same scale, but shifted the RVs from P15 stars to bring them to the same RV scale of the present GMOS/Gemini data. The joint sample revealed a total of four member stars, one of them in common between the two samples. 
The average results using only the two stars from GMOS/Gemini and using the four stars from the joint GMOS/Gemini+FORS2/VLT-ESO sample are reported in Table \ref{tab:results}. We adopt the average of the four stars for the sake of using a larger sample, which is also the input to get the average Gaia PMs.

\begin{figure*}
    \centering
    \includegraphics[width=0.8\columnwidth]{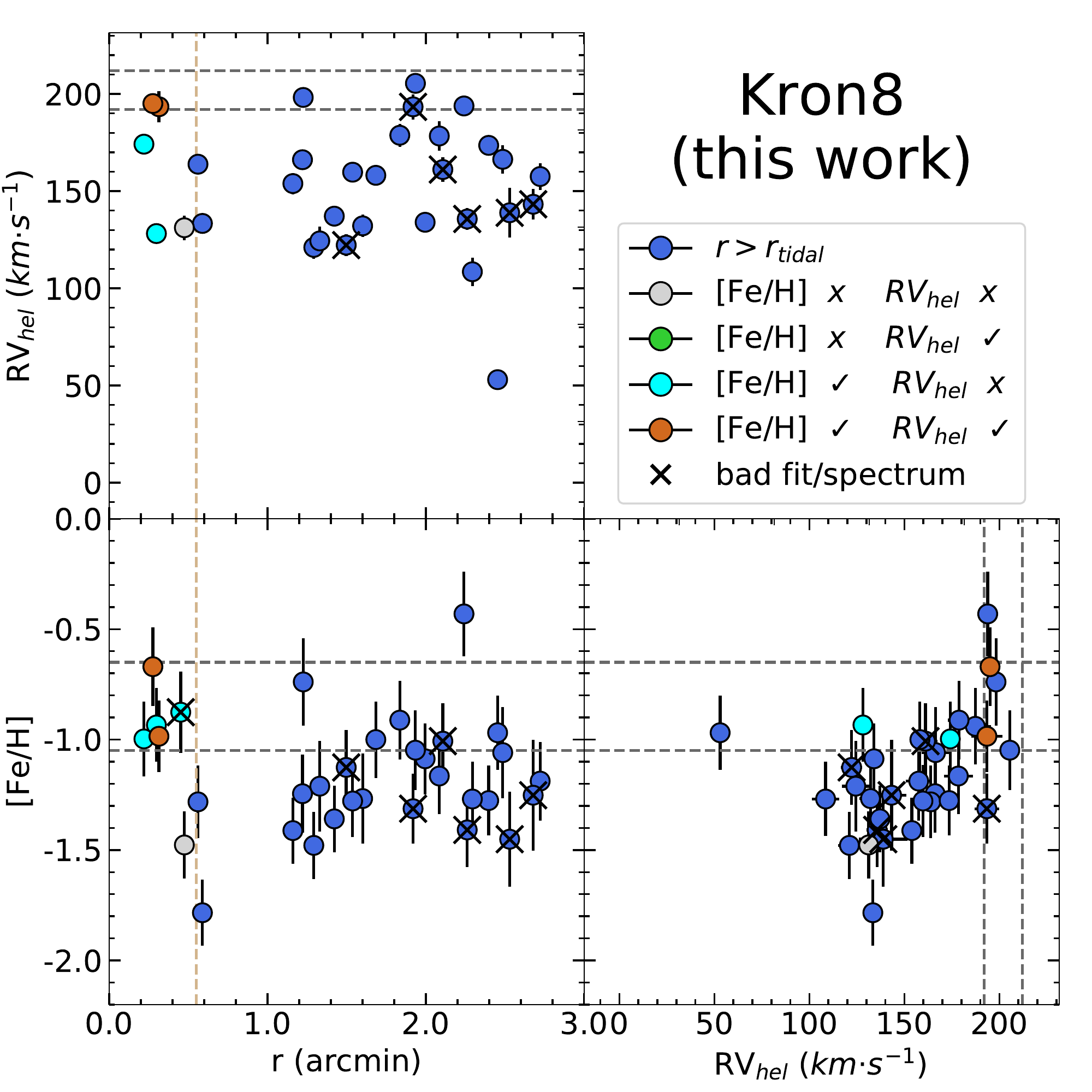}
    \includegraphics[width=0.8\columnwidth]{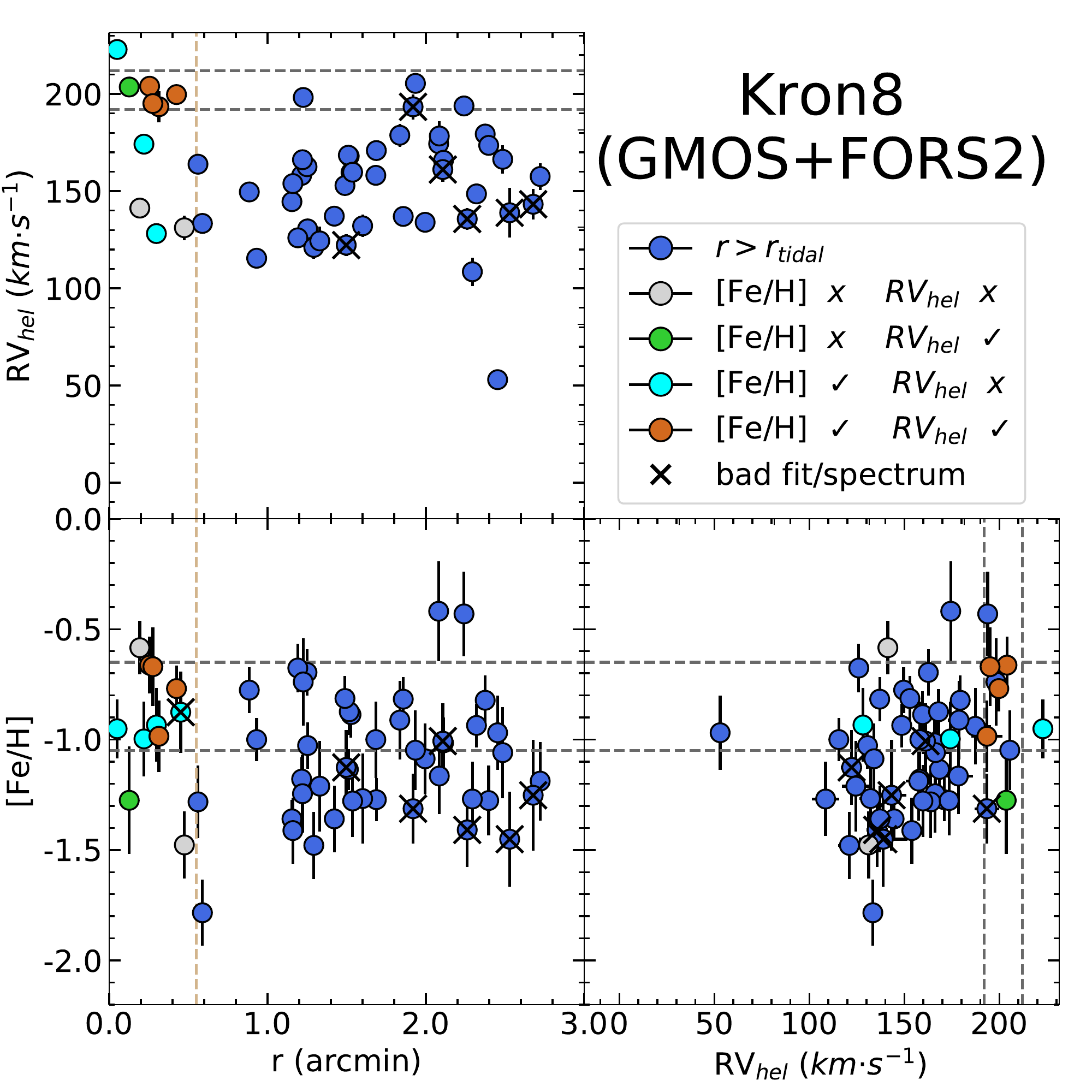}
    \caption{{\it (Left panel):} Same as Fig. \ref{fig:RVsel_appendix} for Kron\,8. {\it (Right panel):} Results from \citet[][P15]{parisi+15} using FORS2 are combined to our GMOS/Gemini results in the same RV and [Fe/H] scale. The number of member stars increase to four stars.}
    \label{fig:k8cat}
\end{figure*}

\section{SIRIUS code results}
\label{app:sirius}

\begin{figure*}
    \centering
    \includegraphics[width=0.43\textwidth]{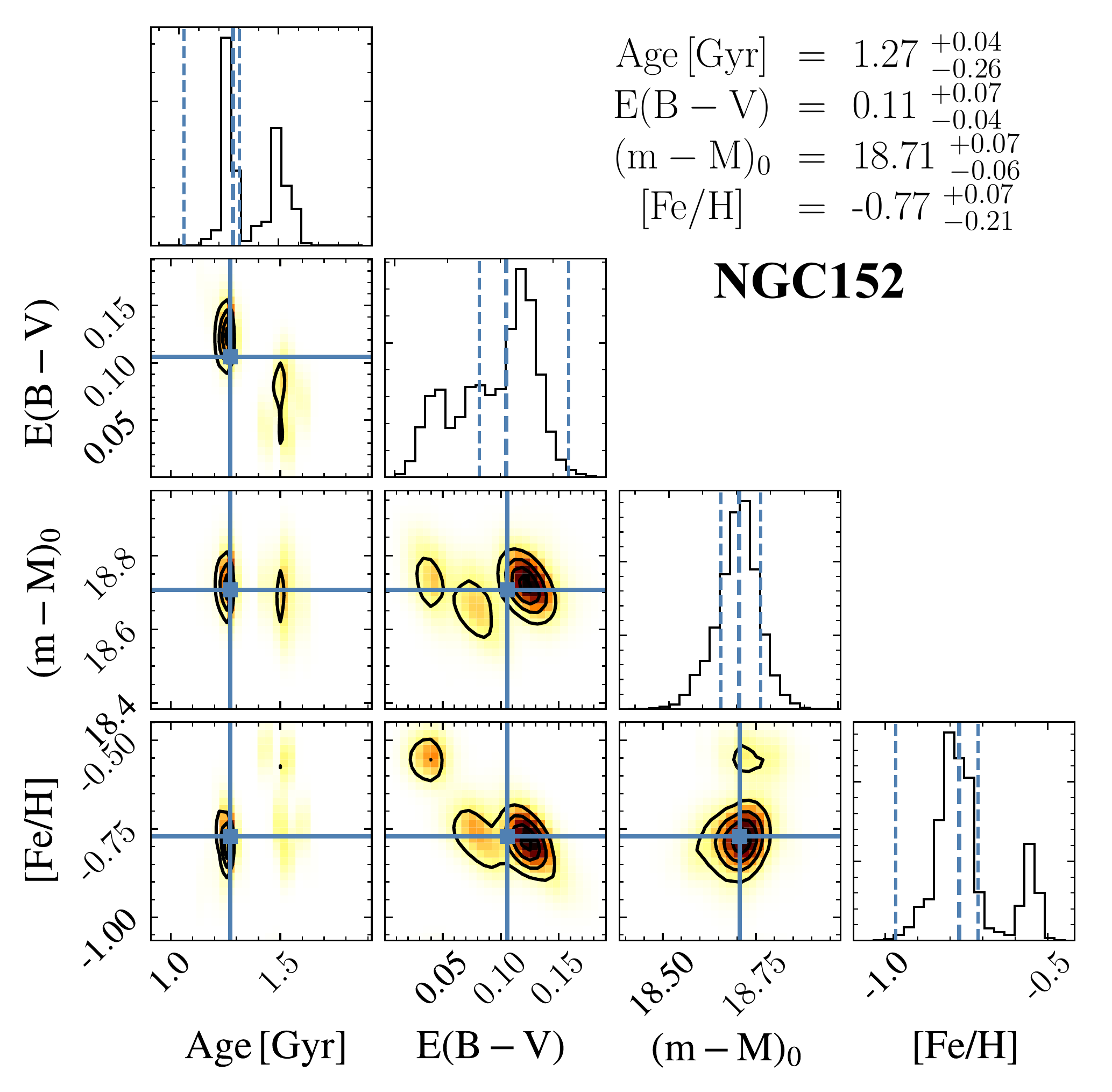}
    \includegraphics[width=0.43\textwidth]{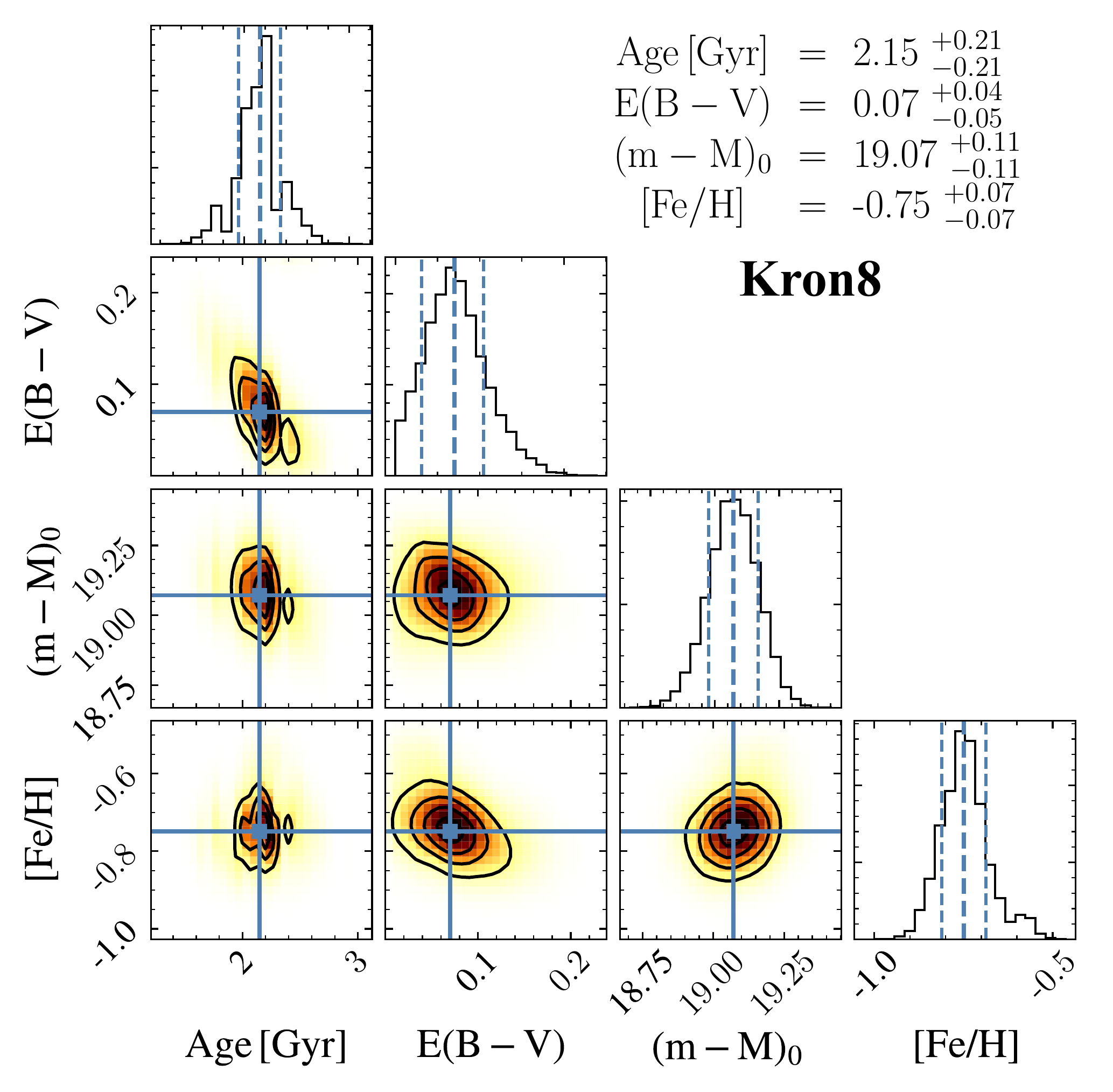}
    \includegraphics[width=0.43\textwidth]{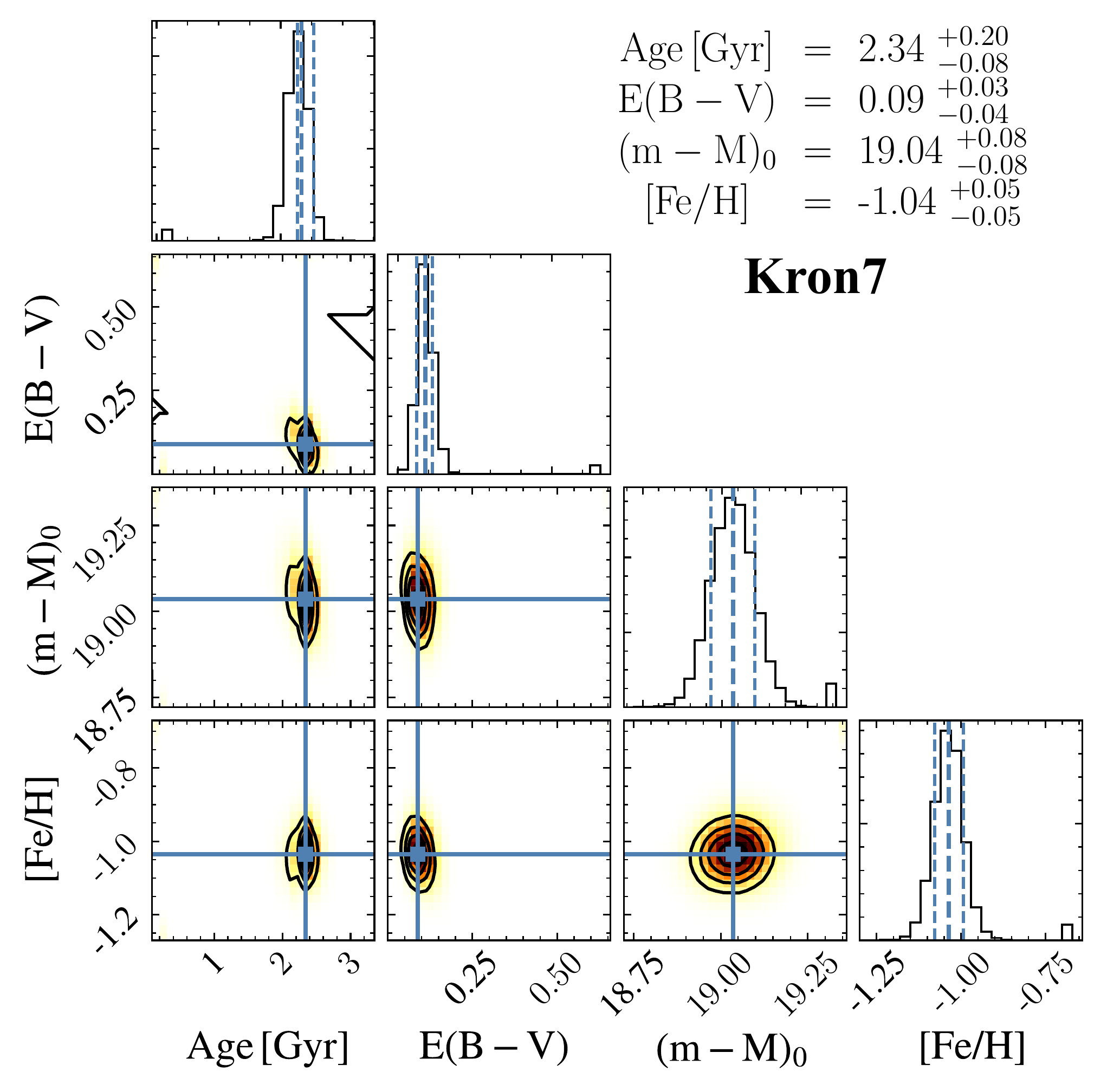}
    \includegraphics[width=0.43\textwidth]{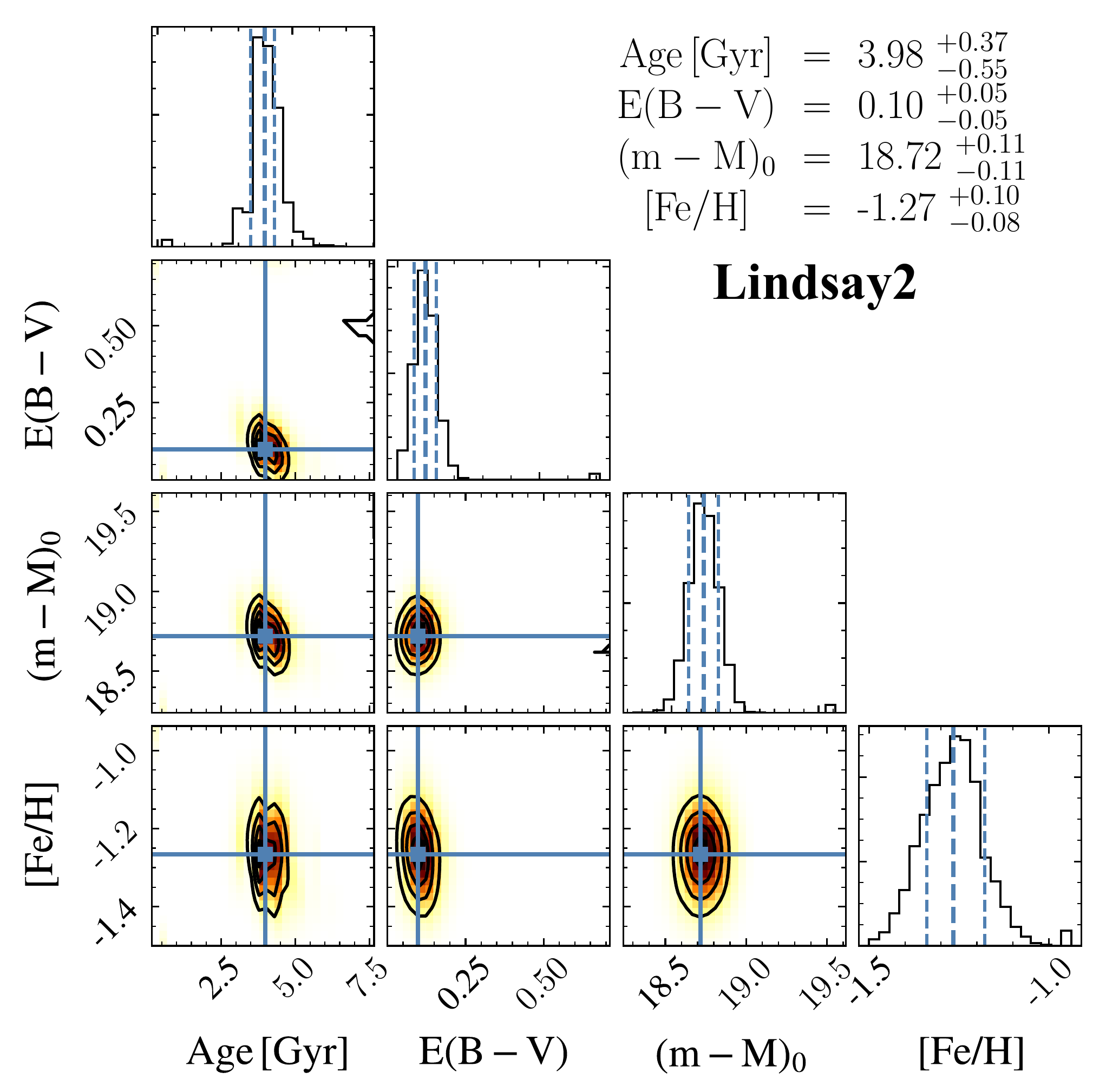}
    \includegraphics[width=0.43\textwidth]{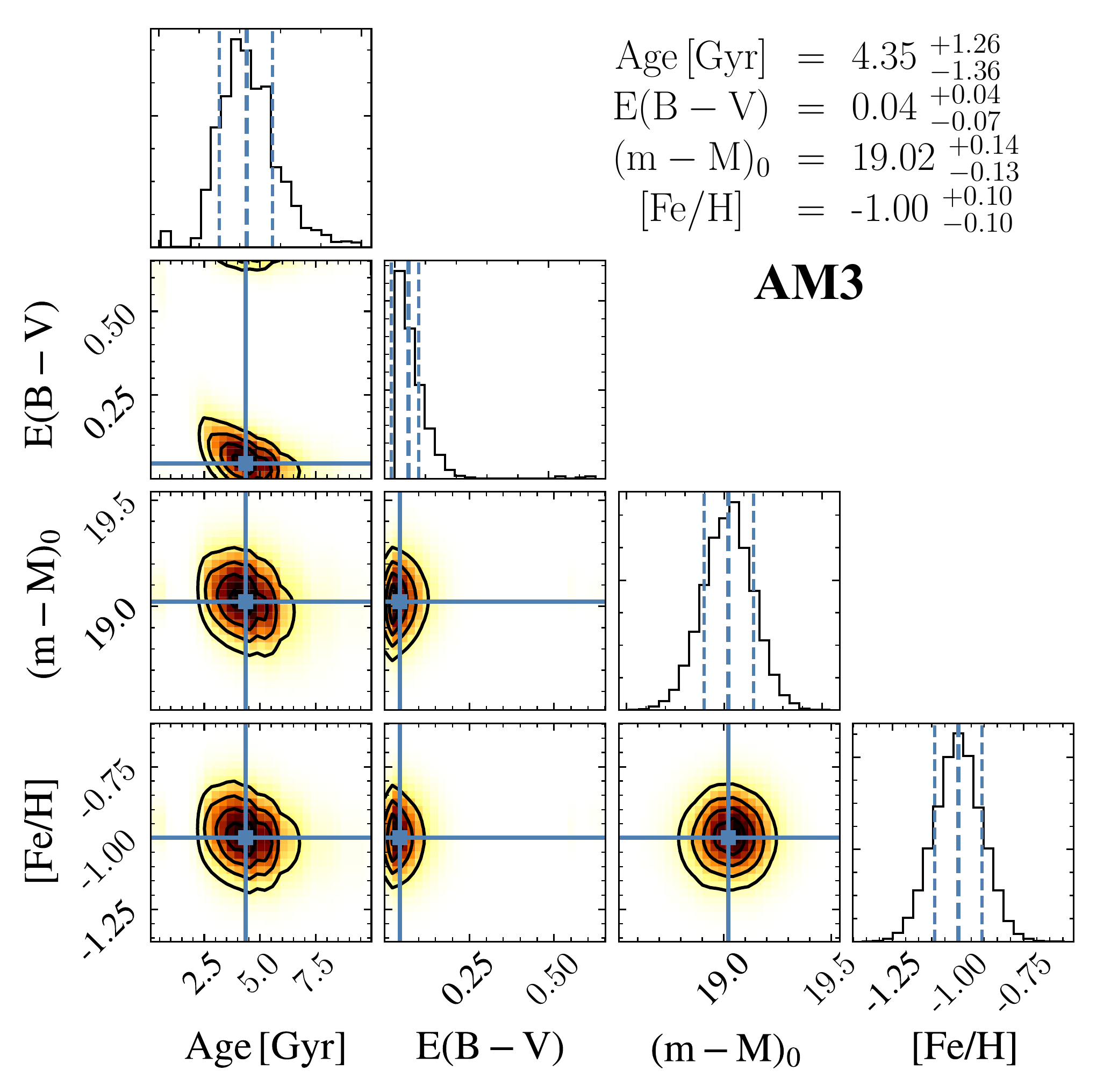}
    \caption{Posterior distributions for the isochrone fits presented in Fig.\ref{fig:CMDfit}.}
    \label{fig:posterior}
\end{figure*}

The posterior distributions of the statistical isochrone fit performed with the SIRIUS code are shown in Fig. \ref{fig:posterior}. All clusters present well behaved distributions of parameters, except NGC\,152. It is known that the phenomenon of extended main sequence turnoff (eMSTO) is more evident in star clusters around the age of NGC\,152 \citep[e.g.][Fig.7]{goudfrooij+14}. In fact, \citet{rich+00} have analysed HST CMD for NGC\,152 and found good isochrone fits for a range of ages between 1.3 and 1.8\,Gyr, for a metallicity [Fe/H]$ = -0.71$ and forcing a good fit at the RC, even though they do not discuss possibilities of eMSTO in this cluster. The CMD of NGC\,152 in Fig.\ref{fig:CMDfit} seems to present eMSTO that is reflected in a range of ages and metallicities peaks shown in the posterior distribution of Fig. \ref{fig:posterior}. This is out of the present scope and will be discussed in a separate paper. The small variations in age and metallicity do not change the conclusions of this paper.

\section{Literature compilation}
\label{app:lit}

In order to compare our total sample from Paper III and the present work in a total of 12 SMC clusters, we have compiled literature parameters on SMC clusters whenever available, even though they are from heterogeneous data sources and different techniques. No sophisticated statistics is employed here because the sources are from heterogeneous data and techniques, and few clusters are in common between different works, therefore it is hard to find a common scale for all parameters. We discuss case by case below.

We took the simple average from seven sources of cluster distance, which were determined using isochrone fitting or red clump magnitude with some corrections. \cite{crowl+01} and \cite{glatt+08} have six clusters in common with a systematic difference of 4~kpc, that we applied to all distances from \cite{crowl+01} before taking the average. We chose \cite{glatt+08} as a reference in this case because it was based on deep HST photometry. Our compiled sample has nine clusters in common with the compilation by \cite{song+21} resulting in a systematic difference of only 1~kpc and dispersion of 3~kpc, therefore we kept our averaged distances, that are reported in Table \ref{tab:lit}.

Radial velocities were obtained via low-resolution (CaT) from three studies and from high-resolution spectroscopy from one study. \cite{dacosta+98} and \cite{parisi+15} have two and four clusters in common with \cite{song+21}, where the differences in RV are 3.6 and -6.6 ${\rm km\ s^{-1}}$. We applied this offset to all RVs from \cite{dacosta+98} and \cite{parisi+15} using as reference the results by \cite{song+21}, based on high-resolution spectroscopy. The RVs from \cite{parisi+09} are in the same scale as \cite{parisi+15}, therefore we applied the same offset. A simple average was calculated for the final RVs from the four studies, as reported in Table \ref{tab:lit}.

Proper motions were estimated by \cite{piatti21} based on Gaia EDR3 data. After applying a quality filter on the selected stars around the cluster centre avoiding foreground contamination, he statistically decontaminated the cluster VPD to finally calculate the mean proper motions for each cluster.

\begin{table*}
\begin{center}
\label{tab:lit}
\caption{Average parameters from the literature as a complement to the samples from Paper III and this work. The errors here are a simple standard deviation when more than one value is available, or the reported uncertainty from the source when only one value is available.}
\begin{tabular}{lccllllccl}
\hline
    \noalign{\smallskip}
    Cluster & $\alpha_{J2000}$ & $\delta_{J2000}$ & d & ref. & $RV_{hel}$ & ref. &   $\mu_{\alpha}\cdot {\rm cos}(\delta)$ & $\mu_{\delta}$ & ref. \\
            & (hh:mm:ss.s)        & (dd:mm:ss)      & (kpc) & & (${\rm km\ s^{-1}}$) & & (${\rm mas\ yr^{-1}}$) & (${\rm mas\ yr^{-1}}$)  & \\
    \noalign{\smallskip}
    \hline
    \noalign{\smallskip}
    Lindsay~1    & 00:03:54.6 & $-$73:28:16 & $56.3\pm 0.9$ & 1,2   & $138.3\pm 4.9$ & 8,9,10 & $0.575\pm 0.011$ & $-1.520\pm 0.014$ & 12   \\
    Lindsay~3    & 00:18:25.2 & $-$74:19:05 & $53.4\pm 1.5$ & 3     & ---            & ---    & ---              & ---               & --- \\
    HW~1         & 00:18:25.9 & $-$73:23:40 & $58.7\pm 1.6$ & 3     & ---            & ---    & ---              & ---               & --- \\
    Bruck~2      & 00:19:17.7 & $-$74:34:26 & $60.8\pm 4.9$ & 4     & ---            & ---    & ---              & ---               & --- \\
    Kron~1       & 00:21:27.3 & $-$73:44:53 & ---           & ---   & $140.2\pm 1.6$ & 11     & $0.437\pm 0.044$ & $-1.290\pm 0.057$ & 12   \\
    Lindsay~5    & 00:22:41.1 & $-$75:04:31 & ---           & ---   & $153.0\pm 3.3$ & 11     & $0.529\pm 0.040$ & $-1.345\pm 0.081$ & 12   \\
    Kron~4       & 00:23:04.1 & $-$73:40:12 & ---           & ---   & $142.3\pm 2.8$ & 11     & $0.510\pm 0.060$ & $-1.280\pm 0.042$ & 12   \\
    Kron~5       & 00:24:43.4 & $-$73:45:14 & ---           & ---   & $131.4\pm 2.6$ & 11     & $0.445\pm 0.031$ & $-1.273\pm 0.079$ & 12   \\
    Kron~3       & 00:24:46.0 & $-$72:47:38 & $60.2\pm 0.5$ & 1,2   & $131.3\pm 1.6$ & 8,9,10 & $0.545\pm 0.023$ & $-1.287\pm 0.024$ & 12   \\
    Bruck~4      & 00:24:54.3 & $-$73:01:50 & $66.6\pm 3.7$ & 4     & ---            & ---    & ---              & ---               & --- \\
    Kron~6       & 00:25:26.3 & $-$74:04:30 & ---           & ---   & $157.4\pm 2.1$ & 9      & $0.419\pm 0.086$ & $-1.159\pm 0.086$ & 12   \\
    NGC~121      & 00:26:48.5 & $-$71:32:05 & $64.7\pm 0.3$ & 1,2   & $144.6\pm 4.0$ & 8      & $0.344\pm 0.025$ & $-1.196\pm 0.022$ & 12   \\
    Bruck~6      & 00:27:59.5 & $-$74:24:04 & $60.0\pm 5.1$ & 4     & ---            & ---    & ---              & ---               & --- \\
    Kron~9       & 00:30:00.3 & $-$73:22:39 & ---           & ---   & $109.5\pm 3.1$ & 9      & ---              & ---               & --- \\
    HW~5         & 00:31:02.6 & $-$72:20:26 & $67.7\pm 3.0$ & 4     & ---            & ---    & ---              & ---               & --- \\
    Lindsay~14   & 00:32:41.0 & $-$72:34:50 & $70.6\pm 1.6$ & 4     & ---            & ---    & ---              & ---               & --- \\
    HW~6         & 00:33:02.5 & $-$72:39:13 & $65.2\pm 3.6$ & 4     & ---            & ---    & ---              & ---               & --- \\
    Kron~13      & 00:35:41.7 & $-$73:35:51 & ---           & ---   & $106.0\pm 1.6$ & 11     & $0.531\pm 0.060$ & $-1.215\pm 0.054$ & 12   \\
    Kron~11      & 00:36:27.2 & $-$72:28:42 & $66.5\pm 4.1$ & 4     & ---            & ---    & ---              & ---               & --- \\
    Lindsay~19   & 00:37:41.8 & $-$73:54:27 & ---           & ---   & $152.7\pm 2.1$ & 11     & $0.634\pm 0.112$ & $-1.331\pm 0.062$ & 12   \\
    Kron~21      & 00:41:24.2 & $-$72:53:27 & ---           & ---   & $175.0\pm 2.6$ & 11     & $0.724\pm 0.043$ & $-1.427\pm 0.066$ & 12   \\
    HW~20        & 00:44:48.0 & $-$74:21:47 & $62.2^{+2.5}_{-1.2}$ & 5     & ---            & ---    & ---              & ---               & --- \\
    Lindsay~32   & 00:47:24.5 & $-$68:55:05 & ---           & ---   & ---            & ---    & ---              & ---               & --- \\
    H86-97       & 00:47:52.2 & $-$73:13:19 & ---           & ---   & $120.9\pm 2.8$ & 9      & ---              & ---               & --- \\
    Lindsay~38   & 00:48:50.4 & $-$69:52:11 & $66.7\pm 1.9$ & 1,2   & ---            & ---    & ---              & ---               & --- \\
    Kron~28      & 00:51:41.7 & $-$71:59:54 & $58.8\pm 3.3$ & 1     & ---            & ---    & ---              & ---               & --- \\
    NGC~294      & 00:53:06.0 & $-$73:22:49 & $57.5\pm 3.0$ & 6*    & ---            & ---    & ---              & ---               & --- \\
    Kron~34      & 00:55:33.2 & $-$72:49:56 & $57.5\pm 3.0$ & 6*    & ---            & ---    & ---              & ---               & --- \\
    NGC~330      & 00:56:18.7 & $-$72:27:48 & $57.5\pm 3.0$ & 6*    & $153.0\pm 0.7$ & 10     & ---              & ---               & --- \\
    NGC~339      & 00:57:47.5 & $-$74:28:17 & $58.7\pm 1.5$ & 1,2   & $118.3\pm 7.6$ & 8,10   & $0.684\pm 0.019$ & $-1.256\pm 0.018$ & 12   \\
    Kron~37      & 00:57:47.8 & $-$74:19:31 & $62.4^{+2.3}_{-1.8}$ & 5     & $121.0\pm 9.3$ & 9      & $0.472\pm 0.083$ & $-1.322\pm 0.075$ & 12   \\
    HW~40        & 01:00:25.7 & $-$71:17:39 & $65.6\pm 1.8$ & 3     & $138.5\pm 2.0$ & 9      & ---              & ---               & ---   \\
    Bruck~99     & 01:00:28.3 & $-$73:05:10 & ---           & ---   & $155.6\pm 2.6$ & 9      & ---              & ---               & ---   \\
    Kron~44      & 01:02:04.0 & $-$73:55:33 & $63.6\pm 2.6$ & 1     & $161.5\pm 1.1$ & 9      & $0.711\pm 0.038$ & $-1.225\pm 0.031$ & 12   \\
    NGC~361      & 01:02:11.0 & $-$71:36:21 & $53.8\pm 1.7$ & 1     & $170.5\pm 0.2$ & 8,10   & $0.796\pm 0.039$ & $-1.221\pm 0.035$ & 12   \\
    OGLE~133     & 01:02:31.3 & $-$72:19:06 & ---           & ---   & $145.4\pm 3.2$ & 9      & ---              & ---               & ---   \\
    NGC~376      & 01:03:53.7 & $-$72:49:32 & ---           & ---   & $145.6\pm 3.9$ & 11     & ---              & ---               & ---   \\
    HW~47        & 01:04:04.9 & $-$74:37:04 & ---           & ---   & $122.9\pm 3.4$ & 11     & ---              & ---               & ---   \\
    BS~121       & 01:04:23.8 & $-$72:50:48 & ---           & ---   & $164.1\pm 4.2$ & 11     & $0.654\pm 0.081$ & $-1.143\pm 0.046$ & 12   \\
    NGC~411      & 01:07:55.3 & $-$71:46:04 & $55.3\pm 4.0$ & 1,7   & $163.8^{+4.5}_{-0.3}$ & 10     & ---              & ---               & ---   \\
    NGC~416      & 01:07:59.1 & $-$72:21:18 & $60.5\pm 0.2$ & 1,2   & $155.0^{+1.0}_{-0.5}$ & 10     & ---              & ---               & ---   \\
    NGC~419      & 01:08:18.0 & $-$72:53:02 & $54.5\pm 6.1$ & 2,7   & $189.9^{+0.3}_{-0.2}$ & 10     & $0.783\pm 0.063$ & $-1.230\pm 0.029$ & 12   \\
    HW~67        & 01:13:01.6 & $-$70:57:47 & ---           & ---   & $106.4\pm 3.1$ & 9      & ---              & ---               & ---   \\
    NGC~458      & 01:14:52.8 & $-$71:33:01 & ---           & ---   & $149.0^{+0.8}_{-0.9}$ & 10     & ---              & ---               & ---   \\
    Lindsay~106  & 01:30:38.0 & $-$76:03:16 & ---           & ---   & $165.8\pm 3.3$ & 11     & $1.125\pm 0.078$ & $-1.313\pm 0.041$ & 12   \\
    Lindsay~108  & 01:31:38.2 & $-$71:56:50 & ---           & ---   & $ 95.0\pm 4.0$ & 11     & ---              & ---               & ---   \\
    Lindsay~110  & 01:34:26.0 & $-$72:52:28 & ---           & ---   & $178.8\pm 3.0$ & 11     & $0.816\pm 0.033$ & $-1.182\pm 0.020$ & 12   \\
    NGC~643      & 01:35:01.0 & $-$75:33:23 & ---           & ---   & $172.0\pm 1.9$ & 11     & $1.259\pm 0.099$ & $-1.301\pm 0.034$ & 12   \\
    Lindsay~112  & 01:36:00.3 & $-$75:27:28 & ---           & ---   & $172.2\pm 2.3$ & 9      & $1.133\pm 0.058$ & $-0.971\pm 0.043$ & 12   \\
    HW~84        & 01:41:41.6 & $-$71:09:39 & ---           & ---   & $135.6\pm 1.5$ & 11     & $1.210\pm 0.034$ & $-1.201\pm 0.059$ & 12   \\
    HW~86        & 01:42:23.3 & $-$74:10:28 & ---           & ---   & $143.8\pm 1.6$ & 11     & $1.191\pm 0.112$ & $-1.276\pm 0.154$ & 12   \\
    Lindsay~113  & 01:49:30.3 & $-$73:43:40 & $52.4\pm 1.7$ & 1     & $171.8\pm 4.5$ & 8,9    & $1.287\pm 0.033$ & $-1.221\pm 0.022$ & 12   \\
    Lindsay~116  & 01:55:34.5 & $-$77:39:15 & ---           & ---   & ---            & ---    & ---              & ---               & ---   \\
    NGC~796      & 01:56:44.6 & $-$74:13:10 & $60.3^{+2.7}_{-2.4}$ & 5     & ---            & ---    & ---              & ---               & ---   \\
    \noalign{\smallskip}
    \hline
\end{tabular}
\end{center}
\raggedright
References: 
($1$) \cite{crowl+01}; 
($2$) \cite{glatt+08};
($3$) \cite{dias+14};
($4$) \cite{dias+16};
($5$) \cite{maia+19};
($6$) \cite{milone+18};
($7$) \cite{goudfrooij+14};
($8$) \cite{dacosta+98};
($9$) \cite{parisi+15};
($10$) \cite{song+21};
($11$) \cite{parisi+09};
($12$) \cite{piatti21}.\\
Notes: (*) Reference \#6 does not provide uncertainties, therefore we assigned 3.0~kpc.
\end{table*}

\section{Comparison of observed and simulated kinematics}
\label{app:kinematics}

\begin{figure*}
    \centering
    \includegraphics[width=0.9\textwidth]{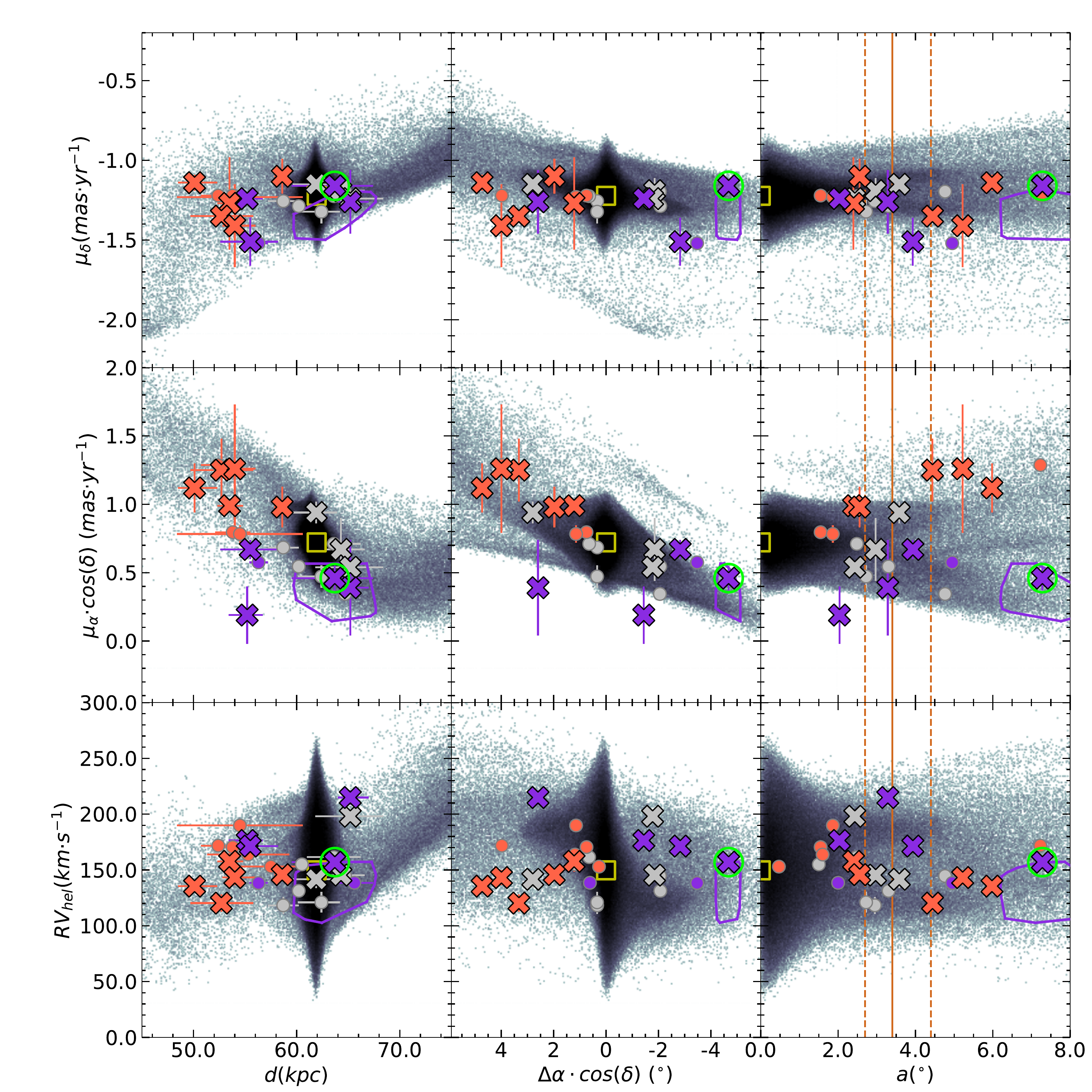}
    \caption{Same as Fig.\ref{fig:3Dmov} but with the simulated particles showing the disc component (gas).}
    \label{fig:3Dmovdisc}
\end{figure*}

\begin{figure*}
    \centering
    \includegraphics[width=0.9\textwidth]{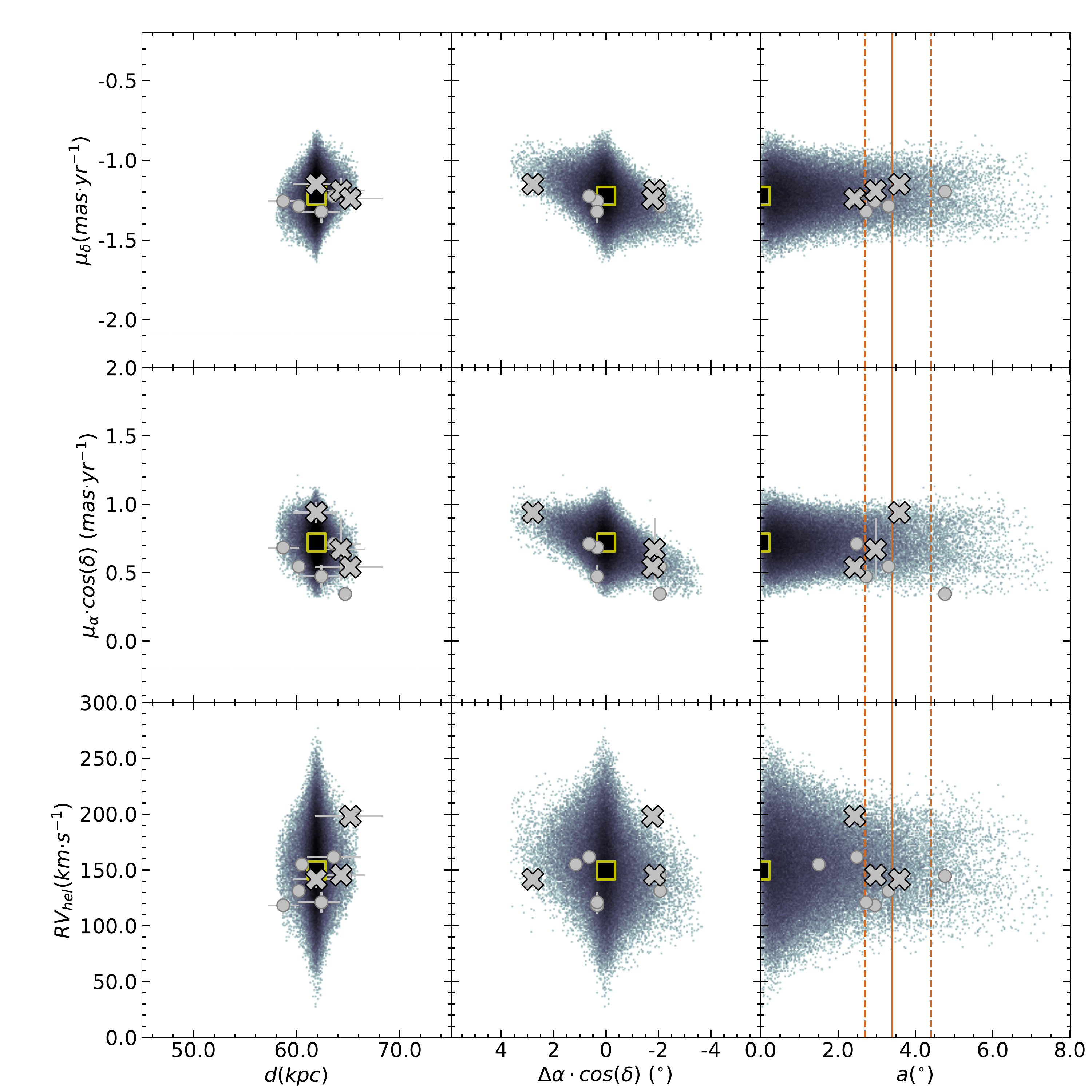}
    \caption{Same as Fig.\ref{fig:3Dmov} but with the simulated particles and clusters showing only the particles within a 3D sphere in Cartesian coordinates with the putative SMC tidal radius.}
    \label{fig:3Dmovinner}
\end{figure*}

\begin{figure*}
    \centering
    \includegraphics[width=0.9\textwidth]{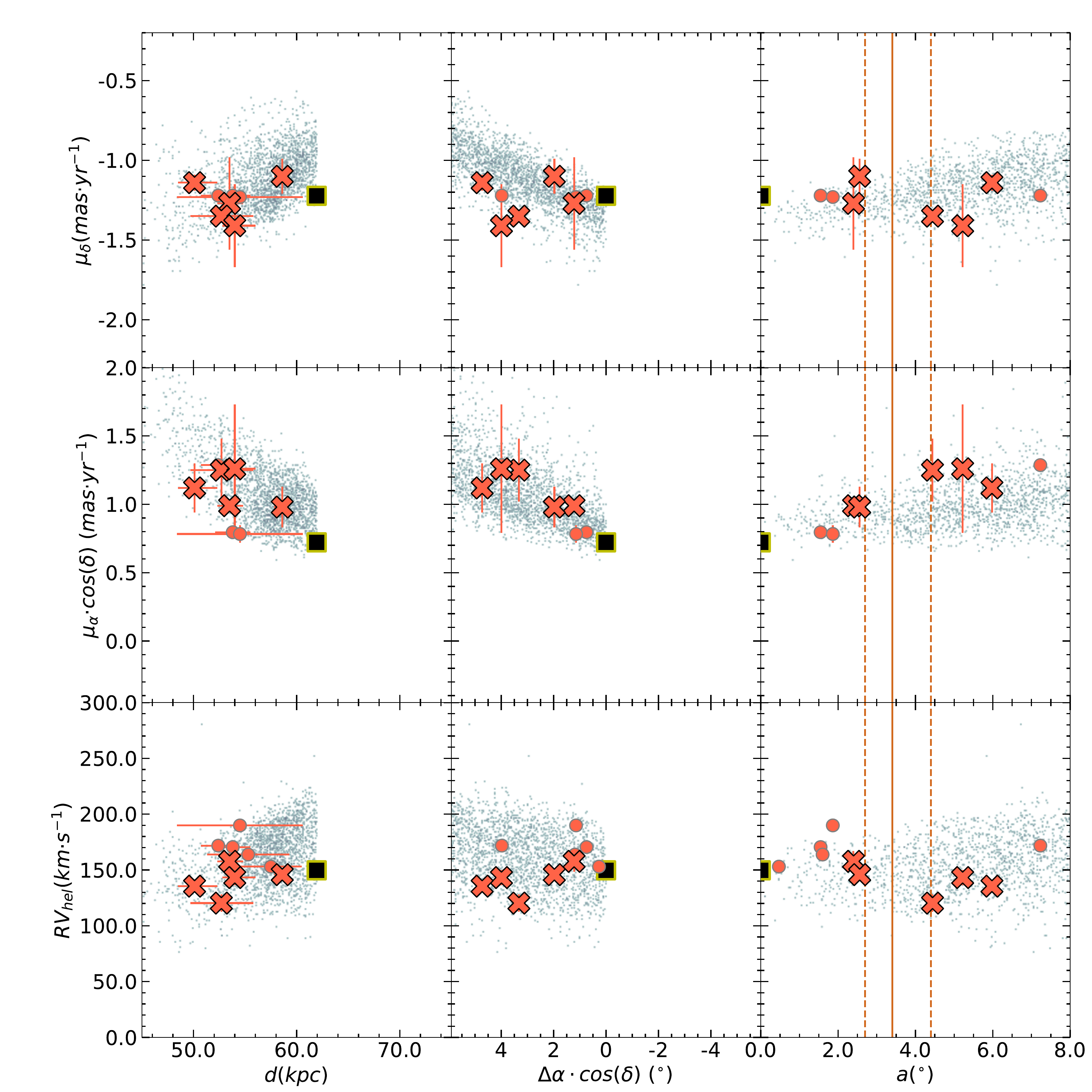}
    \caption{Same as Fig.\ref{fig:3Dmov} but with the simulated particles and clusters showing only the particles outside a 3D sphere in Cartesian coordinates with the putative SMC tidal radius and located on the foreground towards East, i.e., the Bridge.}
    \label{fig:3Dmovfore}
\end{figure*}

\begin{figure*}
    \centering
    \includegraphics[width=0.9\textwidth]{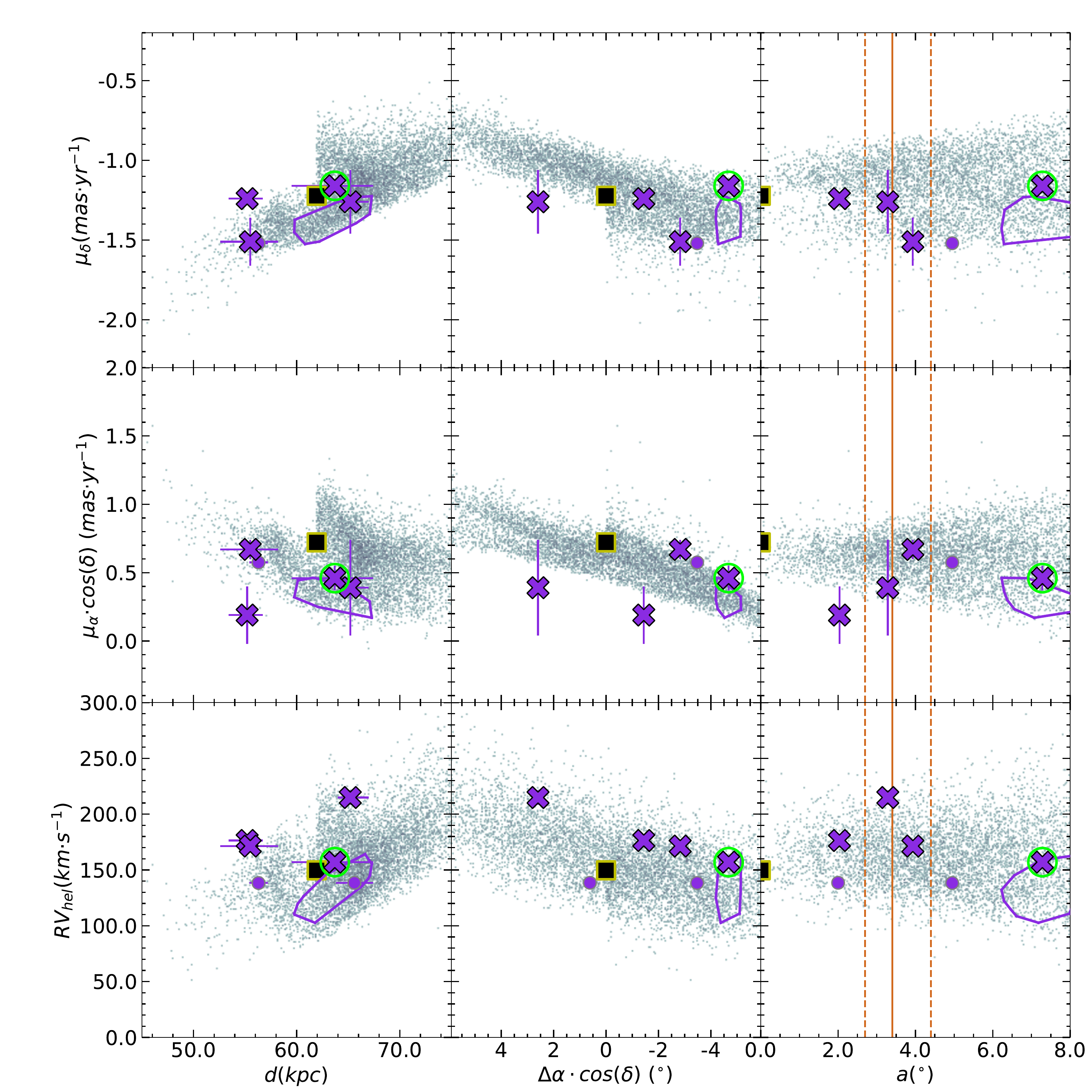}
    \caption{Same as Fig.\ref{fig:3Dmov} but with the simulated particles and clusters showing only the particles outside a 3D sphere in Cartesian coordinates with the putative SMC tidal radius and located on the background, joined with the outer particles located on the foreground towards West, i.e., the Counter-bridge.}
    \label{fig:3Dmovback}
\end{figure*}

We present extra figures as a complement to Fig. \ref{fig:3Dmov} to compare our observational results with the simulations by \cite{diaz+12} in more detail. Fig. \ref{fig:3Dmovdisc} shows the equivalent of Fig. \ref{fig:3Dmov} but for simulated disc (gas) particles. The trends in the simulation points are much better defined and extended with gas. Figures \ref{fig:3Dmovinner} to \ref{fig:3Dmovback} shows simulated spheroid (stars) particles and star clusters region by region, namely, everything within the SMC putative tidal radius, everything outside the putative SMC tidal radius that belongs to the Bridge and Counter-bridge. These figures are useful to make a clear comparison between observations and simulations because the simulated particles from different regions overlap with each other in these panels.

\bsp	
\label{lastpage}
\end{document}